\documentclass[aip,nofootinbib,preprint,pof]{revtex4-1}
\usepackage[T1]{fontenc}
\usepackage{mathptmx}
\usepackage{etoolbox}

\normalsize  
\usepackage{booktabs}
\usepackage{graphicx}
\usepackage{xcolor}
\usepackage{multirow}
\usepackage{bm}
\usepackage{soul}
\usepackage{yhmath}
\usepackage{easyReview}
\usepackage{array}
\usepackage{mathtools}
\usepackage{amsmath,amssymb,amsfonts,amsthm,bm}
\usepackage{hyperref}
\usepackage{subfigure}

\usepackage{xcolor}
\usepackage{fancyhdr}

\makeatletter
\def\@email#1#2{%
	\endgroup
	\patchcmd{\titleblock@produce}
	{\frontmatter@RRAPformat}
	{\frontmatter@RRAPformat{\produce@RRAP{*#1\href{mailto:#2}{#2}}}\frontmatter@RRAPformat}
	{}{}
}%
\makeatother

\fancypagestyle{plain}{%
	\fancyhf{}%
	\fancyhead[C]{\textcolor{red}{Submitted Manuscript: This version has been submitted to Physics of Fluids journal.}}%
}

\begin{document}
	
	
	\title{Super-resolution of turbulent velocity fields in two-way coupled particle-laden flows} 
	
	

	\author{Ali Shamooni}
	\email[]{ali.shamooni@irst.uni-stuttgart.de}
	\affiliation{Institut for reactive flows (IRST), University of Stuttgart, Pfaffenwaldring 31, 70569 Stuttgart, Germany.}

	\author{Ruyue Cheng}
	\affiliation{Institut for reactive flows (IRST), University of Stuttgart, Pfaffenwaldring 31, 70569 Stuttgart, Germany.}
	
	\author{Thorsten Zirwes}
	\affiliation{Institut for reactive flows (IRST), University of Stuttgart, Pfaffenwaldring 31, 70569 Stuttgart, Germany.}

	\author{Hesam Tofighian}
	\affiliation{Scientific Computing Center, Department of Scientific Computing and Mathematics, Karlsruhe Institute of Technology, 76131 Karlsruhe, Germany.}

%

	\author{Oliver T. Stein}
	\affiliation{Engler-Bunte-Institut, Simulation of Reacting Thermo-Fluid Systems, Karlsruhe Institute of Technology, Engler-Bunte-Ring 7, 76131 Karlsruhe, Germany.}
	
	\author{Andreas Kronenburg}
	\affiliation{Institut for reactive flows (IRST), University of Stuttgart, Pfaffenwaldring 31, 70569 Stuttgart, Germany.}

	
	
	\date{\today}
	
	\begin{abstract}
		
		This paper introduces a  deep learning-based super-resolution (SR) framework specifically developed for accurately reconstructing high-resolution  velocity fields in two-way coupled particle-laden turbulent flows. 
		Leveraging conditional generative adversarial networks (cGANs), 
		the generator network architecture incorporates explicit conditioning on physical parameters, such as effective particle mass density and subgrid kinetic energy, while the discriminator network is conditioned on low-resolution data as well as high-frequency content of the input data.  
		High-fidelity direct numerical simulation (DNS) datasets, covering a range of particle Stokes numbers, particle mass loadings,  and carrier gas turbulence regimes, including forced- and decaying-turbulence,  serve as training and testing datasets. 
		Extensive validation studies, including detailed analyses of energy spectra, probability density functions (PDFs), vorticity distributions, and wavelet-based decomposition demonstrate the model’s accuracy and  generalization capabilities across different  particle parameters. 
		{The results show that the network utilizes particle  data, mainly in the reconstruction of high-frequency details modulated by particles. }
		Additionally, systematic assessment of the model’s performance in capturing previously unseen flow regimes further validates its predictive capabilities.
		
	\end{abstract}
	
	
	\maketitle 
	\thispagestyle{plain}

\section{Introduction}
\label{sec:2}

Particle-laden turbulent flows are prevalent in numerous applications such as spray and solid fuel combustion. 
The accurate prediction and comprehensive understanding of these complex flows require capturing the multi-scale interactions occurring between the fluid and dispersed particles. 
Direct numerical simulation (DNS) is a powerful numerical method that resolves all turbulent length and time scales without requiring a turbulence model closure. 
However, due to its extensive computational cost, DNS is typically limited to relatively simple and small-scale problems. 
In contrast, large-eddy simulation (LES) addresses these computational limitations by explicitly resolving the large-scale flow structures while employing subgrid-scale (SGS) models to account for smaller turbulent scales~\cite{SagautP2006_BOOK}. 

Deconvolution and scale-similarity models~\cite{StolzS1999_PF,ShotorbanB2005_PF,	DomingoP2015_PCI,DomingoP2017_CF,BassenneM2019_IJMF,OberleD2020_PF,WangQ2017_CF,WangQ2019_CFa,WangQ2019_CF,BARDINAJ1980_CONF,DesJardinP1998_PF,ShamooniA2020_FTCa,ShamooniA2020_FTC} have been  developed to reconstruct the under-resolved and unresolved variables in LES.  Approximate deconvolution models  rely on fundamental mathematical arguments to obtain an approximate inverse to the LES filtering operator such that the unresolved terms in the governing equations can be computed explicitly from the deconvolved variables~\cite{StolzS1999_PF,ShotorbanB2005_PF,	DomingoP2015_PCI,DomingoP2017_CF,BassenneM2019_IJMF,OberleD2020_PF,WangQ2017_CF,WangQ2019_CFa}. Scale-similarity models leverage the resemblance between resolved and subgrid-scale structures, assuming that the smallest resolved scales are statistically similar to the largest unresolved ones~\cite{BARDINAJ1980_CONF,DesJardinP1998_PF,ShamooniA2020_FTCa,ShamooniA2020_FTC}. 
However, an issue still exists: subgrid information is lost in LES and  approximate reconstructions 
may violate physics such that classical assumptions, e.g. scale similarity need to be invoked to ensure physically realizable solutions~\cite{WangQ2019_CF}. 
{Furthermore, despite its computational efficiency, with few exceptions~\cite{ShotorbanB2005_PF,WangQ2019_CF,BassenneM2019_IJMF}, traditional LES closures have been derived for gas-phase flows only. 
The modeling challenges become, however, particularly pronounced in two-way coupled scenarios, where particles not only respond to the fluid motion but also actively modulate the turbulence characteristics by exchanging momentum with the fluid~\cite{FerranteA2003_PF,AbdelsamieA2012_PF,ElghobashiS2019_ARFM}. 
}

Recent advancements in deep learning (DL) methods have shown promising results in tackling turbulence closure and flow reconstruction problems~\cite{BruntonS2020_ARFM,FukamiK2019_JFM,DuraisamyK2021_PRF}.  Data-driven computer-vision models learn the deconvolution operation directly from high-fidelity data. 
In particular, super-resolution (SR) techniques based on DL have demonstrated  significant potential in serving as deconvolution operators to reconstruct small-scale turbulent flow features from low-resolution (LR) input data, and enhancing the resolution and fidelity of turbulent flow simulations obtained from LES~\cite{FukamiK2019_JFM,KimH2020_JFM,BodeM2021_PCI,BodeM2023_PCI,FukamiK2023_TCFD,TofighianH2024_PF,NistaL2025_CF,ChengR2025_PF}.  
Among these techniques, generative adversarial networks (GANs)~\cite{GoodfellowI2020_CA} and conditional GANs (cGANs)~\cite{MirzaM2014_arXiv} have attracted considerable attention due to their capability in the reconstruction of high-frequency details  by using adversarial training processes~\cite{TofighianH2024_PF,NistaL2025_CF}. 
Despite the progress made in applying DL-based SR techniques to single-phase turbulent flows, research on particle-laden turbulence remains relatively unexplored. The complexity of two-way coupled particle-fluid interactions, characterized by momentum exchange and thus turbulence modulation, introduces significant challenges that existing SR methodologies have  to address. Accurately capturing particle-induced modifications to turbulent kinetic energy (TKE) and dissipation rates  demands DL architectures that explicitly integrate physical conditions and particle characteristics into the model. 

In this work, we present an  SR framework specifically developed to address the complexities inherent to two-way coupled particle-laden turbulent flows. The proposed framework follows a conditional GAN architecture~\cite{TofighianH2024_PF} and is designed explicitly to reconstruct turbulence characteristics in the presence of two-way coupled particles  with high fidelity. 
To the best of our knowledge, this study represents the first investigation applying DL-based SR methods to both forced and decaying  particle-laden turbulent conditions.

\section{Methodology}\label{sec:method}

{A deep convolutional neural network (CNN) is designed to map the three velocity components, viz. ${\bm{\xi}} =\{u, v, w\}$ of the carrier phase from Euler-Lagrange simulations of turbulent particle-laden  flows. The mapping is from low resolution, i.e. ${\bm{\xi}_{LR}} \in \mathbb{R}^{N_{HR}/\Delta \times N_{HR}/\Delta}$, to high resolution, i.e. ${\bm{\hat{\xi}}} \in \mathbb{R}^{N_{HR}\times N_{HR}}$, with an  up-sampling ratio of factor $\Delta = 4$ (denoted in the remainder of the paper as 4x). Here,  $N_{HR}$ is the number of grid points at high resolution which is the same as the DNS grid resolution. 
The mapping is performed in such a way that $\bm{\hat{\xi}}\approx {\bm{\xi}_{GT}} \in \mathbb{R}^{N_{HR}\times N_{HR}}$, where ${\bm{\xi}_{GT}}$ is the ground-truth. 
Deep CNNs such as residual-in-residual dense blocks (RRDBs) with GAN training~\cite{GoodfellowI2020_CA,WangX2019_}  are state-of-the-art DL  approaches  to deconvolve turbulent velocity vectors~\cite{KimH2020_JFM,BodeM2021_PCI,ChungW2023_CONF}. } 
In GANs, two competing CNNs are trained, where the \textit{generator} network learns to generate fake outputs resembling real data and the \textit{discriminator} network learns to distinguish between fake and real data which in turn forces the generator to improve its output to keep fooling the discriminator. Eventually, this process reaches an equilibrium in which the generated data is indistinguishable from real data. 
This is  a well established approach for SR of turbulent flows~\cite{KimH2020_JFM, BodeM2021_PCI,NistaL2023_PCI,TofighianH2024_PF}.

\begin{figure}[!ht] 
	\centering
	\subfigure{\label{fig:generator}\includegraphics[width=\textwidth]{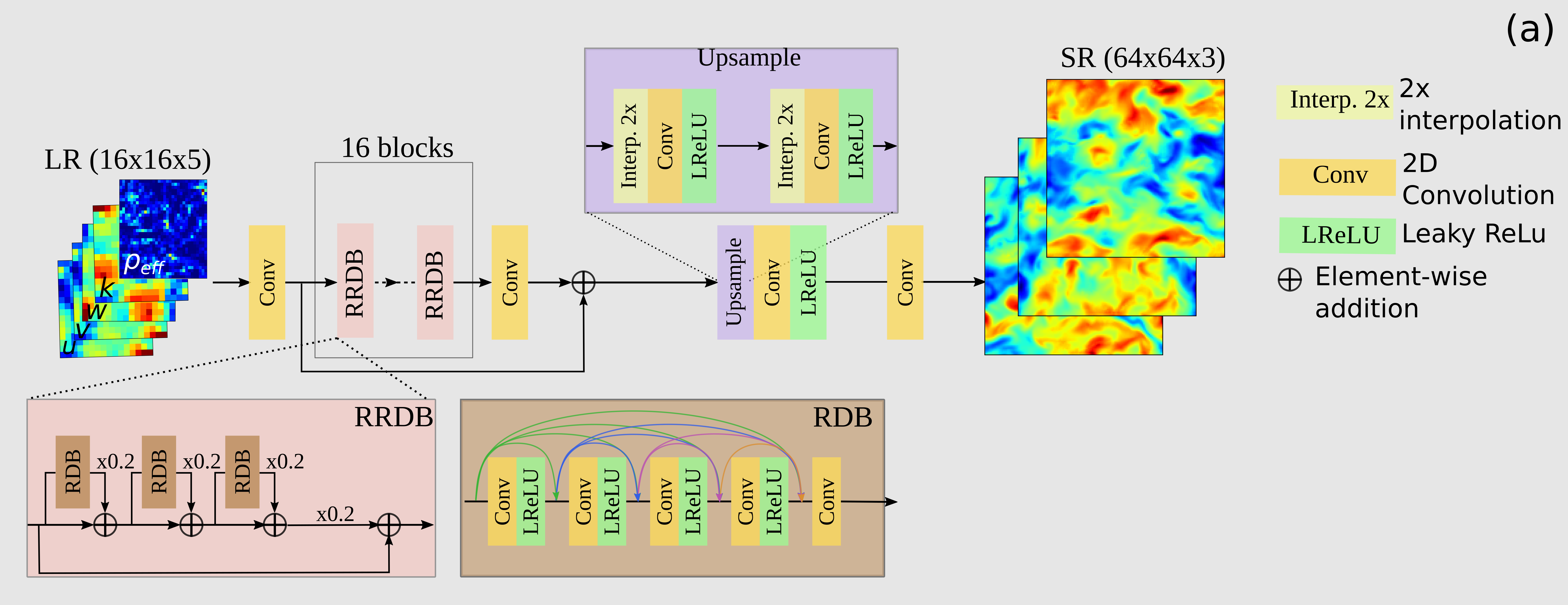}}
	
	\subfigure{\label{fig:disc}\includegraphics[width=\textwidth]{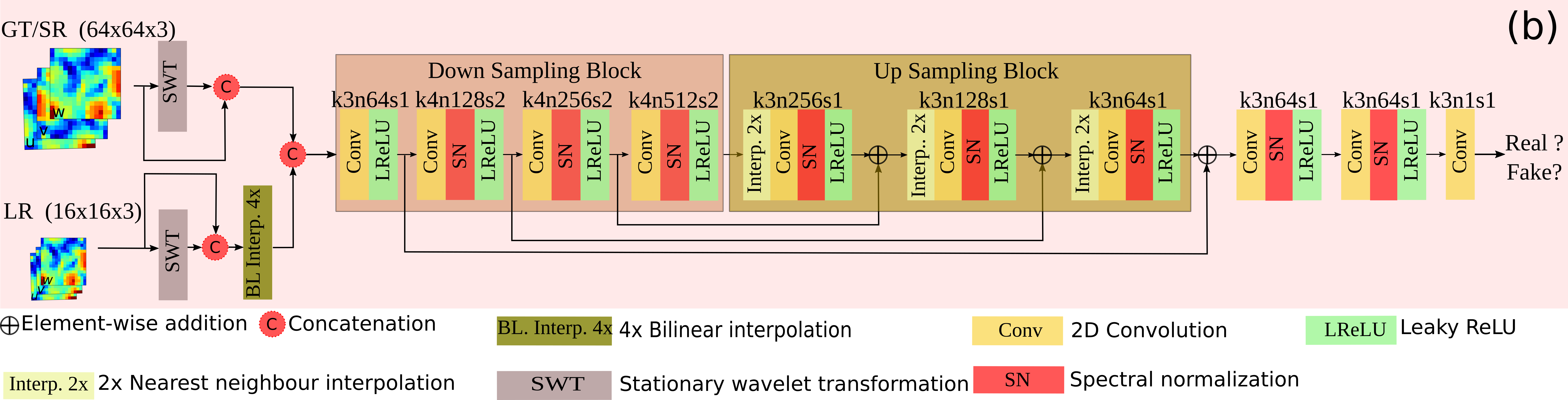}}

	\caption{{The architecture of the generator (\textbf{a}) and the discriminator (\textbf{b}) in the conditional GAN framework.  }}
	\label{fig:RRDB2} 
\end{figure}

The underlying architectures employed in this study are a deep  RRDB architecture for the generator and a U-Net architecture for the discriminator. 
The networks are shown schematically in Fig.\,\ref{fig:RRDB2}. 
The generator $G$ creates SR fields from LR inputs,
\begin{equation}\label{eq:generator}
	\hat{\bm{\xi}} =\{\hat{u}, \hat{v}, \hat{w}\} =G(\bm{\xi}_{{LR}}, k_{sgs}, \rho_{p,eff}),
\end{equation}
where we adopt the conditional GAN~\cite{MirzaM2014_arXiv,TofighianH2024_PF} approach and condition the generator on subgrid kinetic energy,
$k_{sgs} = 0.5\left(\overline{ u_iu_i}-\overline{u_i}\,\overline{u_i}\right)$,
which has been shown to improve the generator's prediction of the   gas fields~\cite{TofighianH2024_PF}. 
Additionally in this study, the local particle information is used as a conditioning parameter 
to let the network learn particle-turbulence interactions. Here we use the effective particle mass density ($\rho_{p,eff}$)  which is defined as
\begin{equation}
	\rho_{p,eff}=N_pm_p/V_{cell}, 
\end{equation}
with  $N_p$ the number of particles residing in each Eulerian coarse cell with a volume of $V_{cell}$, and $m_p$ the mass of each particle in the cell. Since the diameter of particles and carrier gas field density do not change during the simulation in each case in the present study, a single parameter enveloping the total local particle information is used to take into account local particle-turbulence information.

{The generator network has a conventional RRDB~\cite{WangX2019_} architecture with slight modifications. It is composed of 16 RRDBs, each block with 3 residual dense blocks (RDBs), as shown in  Fig.\,\ref{fig:generator}. 
The RDBs are composed of five 2D convolutional layers  with kernel size of 3, stride of 1 and a \textit{circular} padding. All other convolutional layers use the same settings. 
The input channel size of the first  convolutional layer of RDB  is 256 and is increased by a growth rate of 128 after each layer such that the fifth layer of RDB has 768 input and 256 output channels. After each convolutional layer, a leaky ReLu activation function is imposed. 
Skip connections  are employed between RRDBs and within each RDB with a mixing factor of 0.2. 
Two up-sampling layers with fixed channel size of 256 are placed at the end of the network. The up-sampling layers are composed of a convolutional layer followed by nearest neighbor interpolation and the activation function. 
Finally, the last convolutional layer has 3 output channels, i.e. the three velocity components. 
}

An additional network, the discriminator,  is used for the GAN training. The discriminator $D$ decides whether input high-resolution fields are real or fake, 
\begin{equation}\label{eq:criminator}
	D = D\left(\bm{\xi}_{HR}, \bm{\xi}_{LR}, SWT({\bm{\xi}_{HR}})   , SWT({\bm{\xi}_{LR}})\right),
\end{equation}
where HR is high resolution data, i.e. either ground-truth (GT) DNS fields, denoted by $\bm{\xi}_{GT}$, or SR fields ($\hat{\bm{\xi}}$). 
It has been shown that conditioning on LR data improves the stability of the adversarial training~\cite{TofighianH2024_PF}.
The conditioning data, i.e. LR data, are upsampled by the SR factor (4x) using an 
upsampling layer with bilinear interpolation. 
The discriminator is further explicitly conditioned on high-frequency contents of the input data. 
The high-frequency information 
are extracted using stationary wavelet transformation  (SWT)~\cite{SundararajanD2016_BOOK,CotterF2019_THESIS}. This is done to make sure that the discriminator pays attention to the high-frequency content~\cite{WangQ2023_CONF,XuY2025_KS}. Note that this is performed only in the discriminator during the training with no extra cost at the inference because it is not used in the generator. Furthermore, compared to directly enforcing spectral information through spectral loss~\cite{ChengR2025_PF}, the SWT has an advantage of preserving the spatial information while extracting frequency bands. 
{ 
The SWT is a translation-invariant version of the Discrete Wavelet Transform (DWT). Unlike DWT, SWT does not downsample the signal and keeps the output size the same as the input. This makes SWT particularly useful for tasks where shift-invariance is important such as turbulence. }
Haar wavelets~\cite{SundararajanD2016_BOOK} have been employed and only the three sub-bands at the second level of transformation have been used to limit the computational costs, viz.
$SWT({\bm{\xi}})=\{LH_2({\bm{\xi}}), HL_2({\bm{\xi}}), HH_2({{\bm{\xi}}})\}$, with $LH_2(\bm{\xi})$, $HL_2(\bm{\xi})$, $HH_2(\bm{\xi})$ the horizontal, vertical and diagonal detail coefficients  at the second level of the SWT~\cite{SundararajanD2016_BOOK}. 
The inputs {of the discriminator} are concatenated and passed through a series of convolution layers as shown in  Fig.\,\ref{fig:disc}. Note that in Fig.\,\ref{fig:disc}, $k$, $n$, $s$ stand for kernel size, number of channels and the stride of the convolutional layers.  
The  internal architecture is   adopted from~\cite{WangX2021_CONF}. It is a U-Net discriminator with spectral normalization regularization~\cite{MiyatoT2018_6ICLRI2-CTP,WangX2021_CONF} which has an improved stability over conventional CNN-based discriminators~\cite{WangX2019_}. 
The output of \(D\) has a similar resolution as HR and each pixel contains information about the decision of the discriminator on that pixel. 

The training procedure  is carried out by simultaneous minimization of the generator and discriminator loss functions. The generator's loss function  is composed of pixel-wise, gradient, and adversarial losses, viz.  
	\begin{equation}\label{eq:Lgan}
	\mathcal{L}_{G} = \beta_{p} \mathcal{L}_{{pixel}} + \beta_{g} \mathcal{L}_{grad} + \beta_{a} \mathcal{L}_{AD}^{G},
\end{equation}
{where  $ \mathcal{L}_{{pixel}}$ is the pixel-wise 1-norm distance between GT and SR data, viz.
	\begin{equation}
		\mathcal{L}_{{pixel}} =
		\mathbb{E}_{(\bm{\xi}_{LR},\,\bm{\xi}_{GT}) \sim p_{{data}}}
		\left[
		\frac{1}{N}
		\left\|\, \hat{\bm{\xi}} - \bm{\xi}_{GT} \,\right\|_1
		\right]. 
	\end{equation}
	Here, $N$ is the total number of grid points times the number of components (channels)  per sample. 
	Here, $\mathbb{E}_{x\sim p_{x}}[f(x)]$ denotes the expected value  of a function $f(x)$ when $x$ is sampled from the data with a distribution $p_{x}$. Thus, 
	$(x,y)\sim p_{data}$ denotes sampling paired $(x,y)$ data from the empirical data distribution. Since all data pairs are sampled from the training data, we have used the notation $\sim p_{data}$.  The expectation $\mathbb{E}_{(x,y)\sim p_{data}}$ represents an average over all such pairs in the training dataset. 
	The L$_1$ norm, $\left\| . \right\|_1$, denotes the sum of absolute values over all grid points and velocity components. 
	Since the variables have been always sampled from the data, we will drop the  distribution notation in the subsequent equations for the sake of clarity of presentation.  	  
} 

To improve the reconstruction of sharp features and gradients, a gradient loss is used~\cite{ChungW2023_CONF}, 
{
	\begin{equation}
		\mathcal{L}_{{grad}} =
		\mathbb{E}_{(\bm{\xi}_{LR},\,\bm{\xi}_{GT}) }
		\left[
		\frac{1}{N}
		\left\|\, \nabla \hat{\bm{\xi}} - \nabla \bm{\xi}_{GT} \,\right\|_2^2
		\right]. 
	\end{equation}
Here, $\nabla \hat{\bm{\xi}}$ and $\nabla \bm{\xi}_{GT}$ are the spatial gradients (computed with second-order central differences), and the squared L$_2$ norm, $\left\| . \right\|_2^2$, is taken over all grid points and velocity components. Finally, the adversarial loss for the generator is, 
	\begin{equation}
		\mathcal{L}_{{AD}}^{G} =
		- \mathbb{E}_{(\bm{\xi}_{LR}) }
		\left[
		\log D\left(\hat{\bm{\xi}},\, SWT(\hat{\bm{\xi}}),\, \bm{\xi}_{LR},\, SWT(\bm{\xi}_{LR})\right)
		\right]. 
	\end{equation}
}

The discriminator \(D\) is trained  by minimizing the  following adversarial loss function, 
{\begin{align}
	\mathcal{L}_{{AD}}^D 	=
	& - \mathbb{E}_{(\bm{\xi}_{GT}, \bm{\xi}_{LR}) }
	\Big[
	\log D\left(\bm{\xi}_{GT}, SWT({\bm{\xi}_{GT}}), \bm{\xi}_{LR}, SWT({\bm{\xi}_{LR}})\right)
	\Big] \notag \\
	& - \mathbb{E}_{(\bm{\xi}_{LR}) }
	\Big[
	\log \left(1 - D\left(\hat{\bm{\xi}}, SWT({\hat{\bm{\xi}}}), \bm{\xi}_{LR}, SWT({\bm{\xi}_{LR}})\right) \right)
	\Big] \notag \\
	& - \mathbb{E}_{(\bm{\xi}'_{GT}, \bm{\xi}_{LR})}
	\Big[
	\log \left(1 - D\left({\bm{\xi}'_{GT}}, SWT({\bm{\xi}'_{GT}}), \bm{\xi}_{LR}, SWT({\bm{\xi}_{LR}})\right)\right)
	\Big],
\end{align}
where $\bm{\xi}^\prime_{GT}$ is the shuffled version of ground-truth   data~\cite{TofighianH2024_PF} in the training batch. This means that by shuffling, the HR-LR pairs are not matched any more  and
the discriminator is penalized  whenever it identifies $\bm{\xi}^\prime_{GT}$ as ground truth because it is inconsistent with the LR conditions. }

The adversarial training is carried out in one shot with  no pre-training with cropped 64x64, and 16x16 samples for HR and LR data, respectively. The batch size per GPU is set to 16. The input data are randomly flipped and rotated to increase data diversity, thereby reducing over-fitting. Each input channel is normalized by its own mean and standard deviation during the training. 
The Adam optimizer is used with the learning rates of $1\cdot10^{-4}$ and   $1\cdot10^{-5}$
for generator and discriminator, respectively. {The lower learning rate of the discriminator avoids GAN training issues such as mode collapse~\cite{NistaL2022_ASTFEASF2}. 
The hyper-parameters in the total loss function are $\beta_{p}=1$,
$\beta_{g}=3.95 $, $\beta_{a}=1.24$, in such a way that after the first iteration the losses have comparable values within  the same order of  magnitude. 

The super-resolution code is implemented in the basicSR~\cite{WangX2021_PIICCV} framework using PyTorch with DataParallel technique. Two separate trainings of the proposed network were performed, one using particle-free data, and another one using combined particle-free and particle-laden data. The datasets are discussed in Sec.\,\ref{sec:datasets}. 
The trainings are performed on one GPU node with four A100 GPUs using the \textit{Horeka} platform at SCC at KIT. 
}

\section{Data generation}\label{sec:dataGen}
\subsection{Direct numerical simulation}\label{sec:DNS}

\subsubsection{Governing equations}
{Forced homogeneous isotropic turbulence (HIT) and decaying-turbulence DNS were carried out to build the  training/testing databases. Both particle-free and particle-laden scenarios were considered. }   
The  governing equations for particle-free cases are conventional non-reacting incompressible Navier-Stokes equations with a linear forcing term which will be defined later. The particle-laden cases are simulated in an Euler-Lagrange  framework with point-particle source assumption where Lagrangian particles are carried by the carrier gas and exchange momentum with the gas. The governing equations read
\begin{align}
	\frac{\partial{u_i} }{\partial{x_i}}                                           & = 0,  \label{eq:mass}                                                                                                                             \\
	\frac{\partial{ u_i} }{\partial{t}} + \frac{\partial{ u_iu_j} }{\partial{x_j}} & = -\frac{1}{\rho}\frac{\partial{p} }{\partial{x_i}} + \frac{\partial{\tau_{ij}} }{\partial{x_j}} +  \dot{S}_{u_i}^{p} + f_{u_i}, \label{eq:momentum}
\end{align}
where $\rho$ is the density (kept constant), $p$ is the pressure and $\tau_{ij}$ the stress term, $\tau_{ij}=2\nu\left(S_{ij}-\frac{1}{3}S_{ii}\right)$, with $S_{ij}$ the strain-rate tensor, $	S_{ij} = 0.5\left(\frac{\partial{u_i} }{\partial{x_j}} + \frac{\partial{u_j} }{\partial{x_i}} \right)$. 
In Eq.\,\eqref{eq:momentum} , the last term is the linear forcing term which is used in forced-turbulence simulations to maintain the turbulence level in the carrier gas. In this study, the forcing method by Carrol et al.~\cite{CarrollP2013_PFa} has been used with the forcing term, $f_{u_i} = A \left(\frac{k_T}{k}\right) \left(u_i-\overline{u}_i\right)$,
where $A$ is defined by $ A    = \frac{1}{45}\frac{\nu Re_{\lambda,T}^2}{l_{t,T}^2}$,
with  $k$ 
being the turbulent kinetic energy. 
 $\overline{u}_i$ is the Reynolds-averaged velocity over the whole volume. $k_T$ is the  target kinetic energy which is  defined by the user-defined target Taylor Reynolds number, $Re_{\lambda,T}$, and the user-defined target turbulent length scale $l_{t,T}$. Note that the turbulent length scale is defined as $l_{t}={u_{rms}}^3/\varepsilon$ with $\varepsilon$ being the turbulent  dissipation rate. 

In Eq.\,\eqref{eq:momentum}  $\dot{S}_{u_i}^{p}$ is a momentum source term due to the drag force on particles, viz.

\begin{equation}
	\dot{S}_{u_i}^{p}  =  - \frac{1}{V_{cell}} \sum_{np=1}^{N_p} \left[m^{p} \frac{d(u_{i}^{p})}{dt}\right]_{np},
\end{equation}
where $V_{cell}$ is the volume of the computational cell where the particle resides in. The Lagrangian equations govern the particle position in $i^{th}$ direction, i.e.  $x_{i}^{p}$, and velocity, i.e.    $u_{i}^{p}$ , which read
\begin{align}
	\frac{dx_{i}^{p}}{dt} & ={u_{i}^{p}},  \label{eq:particle_vel}                                                      \\
	\frac{du_{i}^{p}}{dt} & =\frac{{u}_{i}(\mathbf{x}^{p})-u^{p}_{i}}{\tau^{p}}, \label{eq:particle_pmomentum}
\end{align}
%
with ${u}_{i}(\mathbf{x}^{p})$ being  the Eulerian velocity in $i^{th}$ direction interpolated at the particle position. In this study, the particle density is considered  to be much larger than the fluid density ($\rho_p/\rho  \gg 1$). Also it is assumed that only the drag force is acting on the  particle~\cite{MaxeyM1983_PF}. 
In Eq.\,\eqref{eq:particle_pmomentum} $\tau^{p}$ is the  particle relaxation time obtained by, $\frac{1}{\tau^{p}} = \frac{18 \mu(\mathbf{x}^{p})}{\rho^{p} {\left(d^{p}\right)}^2} \left(1+ \frac{{({\mathrm{Re}^{p}})}^{2/3}}{6}\right)$%
with Re$^{p}$ being the particle Reynolds number, which is calculated as, $\mathrm{Re}^{p} = \frac{\rho d^{p}}{\mu} \mid \textbf{u} - \textbf{u}^{p} \mid $,
with $d^{p}$ the  particle diameter.

\subsubsection{Numerical setups}

The governing equations are solved in a cube  with periodic boundary conditions. The details of the base DNS setup and the \textit{target}  turbulent statistics which governs the forcing term in the momentum equation are presented in Tab.\,\ref{tab::OF128setup}.
\begin{table}[!htb]
	\centering
	\caption{{Specifications of the general DNS setups and forcing parameters, also used for Case0. Subscript $T$ stands for \textit{target} values. } }
	\label{tab::OF128setup}
	{
		\begin{tabular}{p{0.5\textwidth}|>{\centering\arraybackslash}p{0.3\textwidth}}
			\hline
			Property                                                                 & Value                                            \\
			\hline
			Target Taylor Reynolds number ${(\mathit{Re}_{\lambda,T})}$              & 52                                               \\
			Gas density $(\rho) [\mathrm{kg/m^3}]$                                   & 0.703                                        \\
			Gas kinematic viscosity $(\nu) [\mathrm{m^2/s}]$                         & $3.96\cdot10^{-5} $                              \\
			\hline
			Domain length  $\left(L_x \times L_y \times L_z \right) [m]$             & $\left(0.0256 \times0.0256 \times 0.0256\right)$ \\
			Computational cell number  $\left(N_x \times N_y \times N_z \right) $                  & $\left(128\times 128 \times 128\right)$          \\
			Target turbulent dissipation rate $({\varepsilon_T}) [\mathrm{m^2/s^3}]$ & $623$                                            \\
			Target turbulent kinetic energy $(k_T) [\mathrm{m^2/s^2}]$               & $3.14$                                           \\
			Target turbulent length scale $({l}_{t,T}) [\mathrm{m}]$                   & 0.0048                                           \\
			Target turbulent time scale   	${\tau_{t,T}} [\mathrm{s}]$                & $0.0033$                                         \\
			Target Kolmogorov length scale   	${L_{\eta,T}} [\mathrm{\mu m}]$         & 100                                              \\
			Target Kolmogorov time scale   	${\tau_{\eta,T}} [\mathrm{s}]$            & $2.5\cdot 10^{-4}$                               \\
			 
			$\Delta t/{\tau_{\eta,T }}$                                              & $\approx 2.5\cdot 10^{-3}$                       \\
			
			\hline
		\end{tabular}
	}
	
\end{table}
These settings have been used for "Case0" which is the particle-free  forced-turbulence case.
The gas velocity field is initialized by a field conforming to  von K\'arm\'an–Pao spectrum~\cite{BaillyC1999_CONF,SaadT2017_AJ}.  An in-house code developed in OpenFOAMv2012 with second order schemes for time derivative, convection and diffusion terms, has been employed.
All cases reach statistically stationary conditions at  $t\approx10\tau_{t,T}$ and  are simulated until $t=40\tau_{t,T}$ to create enough data samples. 
The time step of the simulation is set constant (cf. Tab.\,\ref{tab::OF128setup}) which keeps the Courant–Friedrichs–Lewy number below 0.05. Such a low time step avoids bursts of energy injection with  the linear forcing schemes. 
The longitudinal integral scale~\cite{PopeS2000_BOOK} of the particle-free case, viz.
	$L_{11} = \int_{0}^{\infty} \frac{1}{ \overline{{u^\prime_{x}}^2} } \overline{u_x \left(\mathbf{x+e_1r}\right)u_x\left(\mathbf{x}\right)} d\mathbf{r}$,
is $L_{11}=0.0021\approx0.082 L_x$ after reaching the stationary condition which ensures that enough large-scale structures fit into the domain. The mesh is fine enough to resolve the smallest turbulent scales based on Pope's criteria~\cite{PopeS2000_BOOK}, viz. $\kappa_{max}\eta\geq1.5$, with  $\kappa_{max}=\pi/\Delta_{DNS}$ the maximum resolvable wave-number in the domain. $\Delta_{DNS}=L_x/N_x$ is the mesh spacing which is kept the same in all DNS cases in this work.

The particle-laden cases in Tab.\,\ref{tab::OF128setup-particle} use similar settings as in  Tab.\,\ref{tab::OF128setup} for the carrier gas field. 
\begin{table}[!htb]
	\centering
	\caption{{Specifications of particle-laden forced-turbulence DNS setups  which are used as  training/testing datasets.  Subscript $T$ stands for \textit{target} values.}}
	\label{tab::OF128setup-particle}

	\resizebox{1\textwidth}{!}
	{

		\begin{tabular}{p{0.32\textwidth}|>{\centering}p{0.07\textwidth}>{\centering}p{0.07\textwidth}>{\centering}p{0.07\textwidth}|>{\centering}p{0.07\textwidth}>{\centering}p{0.07\textwidth}>{\centering}p{0.07\textwidth}|>{\centering}p{0.07\textwidth}>{\centering}p{0.07\textwidth}>{\centering\arraybackslash}p{0.07\textwidth}}
			\hline
			Property                                                         & Case1                     & Case$1_1$                 & Case$1_2$                & Case2  & Case$2_1$ & Case$2_2$ & Case3 & Case$3_1$ & Case$3_2$ \\
			
			\hline
			Target Stokes number based on Kolmogorov scale $({St_{\eta,T}})$ & \multicolumn{3}{c|}{0.6}   & \multicolumn{3}{c|}{1}     & \multicolumn{3}{c}{6}                                                                     \\
			Target Stokes number based on Taylor scale $({St_{\lambda,T}})$  & \multicolumn{3}{c|}{0.1}   & \multicolumn{3}{c|}{0.17}  & \multicolumn{3}{c}{1}                                                                     \\
			Target Stokes number  based on turbulence scale $({St_{t,T}})$    & \multicolumn{3}{c|}{0.045} & \multicolumn{3}{c|}{0.075} & \multicolumn{3}{c}{0.45}                                                                  \\
			Volume ratio $(\phi_{vol}) \times 10^4$                          & $5$                       & $5$                       & $7.5$                    & $5$    & $5$       & $7.5$     & $5$   & $5$       & $7.5$     \\
			Particle density $(\rho_p) [\mathrm{kg/m^3}]$                    & 700                       & 1050                      & {700}                    & 700    & 1050      & {700}     & 700   & 1050      & {700}     \\
			Particle mass loading $(ML)$                                              & 0.49                      & 0.75                      & 0.75                     & 0.49   & 0.75      & 0.75      & 0.49  & 0.75      & 0.75      \\
			Number of particles $(N_p) \times 10^6 $                         & 14                        & 26                        & 21                       & 6.5    & 12        & 10        & 0.5   & 0.8       & 0.6       \\
			Number density $\times 10^{11} [\mathrm{m}^{-3}]$
			& $8.5$                     & $15$                      & $13$
			& $3.9$                     & $7.2$                     & $5.9$                    & $0.27$ & $0.49$    & $0.4$                                     \\
			
			Particle diameter $(d_p) [\mu \mathrm{m}]$                       & 10                        & 8                         & 10                       & $13$   & 10        & 13        & $33$  & 27        & 33        \\

			
			\hline
		\end{tabular}

	}

\end{table}
Lagrangian particles are seeded randomly to the stationary turbulence 
with initial velocities equal to the local Eulerian cells' velocities. 
As can be seen in Tab.\,\ref{tab::OF128setup-particle}, nine particle-laden cases  are generated to account for particle-turbulence interactions in isotropic turbulence  by changing the main contributing factors, i.e.  particle Stokes number and mass loading~\cite{FerranteA2003_PF,AbdelsamieA2012_PF}. 
{Particle Stokes number can be defined using different target flow scales, namely target Stokes numbers based on Kolmogorov scale,  $\mathrm{St}_{\eta,T} = \tau_p / \tau_{\eta,T}$, based on Taylor scale, $\mathrm{St}_{\lambda,T} = \tau_p / \tau_{\lambda,T}$, and based on turbulent large scale, $\mathrm{St}_{t,T} = \tau_p / \tau_{t,T}$, with the values mentioned in Tab.\,\ref{tab::OF128setup-particle}. Note that $\tau_{t,T}=l_{t,T}/u_{rms,T}$  is the turbulent time scale and $\tau_{\lambda,T}=(\frac{\lambda^2}{\varepsilon_T})^{1/3}$ is the Taylor scale time scale. Moreover, $u_{rms,T}=\sqrt{2k_T/3}$ is the target  root mean square velocity. }
Particle-laden cases, i.e. Case1, Case2, and Case3 are generated by keeping particle mass loading constant but using different particle diameters to generate different Stokes numbers, viz. Stokes numbers 0.6, 1, and 6, respectively. 
Additional particle-laden cases are  generated for each Stokes number by changing the particle density to generate different mass loadings. Particle sizes are reduced to keep the Stokes number the same. This resulted in cases labeled by  Case$i_1$ with $i \in \{1,2,3\}$. 
A further variation for each Stokes number (Case$i_2$ with $i \in \{1,2,3\}$) is generated  by keeping particle density and diameter the same as for the Case$i_1$ but changing  particle volume fraction to analyze the effect of particle number density when Stokes number and mass loading are constant.  Table\,\ref{tab::OF128setup-particle_test} lists the specifications for two additional cases, cases 4 and 5 that feature different Stokes numbers compared to the ones in Tab.\,\ref{tab::OF128setup-particle} and are  used as independent test datasets for generalization tests.

\begin{figure}[!ht]
	\centering
	
	{\includegraphics[width=0.49\textwidth]{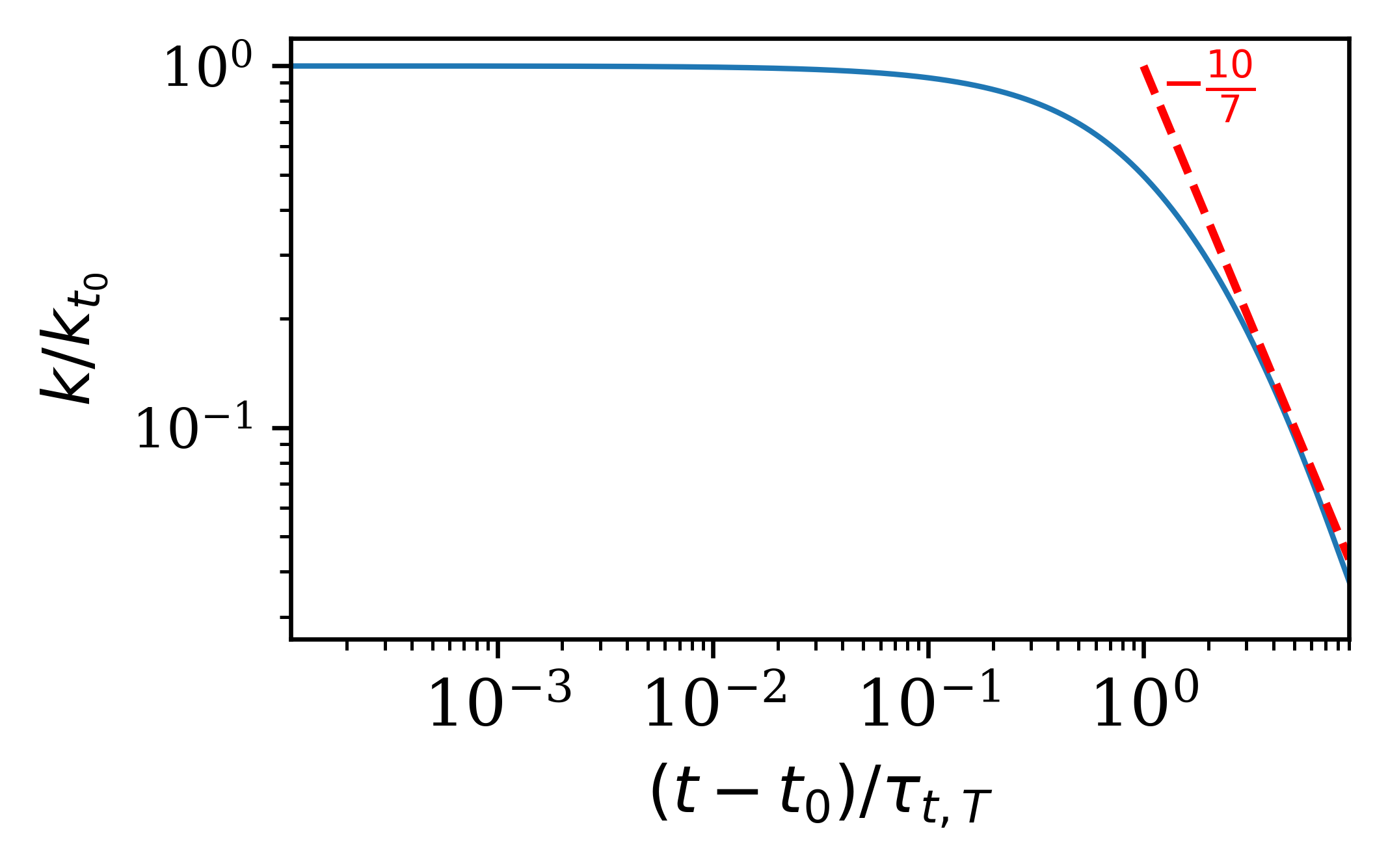}}
	
	\caption{{Temporal evolution of turbulent kinetic energy $k$ in decaying-turbulence mode, normalized by the value at the start time $t_0=40\tau_{t,T}$ vs. normalized time. The dashed line shows the analytical scaling exponent of $-{10}/{7}$ in the self-similar decay regime~\cite{RistorcelliJ2003_PF}. }  }
	\label{fig:self_similar_decay_case0}
\end{figure}

\begin{table}[!htb]
	\centering
	\caption{Specifications of particle-laden forced-turbulence DNS setups which are used as {test} dataset.}
	\label{tab::OF128setup-particle_test}
	{
		
		\begin{tabular}{p{0.72\textwidth}|>{\centering}p{0.07\textwidth}>{\centering\arraybackslash}p{0.07\textwidth}}
			\toprule
			Property                                                         & Case4 & Case$5$  \\

			\midrule
			Target Stokes number based on Kolmogorov scale $({St_{\eta,T}})$ & 3     & 10           \\
			Target Stokes number based on Taylor scale $({St_{\lambda,T}})$  & 0.5   & 1.7         \\
			Target Stokes number  based on turbulence scale $({St_{t,T}})$    & 0.2   & 0.7       \\
			Volume ratio $(\phi_{vol}) \times 10^4$                          & $5$   & $5$     \\
			Particle density $(\rho_p) [\mathrm{kg/m^3}]$                    & 1050  & 1050     \\
			Particle mass loading $(ML)$                                              & 0.75  & 0.75       \\


			
			Number of particles $(N_p)  \times 10^6  $      & 2.3   & 0.38      \\
			Number density $\times 10^{11} [\mathrm{m}^{-3}]$
			& $1.3$ & $0.22$  \\
			
			Particle diameter $(d_p) [\mu \mathrm{m}]$                       & 19    & 34      \\

			

			\bottomrule
		\end{tabular}
	}
	
\end{table}
{
The above-mentioned DNS setups are forced-turbulence setups which mimic particle-turbulence interaction when an external turbulence forcing mechanism exists.
As was mentioned in previous studies~\cite{AbdelsamieA2012_PF},  the physics of particle-turbulence interaction is different in the case of decaying turbulence compared to forced turbulence which will be discussed more  in Sec.\,\ref{sec:dns_features}. 
Hence,  the dataset is expanded by including decaying-turbulence simulations. 
The initial velocity field is generated by letting  Case0 at stationary state decay  over one turbulent time scale, i.e.  from $t_0=40\tau_{t,T}$ to $t_{inj}=t_0+\tau_{t,T}$. Hence, we ensure that the flow field nearly reaches the self-similar decay regime~\cite{RistorcelliJ2003_PF}, viz. $k \propto t^{-10/7}$ (cf. Fig.\,\ref{fig:self_similar_decay_case0}). }
This field is then employed as an initial condition for all decaying-turbulence  particle-laden cases. It should be mentioned that it is still needed for particle-laden cases to develop over one turbulent time scale until $t_{start}=t_{inj}+\tau_{t,T}$ to relax the initial transients after which statistics can be collected and the data can be used for training/testing database generation purposes.

The numerical simulations in this study are carried out on the \textit{Hawk} supercomputer at HLRS.  The maximum  time required for a simulation is related to Case$1_{1}$ which costs about 15,000 CPUh (15 hours on 1024 cores) using AMD EPYC 7742 cores. 

\subsubsection{Database features}\label{sec:dns_features}
\begin{figure}[!ht]
	\centering

	\subfigure{\label{fig:stats128_f0_1_2_3_1}\includegraphics[width=0.49\textwidth,trim=0.4cm 0.4cm 0.4cm 0.4cm, clip]{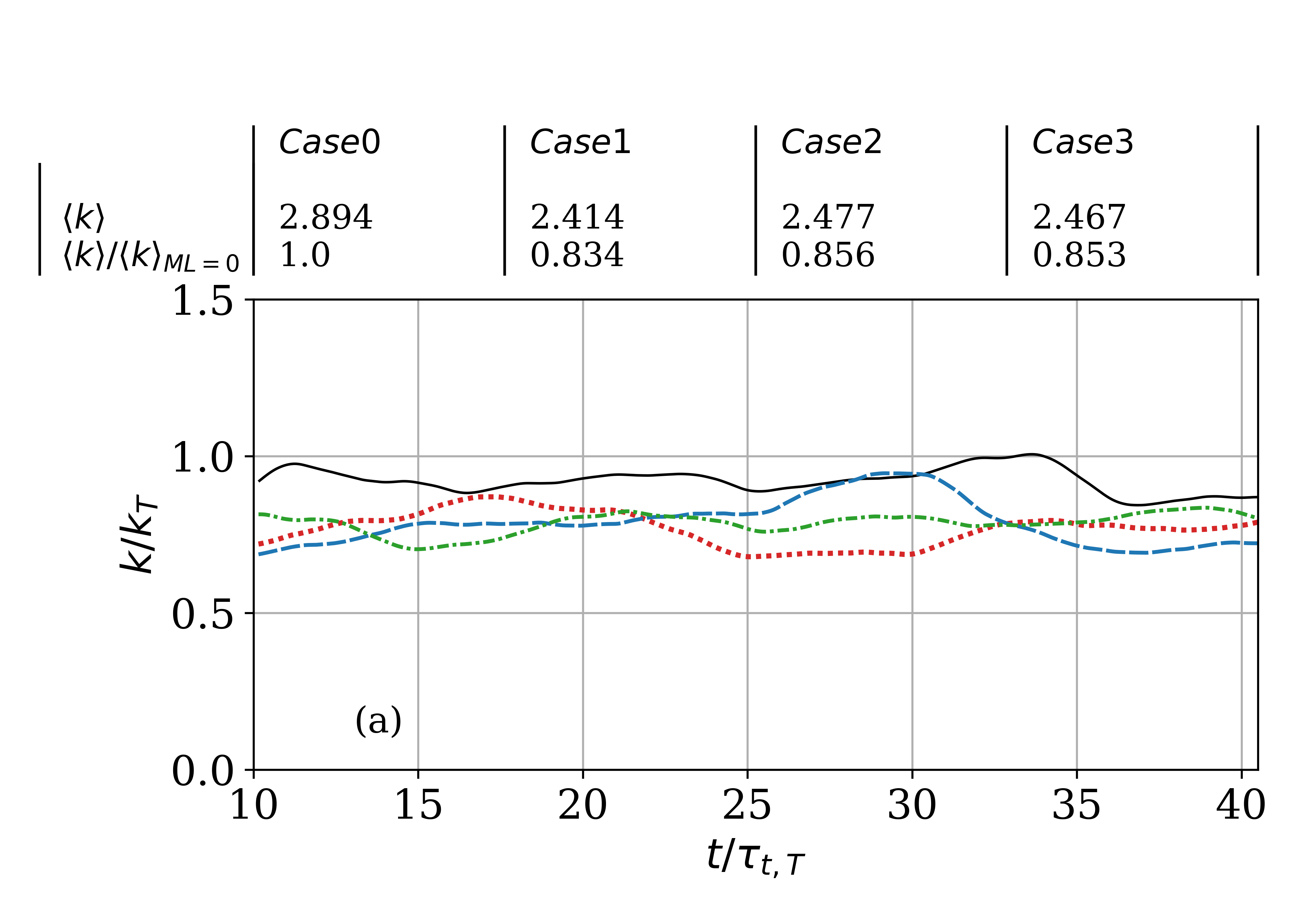}}
	\subfigure{\label{fig:stats128_f0_1_2_3_2}\includegraphics[width=0.49\textwidth,trim=0.4cm 0.4cm 0.4cm 0.4cm, clip]{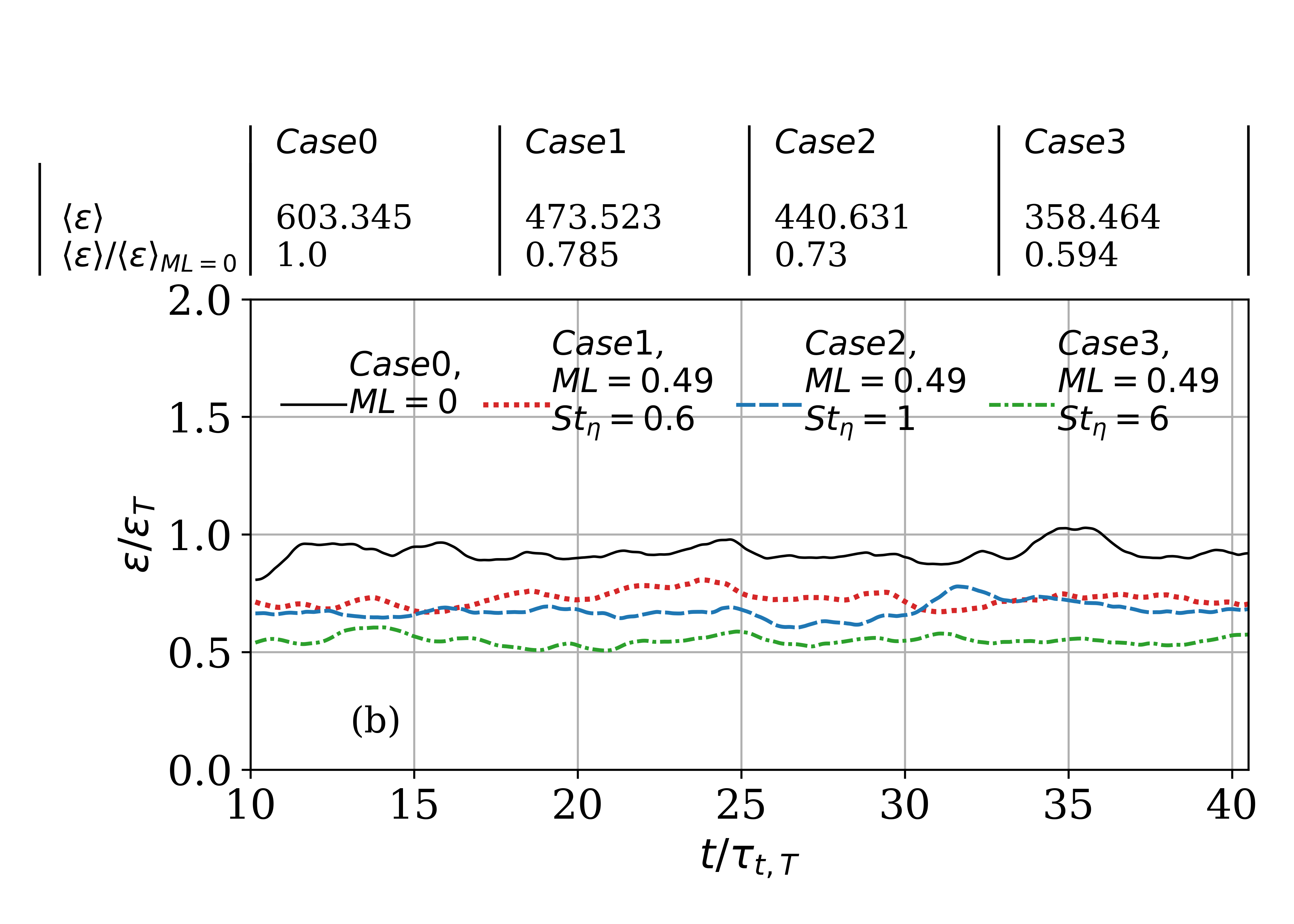}}\\
	\subfigure{\label{fig:stats128_f0_2_21_22_1}\includegraphics[width=0.49\textwidth,trim=0.4cm 0.4cm 0.4cm 0.4cm, clip]{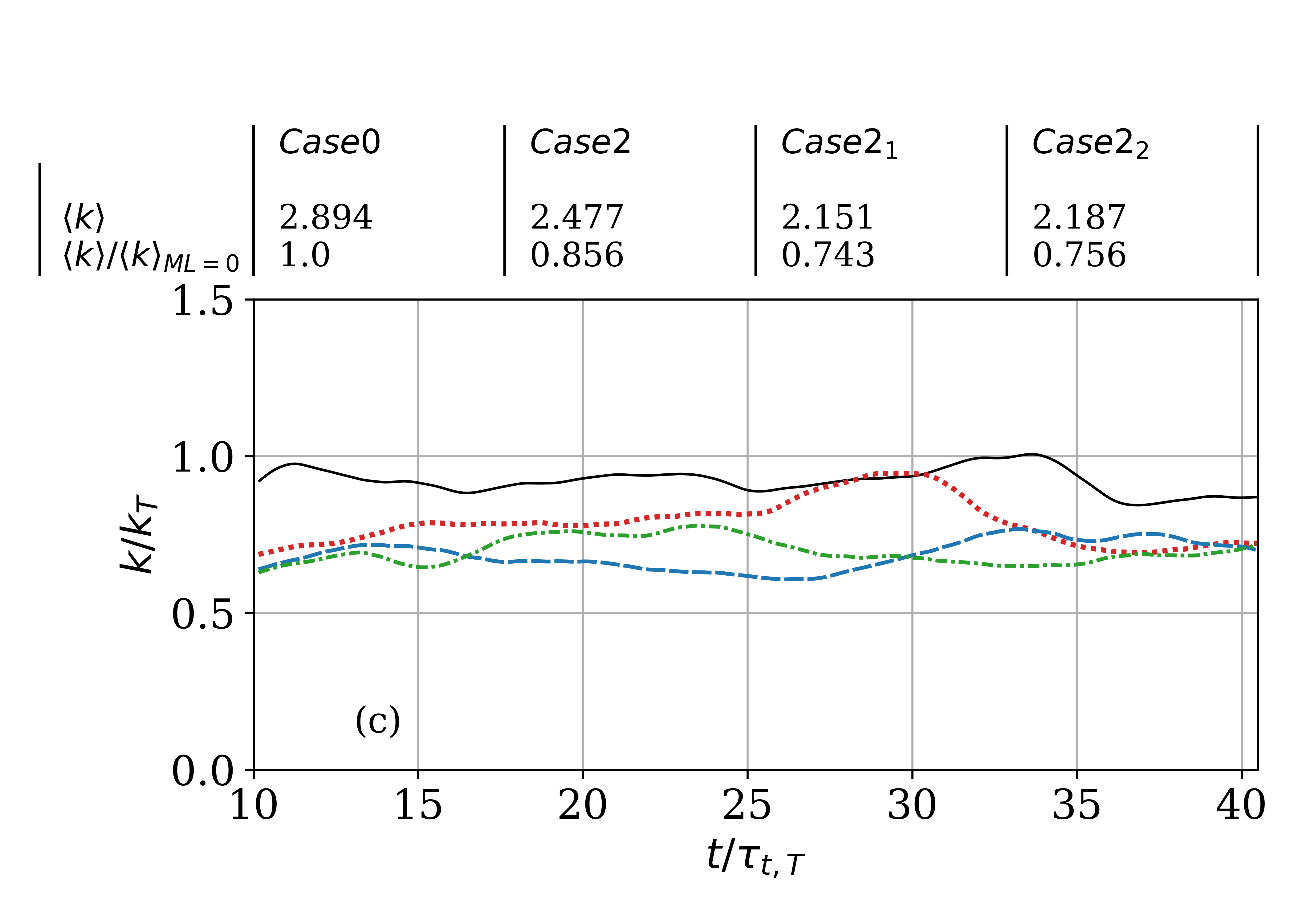}}
	\subfigure{\label{fig:stats128_f0_2_21_22_2}\includegraphics[width=0.49\textwidth,trim=0.4cm 0.4cm 0.4cm 0.4cm, clip]{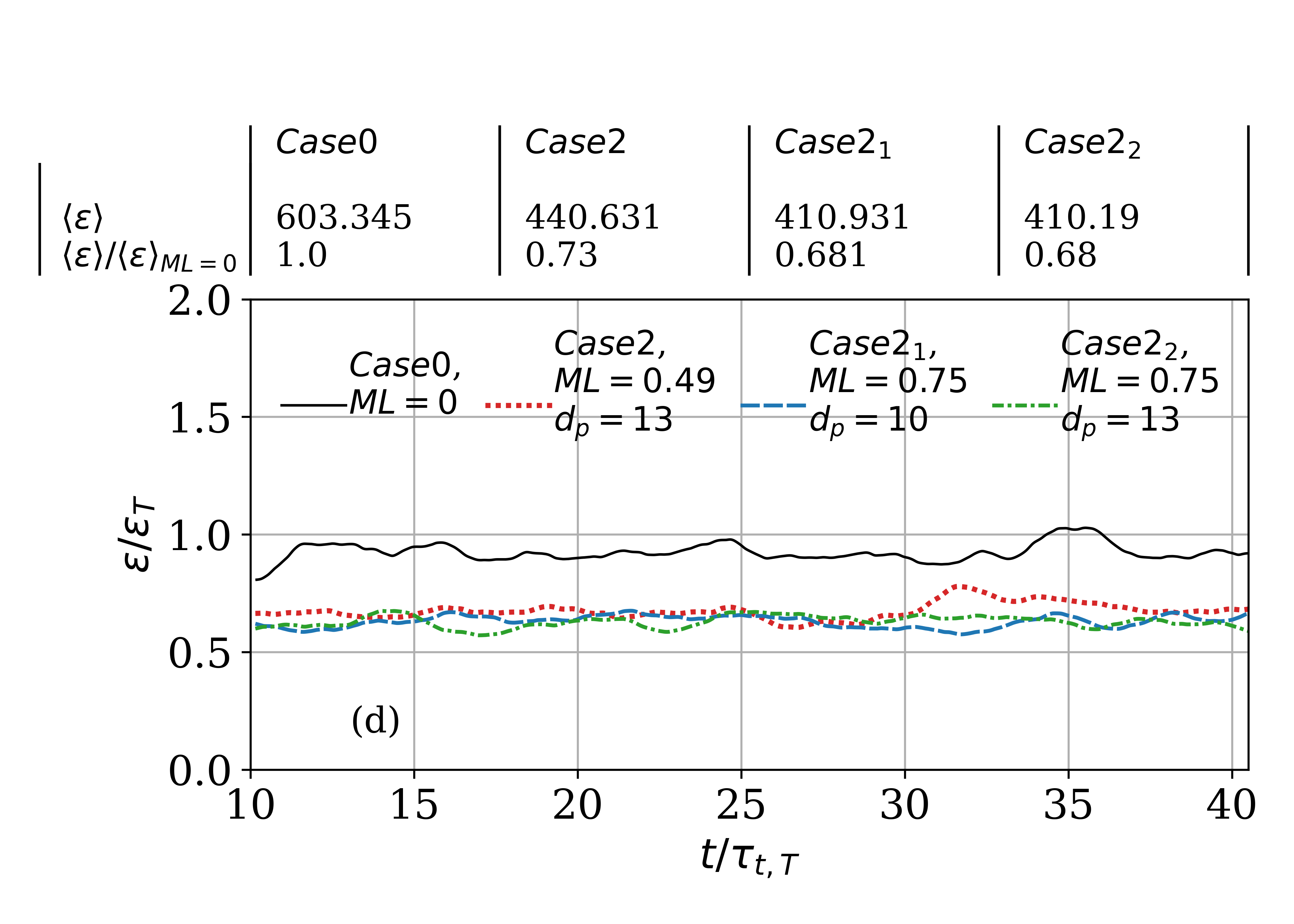}}

	\caption{{Turbulent statistics over time for (\textbf{top}) four  forced-turbulence cases, i.e.  Case0 (particle-free), Case1, Case2, and Case3, and (\textbf{bottom}) Case0 (particle-free), Case2, Case2$_1$, and Case2$_2$. In the figures the time-averaged statistics $\left(\langle . \rangle \right)$ are shown in the top table and compared with the particle-free case, i.e. Case0 with zero particle mass loading ($ML=0$). 
		}
	}
	\label{fig:stats128_fx}
\end{figure}

The DNS databases are designed to include different physics related to particle-turbulence interactions. 
Turbulent statistics over time for different  forced-turbulence cases are shown in Fig.\,\ref{fig:stats128_fx}. Moreover,   the time-averaged statistics $\left(\langle . \rangle \right)$ are shown in the top tables in each sub-figure and compared with the particle-free case, i.e. Case0 with zero particle mass loading ($ML=0$). As can be seen in Figs.\,\ref{fig:stats128_f0_1_2_3_1} and \ref{fig:stats128_f0_1_2_3_2}, by increasing the Stokes number, reduction of $k$ (with a relatively low $St_\eta$ dependency) and $\varepsilon$ (with high $St_\eta$ dependency) are observed. Further, as can be seen in Figs.\,\ref{fig:stats128_f0_2_21_22_1} and \ref{fig:stats128_f0_2_21_22_2}, the mass loading has an effect on turbulence modulation, however, it is much less than the effect of the Stokes number. 

\begin{figure}[!ht]
	\centering
	
	\includegraphics[width=0.85\textwidth,trim=0.3cm 0.3cm 0.3cm 0.3cm, clip]{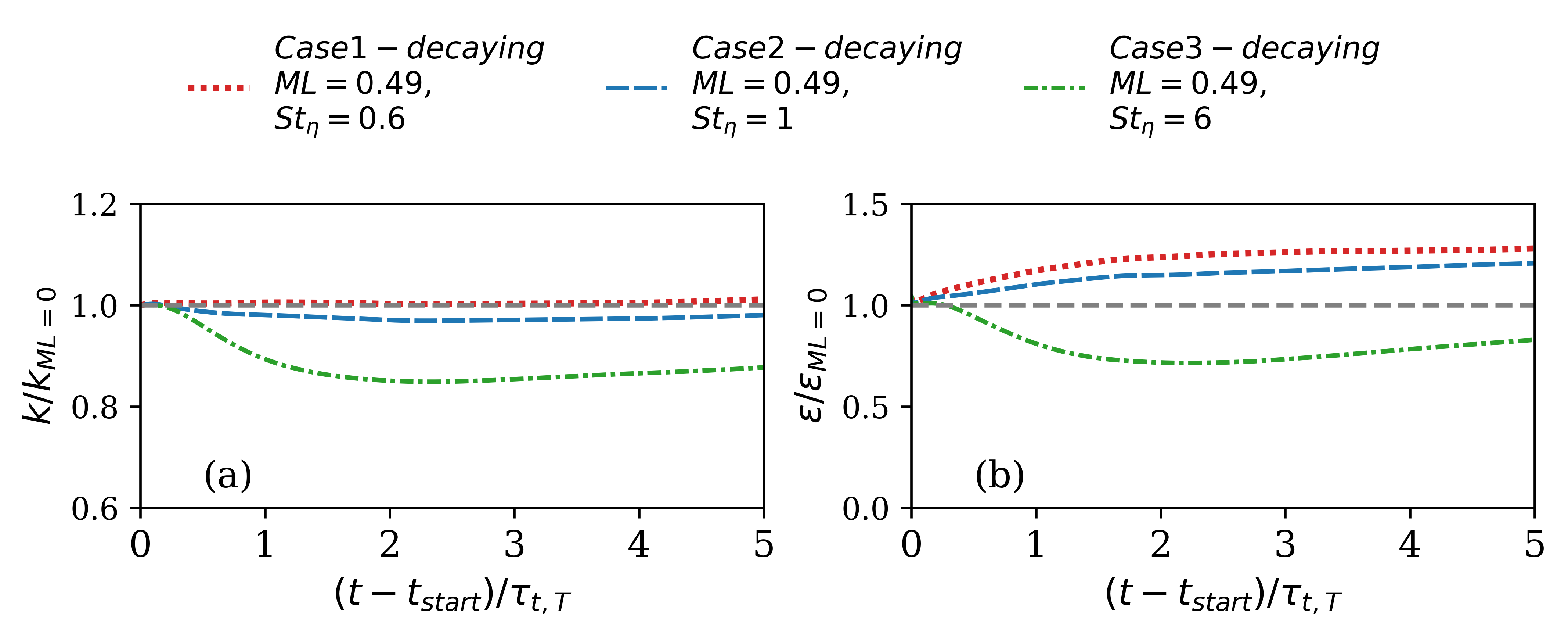}
	
	\caption{{Turbulent kinetic energy (\textbf{a}) and dissipation rate (\textbf{b})   over time for three particle-laden decaying-turbulent cases,  normalized by the values of the particle-free case (Case0).  } }
	\label{fig:stats128_f0_1_2_3_decaying}
\end{figure}

Figure\,\ref{fig:stats128_f0_1_2_3_decaying} shows the evolution of turbulence statistics in particle-laden decaying-turbulence cases (Case1-3) normalized by the particle-free case values. The behavior with respect to Stokes number is consistent with the previously reported data~\cite{AbdelsamieA2012_PF}. 
	It can be seen in Fig.\,\ref{fig:stats128_f0_1_2_3_decaying}a  that  for the decaying turbulence, the  kinetic energy for $St_{\eta}\leq 1$ is close to the particle-free case values,   while at higher Stokes number it is attenuated. Furthermore, in contrast to the stationary turbulence, a non-negligible enhancement of the dissipation rate in Fig.\,\ref{fig:stats128_f0_1_2_3_decaying}b for $St_{\eta} \leq 1$ and attenuation for $St_{\eta} > 1$ can be observed in the decaying-turbulence regime.

\subsection{Training/testing datasets}\label{sec:datasets}
The training/testing datasets are composed of randomly sampled 2-D slices of size $128 \times 128$  from the DNS databases.  
After the forced-turbulence DNS reached a statistically stationary state, 16,000  $x-y$ planes from each case are sampled randomly  from five hundred 3-D instances of the  DNS data in  the time interval $21 \le t/\tau_{t,T} \le 40$   with discrete sampling times separated by $\Delta t\approx 0.038 \tau_{t,T} \approx 0.5 \tau_{\eta,T}$ and used as the training dataset. 
Eight hundred randomly selected 2-D slices  for each  case in  the time interval $10 \le t/\tau_{t,T} \le 19$   with discrete sampling times separated by $\Delta t\approx 0.38 \tau_{t,T} \approx 5 \tau_{\eta,T}$  are used for the testing  dataset. 
Furthermore, the datasets include the decaying-turbulence data. 
The  sample space includes 440 3-D instances of the decaying-turbulence DNS data for each case in the interval between  $42 \le t/\tau_{t,T} \le 45$. Five percent  of the instances are used to build the testing sample space. Thus, in total 20 DNS cases are included in the training/testing sets,  
including the particle-free case (Case0) in Tab.\,\ref{tab::OF128setup}, particle-laden cases in Tab.\,\ref{tab::OF128setup-particle}, and their corresponding decaying-turbulence counterparts. 
The final training and testing datasets include 320,000 and 16,000 2-D $128 \times 128$ samples, respectively. 
These are the main training/testing datasets, however, for the analysis purposes, subsets of the training/testing datasets are used in certain sections. 
{Specifically, a pure particle-free training set with 104,500 samples was generated, containing only forced- and decaying-turbulence particle-free DNS cases. A higher sampling rate was used to increase the number of samples from the two DNS cases. 
This particular dataset was used to investigate the importance of including particle-turbulence interaction physics in the training procedure, as discussed in Sec.\,\ref{sec:prt-free-data}. 
Moreover, subsets of the testing dataset, each including only one or a few specific DNS cases (each with 800 samples), were used in the results section to focus the study on specific aspects such as the generalization capability of the SR model.
}

\section{Results and discussion}
\subsection{Assessment of the model trained on the particle-free dataset}\label{sec:prt-free-data}


The  aim of this section is to investigate the importance of incorporating particle-laden data from Euler-Lagrange simulations in the training process. Bode et al.~\cite{BodeM2021_PCI} used an SR model trained on a particle-free turbulent dataset and applied it on spray combustion LES. However, to the authors' knowledge, no study has been reported   to systematically investigate whether a model that is trained based on single-phase turbulent data can perform high-fidelity super-resolution  for particle-laden flows.  
Hence, the performance of the network which is trained on a subset of the training database which only contains samples from particle-free forced- and decaying-turbulence DNS cases is evaluated. The trained network is first tested on the test dataset which contains only Case$0$. That is, we first evaluate the performance of the proposed architecture and training strategy on single-phase turbulent fields. Further, its performance is evaluated by testing the model on the samples from particle-laden DNS. 

\begin{figure}[!htb]
	\centering

	\subfigure
	{
		\label{fig:00110-01G_PFT-NoPrt_ArchT/inference_trainedOnf0/E_1D_4x_test_4x_gt_128_585000_img0}
		\includegraphics[width=.42\textwidth,trim=0.3cm 0.3cm 0.3cm 0.3cm, clip]{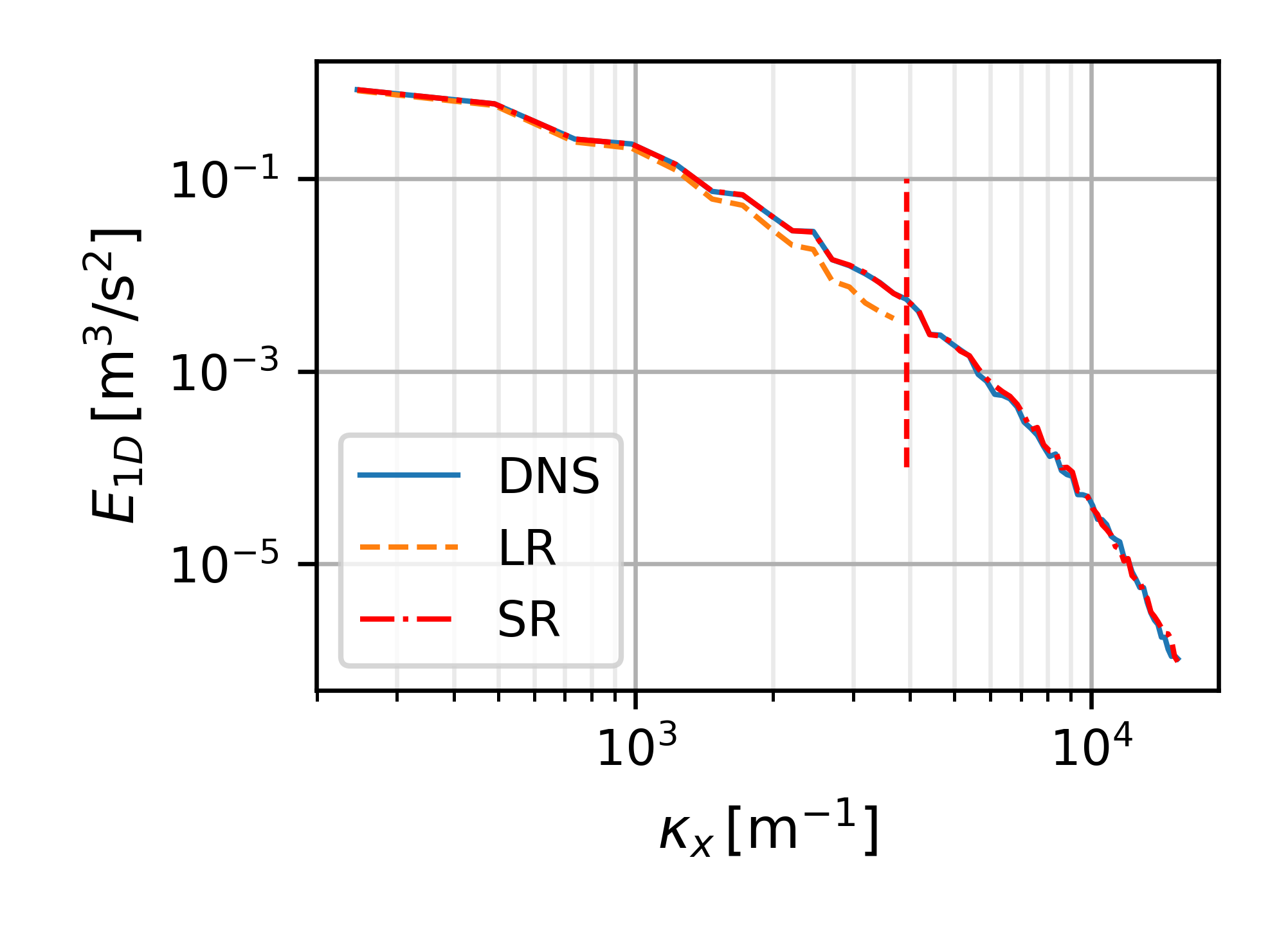}
	}
	\subfigure
	{
		\label{fig:00110-01G_PFT-NoPrt_ArchT/inference_trainedOnf0/pdf_vort_4x_test_4x_gt_128_585000_img0}
		\includegraphics[width=.42\textwidth,trim=0.3cm 0.3cm 0.3cm 0.3cm, clip]{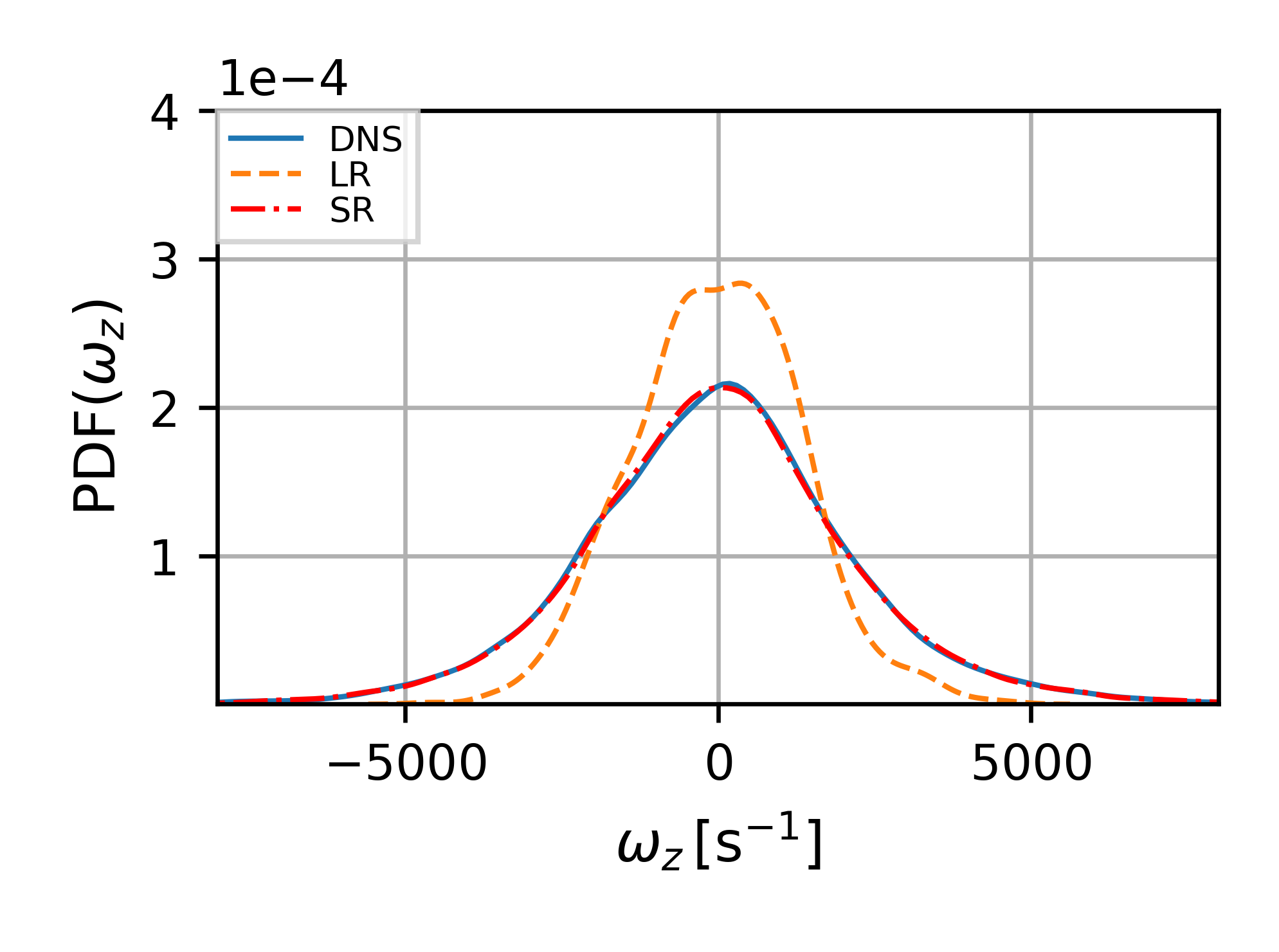}
	}
	\caption{ One-dimensional kinetic energy spectra (\textbf{left}) and  PDF of $z$-component of the vorticity vector (\textbf{right})  for ground-truth (DNS), filtered DNS (LR), and super-resolved (SR) fields for  test  data sampled from the particle-free case ({Case0}) test data. }
	\label{fig:00110-01G_PFT-NoPrt_ArchT/inference_trainedOnf0}
\end{figure}

In Fig.\,\ref{fig:00110-01G_PFT-NoPrt_ArchT/inference_trainedOnf0},  the  1-D  kinetic energy spectra and PDFs of the $z$-component of the vorticity vector are shown. The statistics are calculated from the reconstructed velocity field (SR), {the input low-resolution field (LR),} and the ground-truth DNS data (DNS). It can be seen that both quantities are well reconstructed. 
Specifically, both the under-resolved (left side of the cut-off denoted by vertical red line in the spectra figure) and unresolved (right side of the cut-off) turbulent energy  are  reconstructed with  high accuracy. This demonstrates that the network can perform SR for particle-free turbulent data well. 

\begin{figure}[!htb]
	\centering
	
	\subfigure
	{
		\label{fig:00110-01G_PFT-NoPrt_ArchT/extToPrt_f11/E_1D_4x_test_4x_gt_128_585000_img200}
		\includegraphics[width=.42\textwidth,trim=0.3cm 0.3cm 0.3cm 0.3cm, clip]{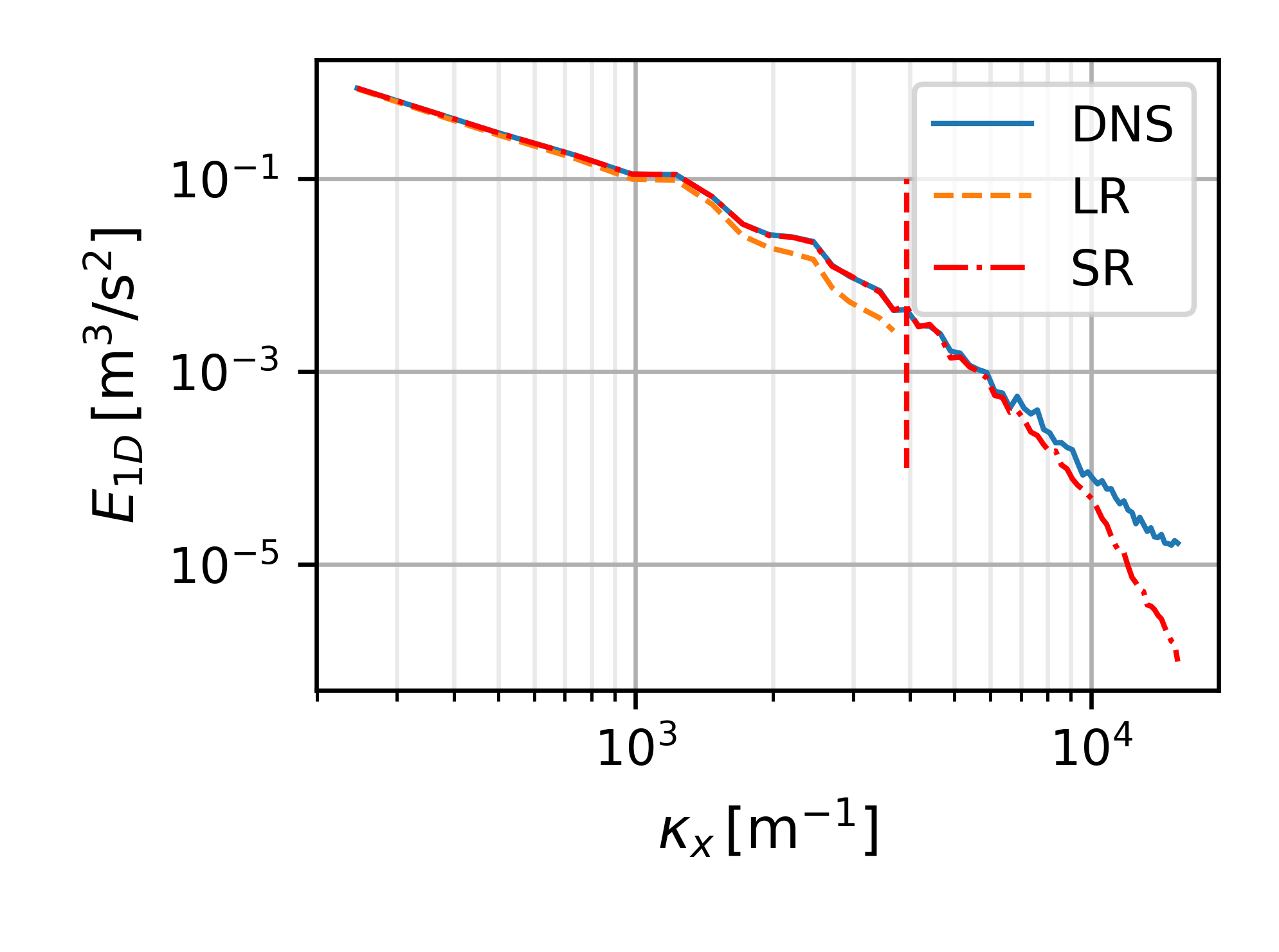}
	}
	\subfigure
	{
		\label{fig:00110-01G_PFT-NoPrt_ArchT/extToPrt_f11/pdf_vort_4x_inset_test_4x_gt_128_585000_img200}
		\includegraphics[width=.42\textwidth,trim=0.3cm 0.3cm 0.3cm 0.3cm, clip]{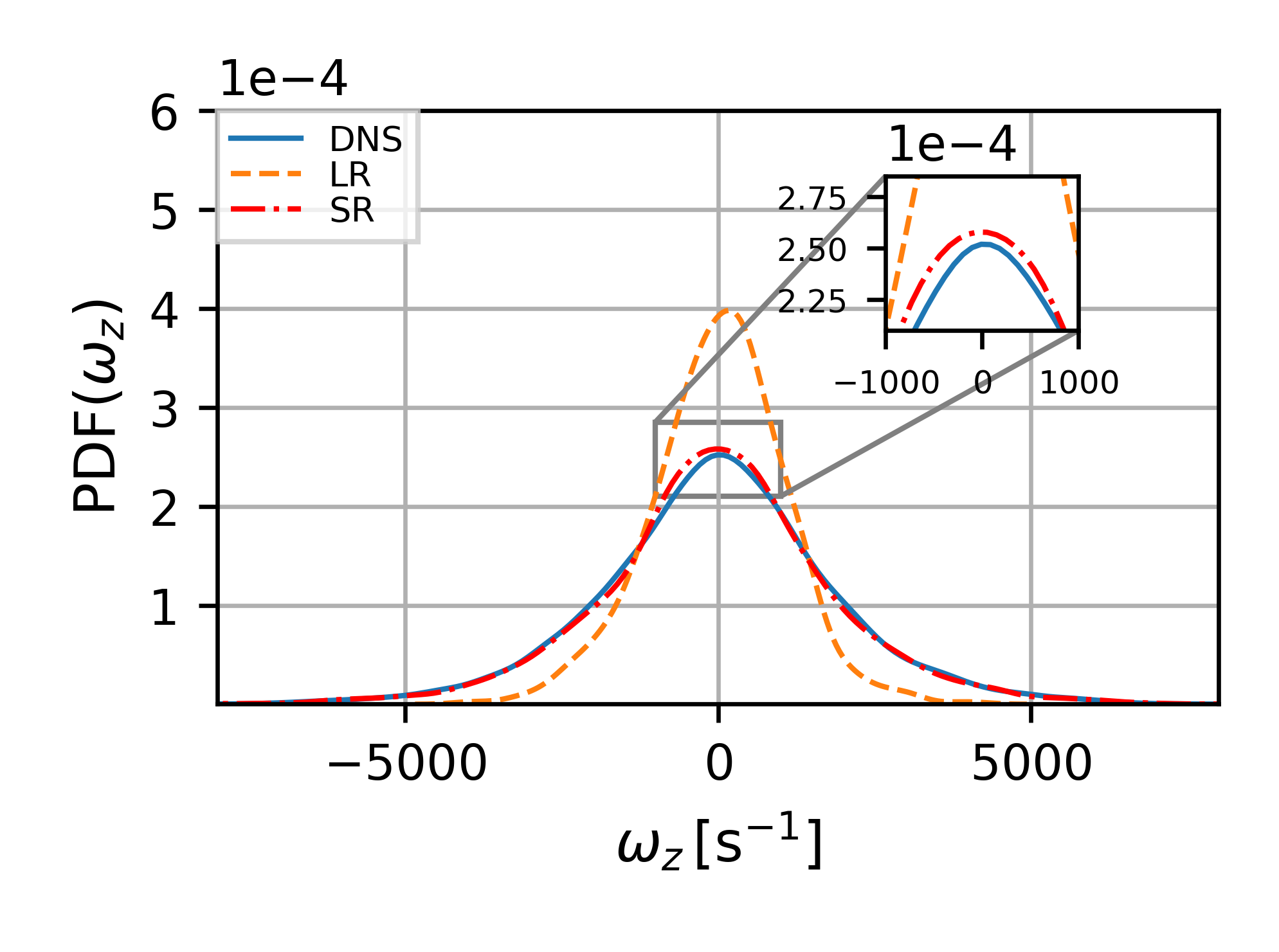}
	}
	
	\caption{One-dimensional kinetic energy spectra (\textbf{left}) and PDF of $z$-component of vorticity  (\textbf{right}) for ground-truth (DNS), filtered DNS (LR), and super-resolved (SR) fields for  test  data sampled from a particle-laden case ({Case1}). }
	\label{fig:extToPrt_f11}
\end{figure}

\begin{figure}[!ht]
	\centering

	\subfigure{\label{fig:Fig/forced_hit_spectra_averaged/energy_spect_f0_11_12_21_22_31_32}\includegraphics[width=0.45\textwidth,trim=0.3cm 0.3cm 0.3cm 0.3cm]{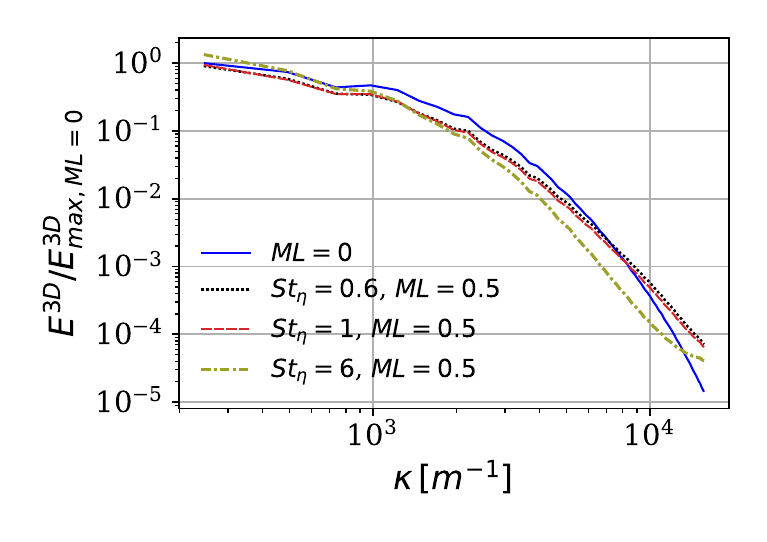}}
	\subfigure{\label{fig:Fig/forced_hit_spectra_averaged/energy_diss_spect_f0_11_12_21_22_31_32}\includegraphics[width=0.45\textwidth,trim=0.3cm 0.3cm 0.3cm 0.3cm]{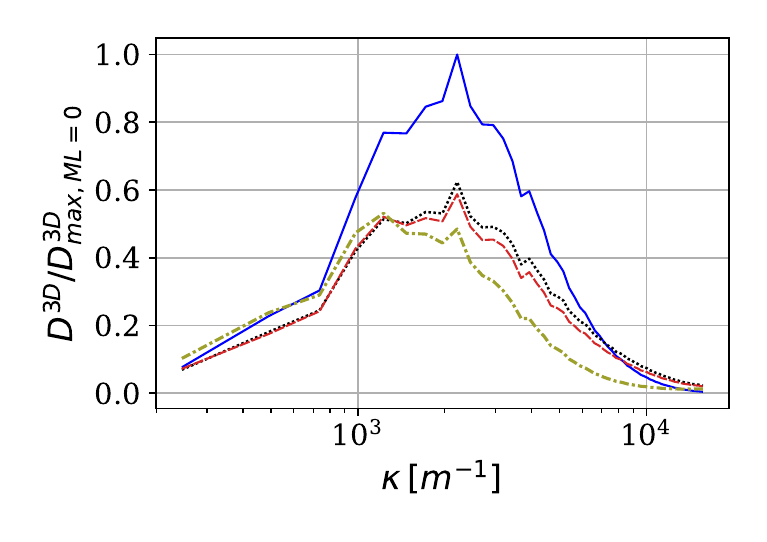}}
	
	\caption{Three-dimensional kinetic energy (\textbf{left}) and dissipation (\textbf{right}) spectra for particle-free (Case0) and selected particle-laden DNS cases, i.e.  Case1, Case2, and Case3 averaged over 100 time instances in the dataset and  normalized by the values of the particle-free case (Case0).   }
	\label{fig:spectra_f0_1_2_3}
\end{figure}

Next, the network is tested on samples from a particle-laden case. 
In  Fig.\,\ref{fig:extToPrt_f11},  the network is tested on samples from Case1 which is a particle-laden forced-turbulence case with $\mathrm{St}_{\eta}=0.6$. As can be seen, the model is unable to accurately reconstruct the energy spectrum at frequencies higher than the cut-off frequency. Moreover, the gradients are also not well reconstructed which manifests itself  in small but noticeable  discrepancies in the reconstruction of the vorticity components.  The reason of these discrepancies can be explained by Fig.\,\ref{fig:spectra_f0_1_2_3}. 
In Fig.\,\ref{fig:spectra_f0_1_2_3}, the 3D kinetic energy and dissipation spectra for the selected samples in the dataset are shown. The spectra are evaluated by 3-D fast Fourier transforms of velocity components 
averaged over the 100 snapshots in the datasets. The spectra are normalized using a maximum value of the particle-free case. This specific type of normalization was proposed by Abdelsamie et al.~\cite{AbdelsamieA2012_PF} to highlight the effect of particle-turbulence interactions. 
As can be seen in Fig.\,\ref{fig:spectra_f0_1_2_3}, particles affect the turbulent energy distribution and its dissipation at a wide range of frequencies. Specifically, the slope of the energy spectra changes at high frequencies in low Stokes number cases ($\mathrm{St}_{\eta}=0.6$, $\mathrm{St}_{\eta}=1$) compared to the particle-free ($ML=0$) case. This results in higher energy and higher dissipation at high frequencies. The modulation of turbulence by particles increases by increasing the  Stokes number as can be inferred by comparing Case3 ($\mathrm{St}_{\eta}=6$) with the other cases. Thus, it can be concluded that the networks trained solely on particle-free data cannot predict the particle-laden physics very well without learning particle-turbulence interactions.

\subsection{Assessment of the model trained on the complete dataset}\label{sec:prt-data}
In this section, we evaluate the model's performance after the training on the complete dataset. This dataset comprises both particle-free and particle-laden flows, featuring carrier gas fields characterized by forced- and decaying-turbulence. We assess the proposed strategy of incorporating particle information during training by evaluating the network on two separate test datasets: one containing unseen test samples with Stokes numbers similar to those used during training, and another with distinct Stokes numbers that were not represented in the training data.

\subsubsection{Tests on the previously observed physics}\label{sec:prt-data-seen}
\begin{figure}[!htb] 
	\centering
	
	\subfigure
	{
		\label{fig:01004-00H_PFT-Prt_ArchTest_condSRGANModel/inference_testOnfx/vel_vort_4x_test_4x_gt_128_720000_img14800_full_256}
		\includegraphics[width=.7\textwidth,trim=0 1.3cm 0.3 0cm, clip]{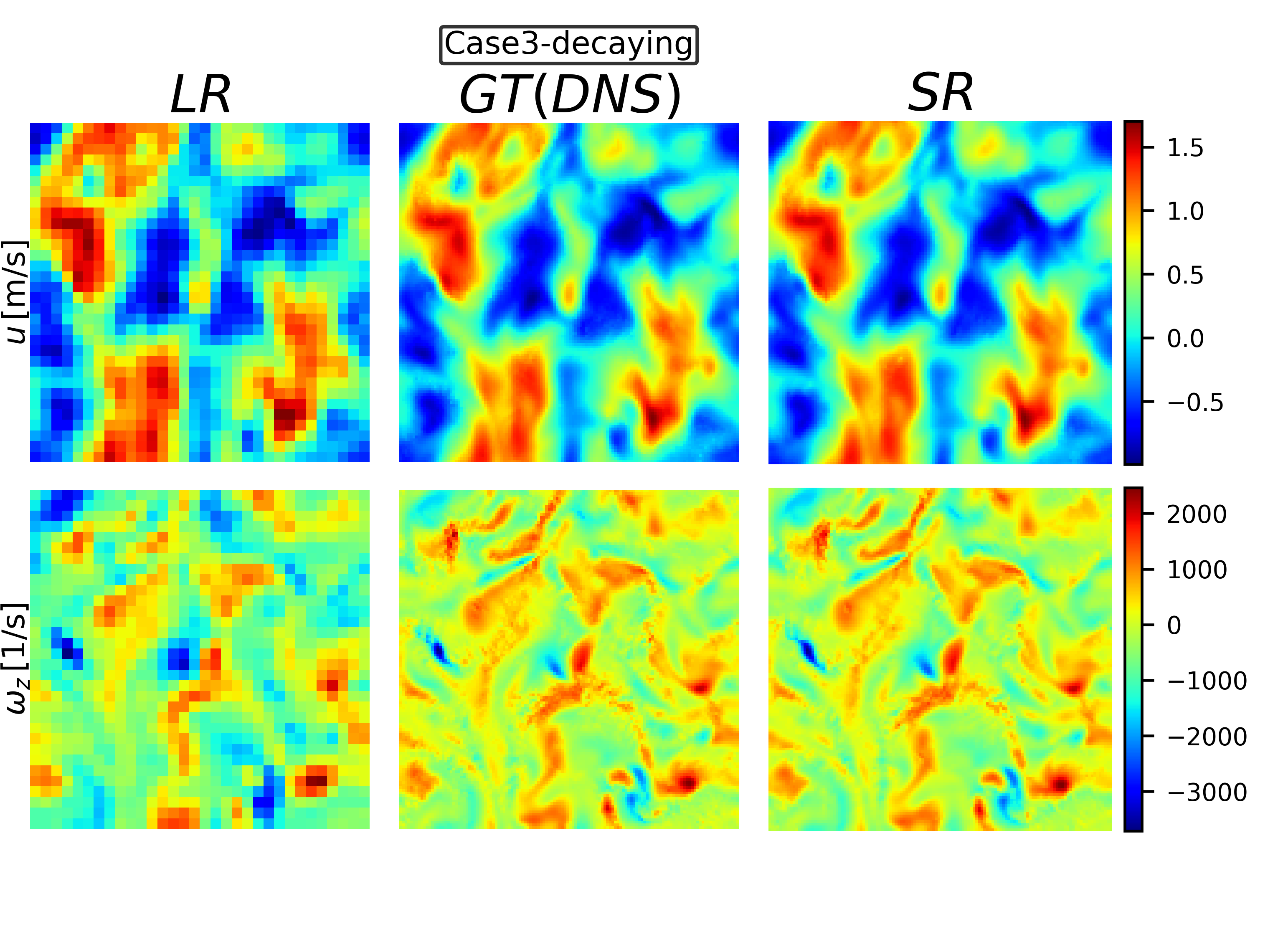}	 
	}\\[.1mm]
	\subfigure
	{
		\label{fig:01004-00H_PFT-Prt_ArchTest_condSRGANModel/inference_testOnfx/vel_vort_4x_test_4x_gt_128_720000_img14000_full_256}
		\includegraphics[width=.7\textwidth,trim=0 1.3cm 0 0cm, clip]{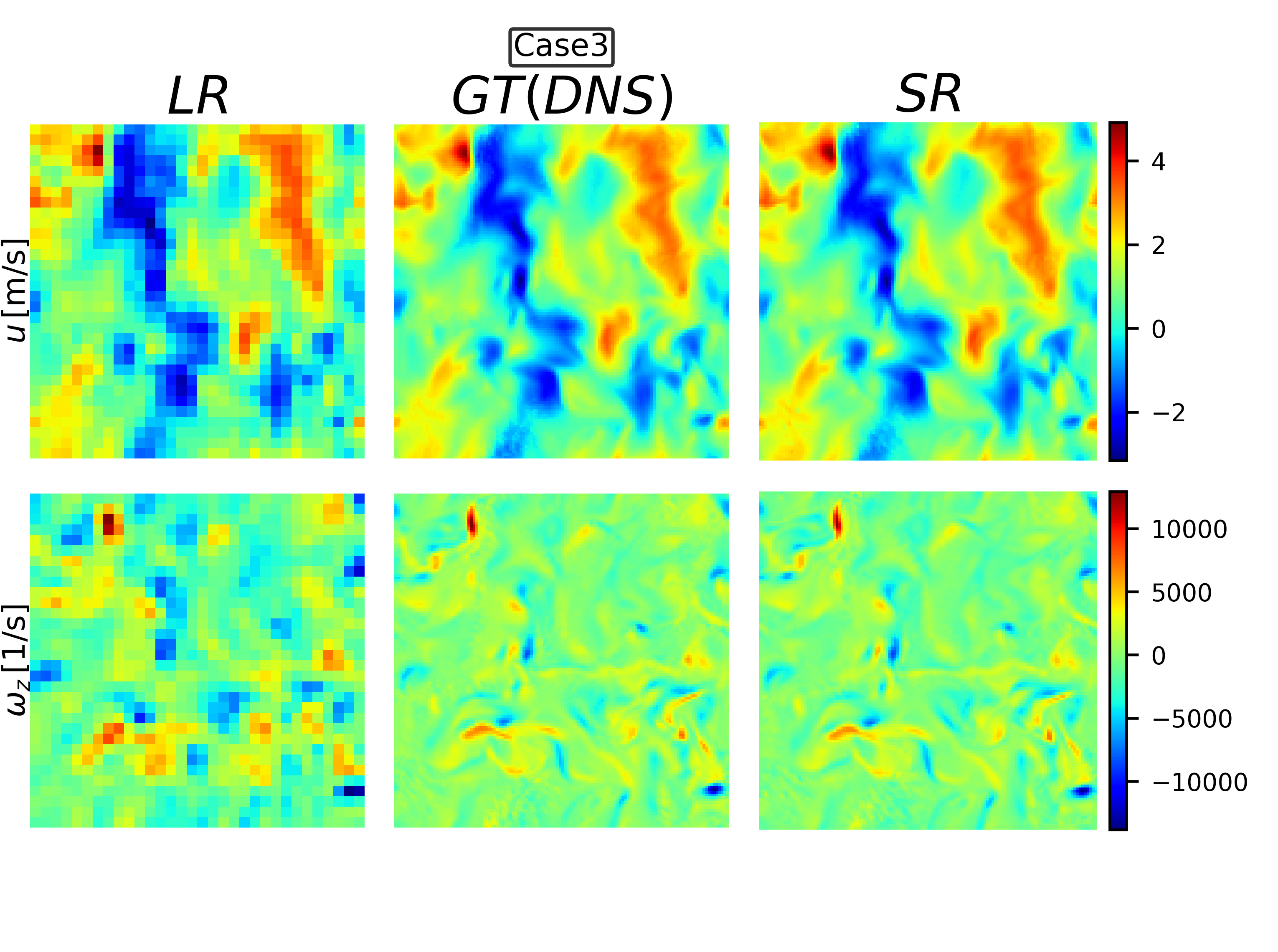}
	}

	\caption{Contours of the first velocity component ($u$) and $z$-component of the vorticity vector ($\omega_z$) for low-resolution input (left) vs. DNS (middle) vs. SR (right) from two randomly selected slices in the test dataset for the decaying- and forced-turbulence Case3. }
	\label{fig:01004-00H_PFT-Prt_ArchTest_condSRGANModel/inference_testOnfx_1_contors}
\end{figure}

In Fig.\,\ref{fig:01004-00H_PFT-Prt_ArchTest_condSRGANModel/inference_testOnfx_1_contors}, the contours of the velocity and vorticity from low-resolution input, ground-truth DNS data, and super-resolved data  from two randomly selected slices in the test dataset for a decaying-turbulence ({labeled "Case3-decaying"}) and a forced-turbulence ({labeled "Case3"}) case are shown. As can be seen, the generated SR fields qualitatively agree very well  with the ground-truth fields, both in decaying- and forced-turbulence physics. 

\begin{figure}[!htb]
	\centering
	
	\subfigure[$ML = 0$]
	{
		\label{fig:01004-00H_PFT-Prt_ArchTest_condSRGANModel/inference_testOnfx/E_1D_4x_test_4x_gt_128_720000_img400}
		\includegraphics[width=.42\textwidth,trim=0.3cm 0.3cm 0.3cm 0.3cm, clip]{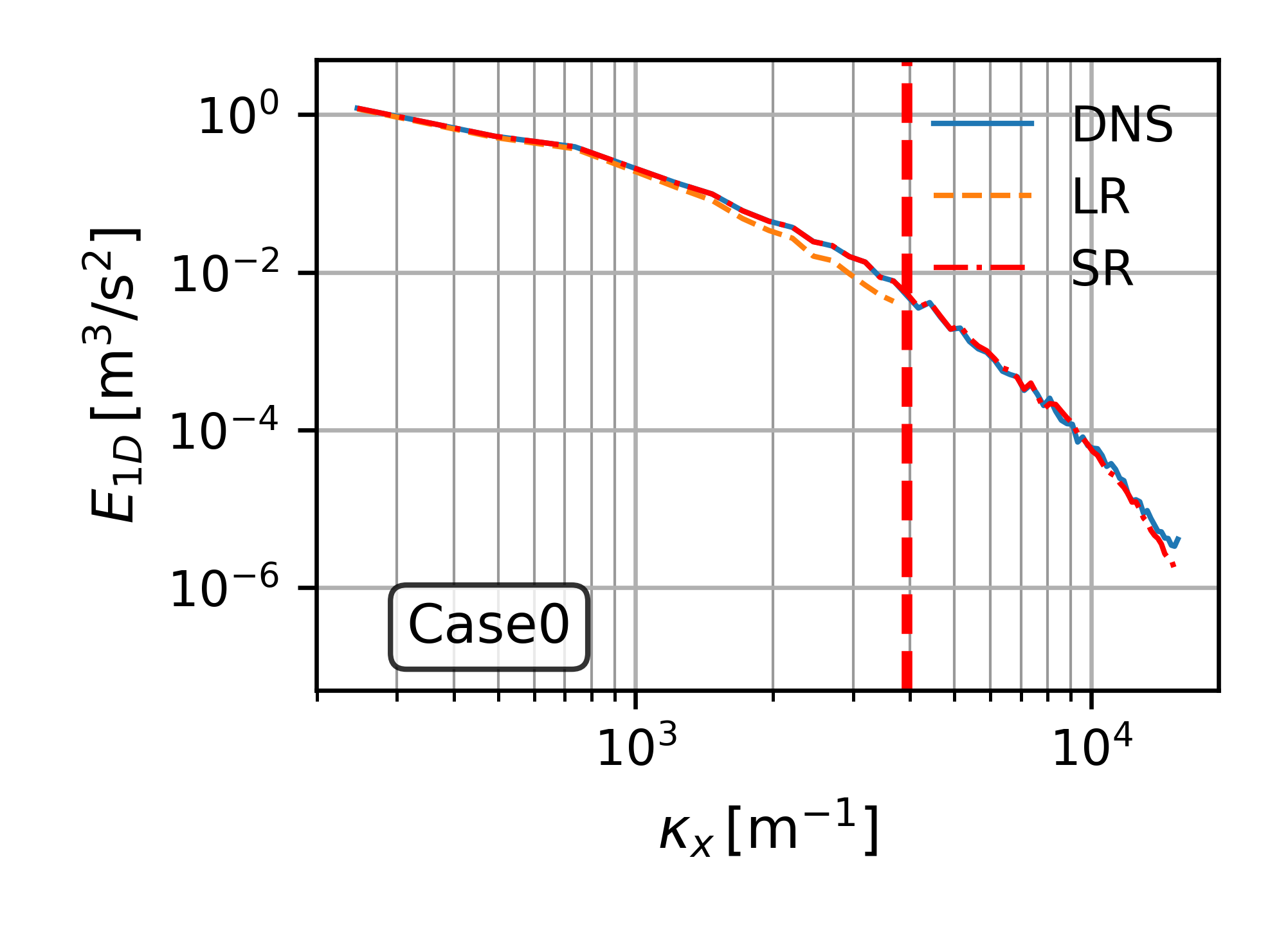}
	}
	\subfigure[$\mathrm{St}_{\eta}=1,~ML=0.49$]
	{
		\label{fig:01004-00H_PFT-Prt_ArchTest_condSRGANModel/inference_testOnfx/E_1D_4x_test_4x_gt_128_720000_img9200}
		\includegraphics[width=.42\textwidth,trim=0.3cm 0.3cm 0.3cm 0.3cm, clip]{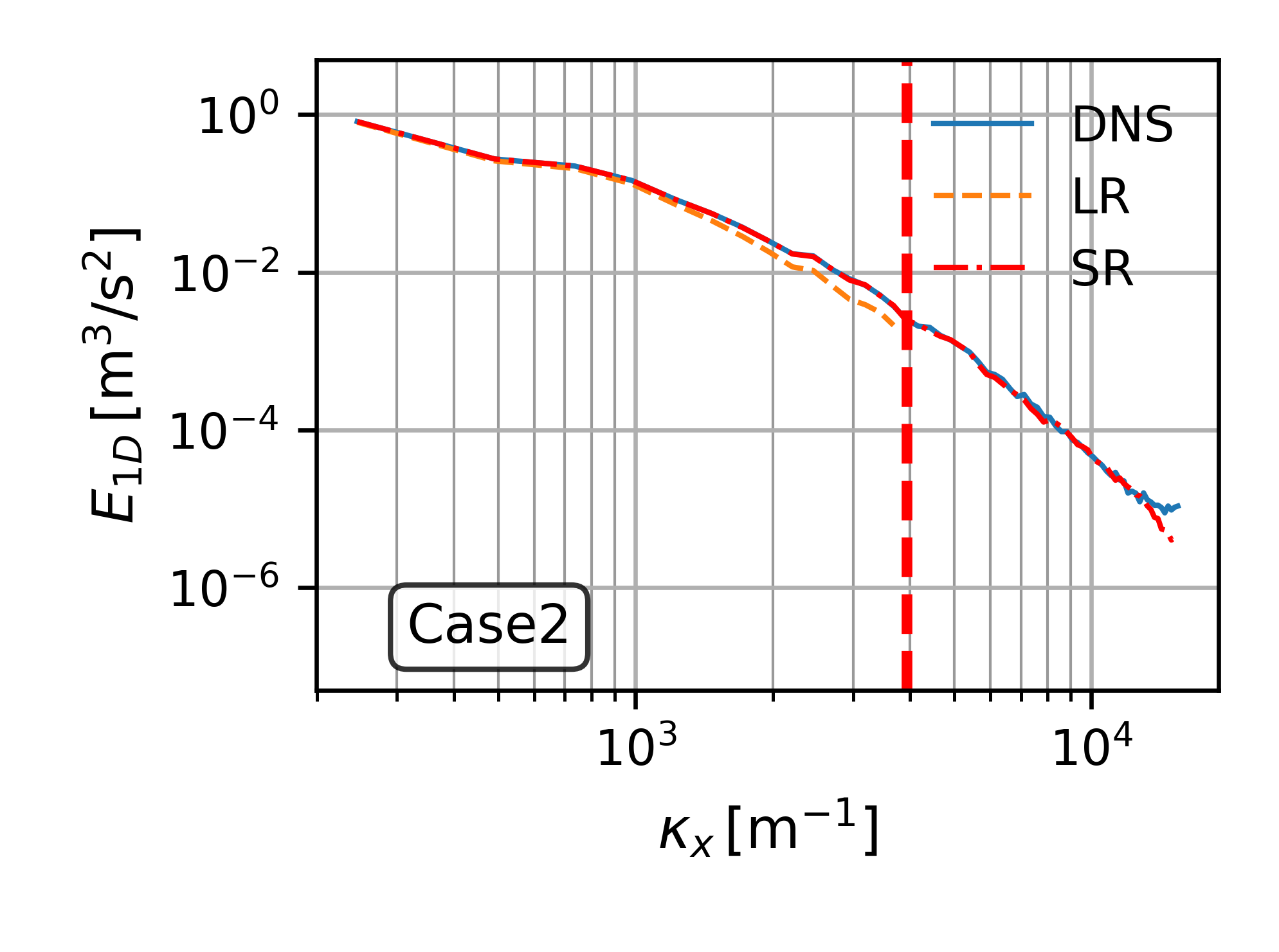}
	}

	\subfigure[$\mathrm{St}_{\eta}=6,~ML=0.49$]
	{
			\label{fig:01004-00H_PFT-Prt_ArchTest_condSRGANModel/inference_testOnfx/E_1D_4x_test_4x_gt_128_720000_img14000}
			\includegraphics[width=.42\textwidth,trim=0.3cm 0.3cm 0.3cm 0.3cm, clip]{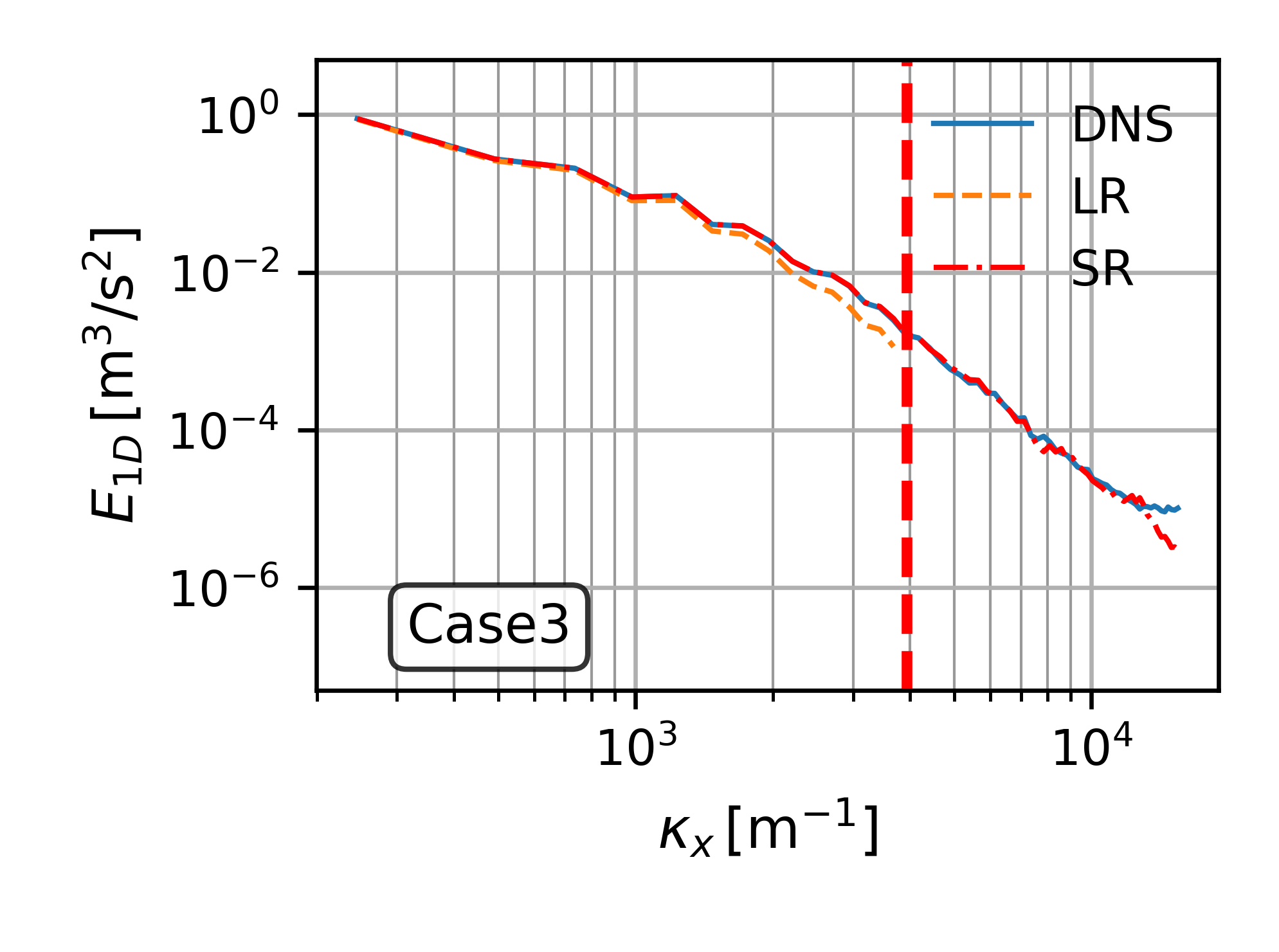}
		}
	\subfigure[$\mathrm{St}_{\eta}=1,~ML=0.75$]
	{
		\label{fig:01004-00H_PFT-Prt_ArchTest_condSRGANModel/inference_testOnfx/E_1D_4x_test_4x_gt_128_720000_img6000}
		\includegraphics[width=.42\textwidth,trim=0.3cm 0.3cm 0.3cm 0.3cm, clip]{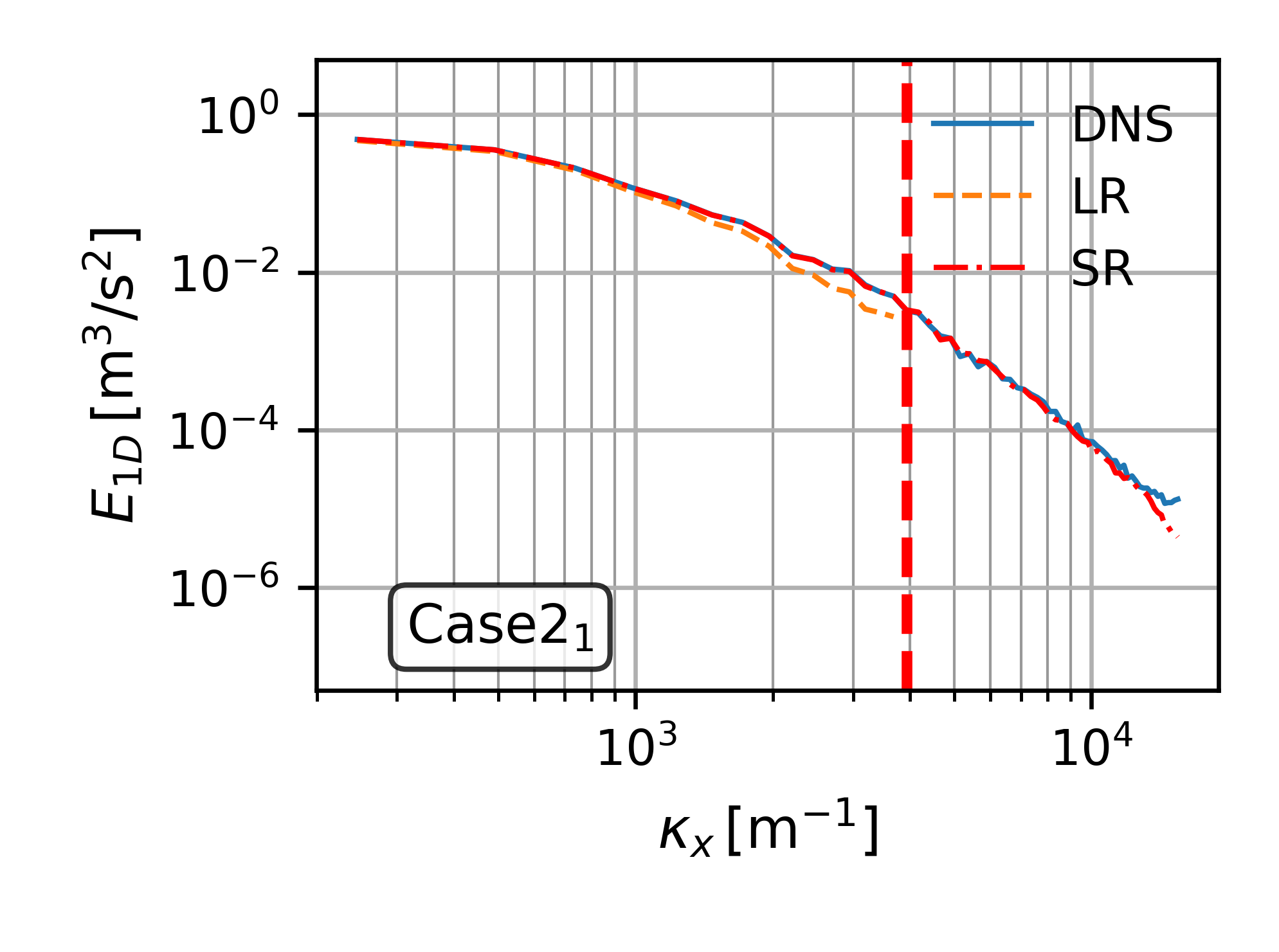}
	}

	\caption{One-dimensional kinetic energy spectra: The model performance tested on cases with different Stokes numbers and mass loadings. }
	\label{fig:01004-00H_PFT-Prt_ArchTest_condSRGANModel/inference_testOnfx_1}
\end{figure}

The kinetic energy spectra of test samples from different cases are shown in Fig.\,\ref{fig:01004-00H_PFT-Prt_ArchTest_condSRGANModel/inference_testOnfx_1} which confirm the good performance of the model in a quantitative manner. Specifically it is shown that the spectra of DNS and SR match in the particle-free (Fig.\,\ref{fig:01004-00H_PFT-Prt_ArchTest_condSRGANModel/inference_testOnfx/E_1D_4x_test_4x_gt_128_720000_img400}) and a particle-laden (Fig.\,\ref{fig:01004-00H_PFT-Prt_ArchTest_condSRGANModel/inference_testOnfx/E_1D_4x_test_4x_gt_128_720000_img9200}) case. Furthermore, the spectra in samples from  different particle Stokes number, i.e.  $\mathrm{St}_{\eta}=6$ (Fig.\,\ref{fig:01004-00H_PFT-Prt_ArchTest_condSRGANModel/inference_testOnfx/E_1D_4x_test_4x_gt_128_720000_img14000}) and different particle mass loading, i.e. $ML=0.75$ (Fig.\,\ref{fig:01004-00H_PFT-Prt_ArchTest_condSRGANModel/inference_testOnfx/E_1D_4x_test_4x_gt_128_720000_img6000}) can be recovered as well. 

\begin{figure}[!htb]
	\centering
	\subfigure
	{\label{fig:01004-00H_PFT-Prt_ArchTest_condSRGANModel/inference_testOnfx/test_4scatt_Eps_sgs_ 2_x_gt_with_particle128_720000}
		\includegraphics[width=.49\textwidth,trim=0.3cm 0.3cm 0.3cm 0.3cm, clip]{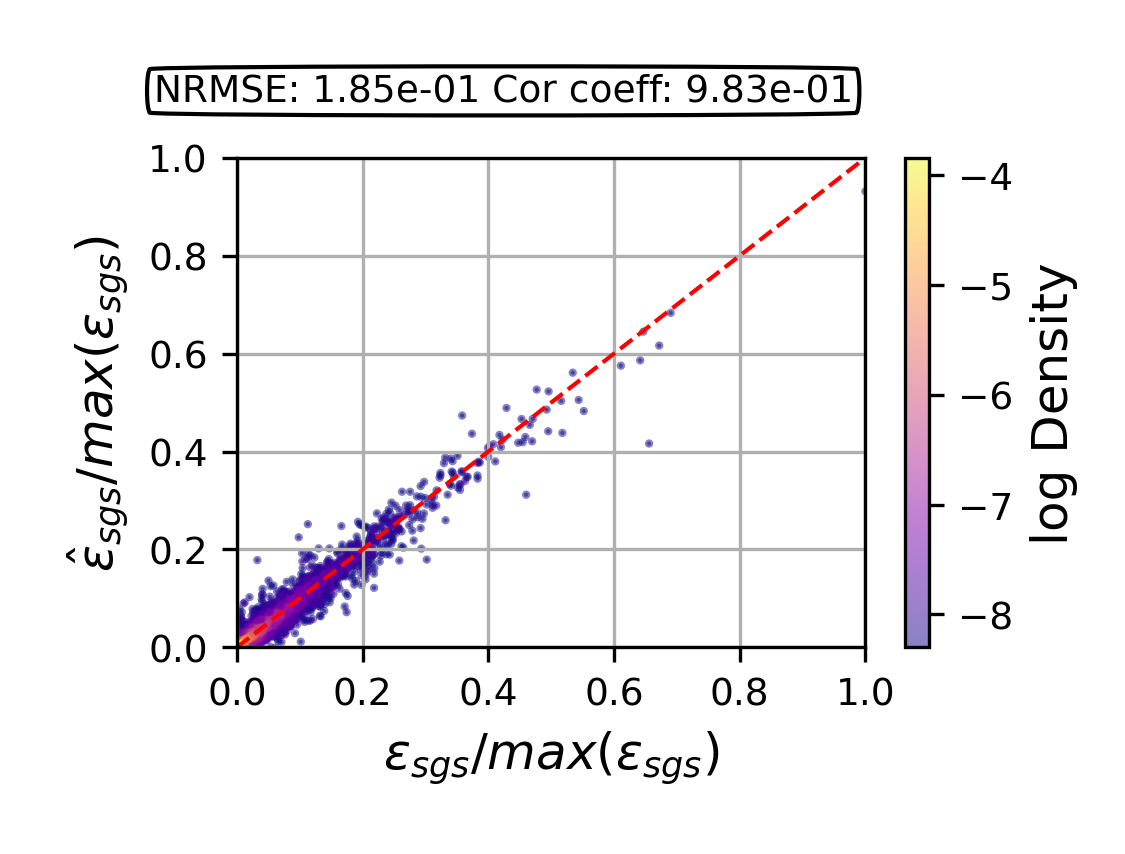}}

	\caption{Scatter plot of subgrid dissipation rate for 50,000 randomly selected points in the whole test datasets. The normalized root-mean-square error (NRMSE)  and the correlation coefficient have been reported as well. }
	\label{fig:01004-00H_PFT-Prt_ArchTest_condSRGANModel/inference_testOnfx_1/scatt_eps_k_sgs}
\end{figure}

In Fig.\,\ref{fig:01004-00H_PFT-Prt_ArchTest_condSRGANModel/inference_testOnfx_1/scatt_eps_k_sgs}, the scatter plot of the subgrid dissipation rate  is depicted. The scatter plots are evaluated using 50,000 randomly selected data points from the entire test dataset that include 16,000 2-D $128 \times 128$ samples from all particle-free and particle-laden cases (decaying- and forced-turbulence). The subgrid dissipation rate is evaluated as
\begin{equation}\label{eq:subgrid_diss}
	{\varepsilon}_{sgs} = -\mathcal{T}_{ij}\overline{{S}}_{ij},
\end{equation}
where {$\mathcal{T}_{ij}$ is the SGS stress tensor,}
\begin{equation}\label{eq:sgs_stress}
\mathcal{T}_{ij}=\overline{{u}_i{u}_j} - \overline{{u}_i} \,\overline{{u}_j},
\end{equation}
with $\overline{{S}}_{ij} = 0.5\left(\partial  \overline{{u}_i}/\partial x_j +\partial \overline{{u}_j}/\partial x_i \right)$ {being the filtered strain rate tensor}. 
The $\overline{(.)}$ represents the top-hat filter with a filter width of 4$\Delta_{DNS}$.  
Similarly, the predicted subgrid dissipation rate ($\hat{\varepsilon}_{sgs}$) can be evaluated using SR velocity fields, i.e. $\hat{u}$ in the evaluation of each tensor in Eq.\,\eqref{eq:subgrid_diss}. 
It can be seen in Fig.\,\ref{fig:01004-00H_PFT-Prt_ArchTest_condSRGANModel/inference_testOnfx_1/scatt_eps_k_sgs} that the predicted subgrid dissipation rates follow the DNS values well with a reasonable correlation coefficient.

\begin{figure}[!htb]
	\centering
		\subfigure[]
	{
		\label{fig:01004-00H_PFT-Prt_ArchTest_condSRGANModel/inference_testOnfx/test4_4PDF_eps_sgs__x_gt_with_particle128_720000}
		\includegraphics[width=.48\textwidth,trim=0.3cm 0.3cm 0.3cm 0.3cm, clip]{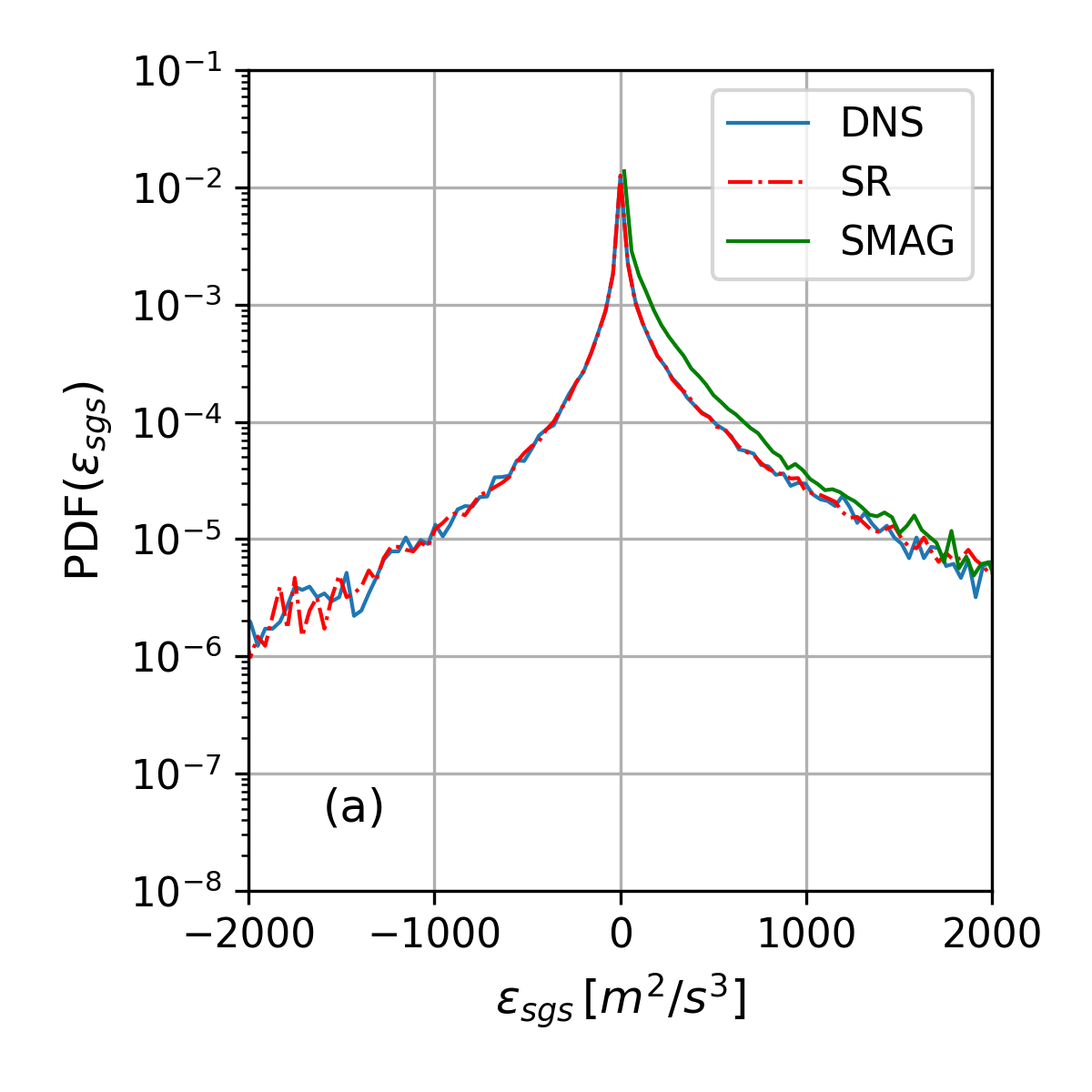}
	}
	\subfigure[]
	{
		\label{fig:01004-00H_PFT-Prt_ArchTest_condSRGANModel/inference_testOnfx/test4_4PDF_eps__x_gt_with_particle128_720000}
		\includegraphics[width=.48\textwidth,trim=0.3cm 0.3cm 0.3cm 0.3cm, clip]{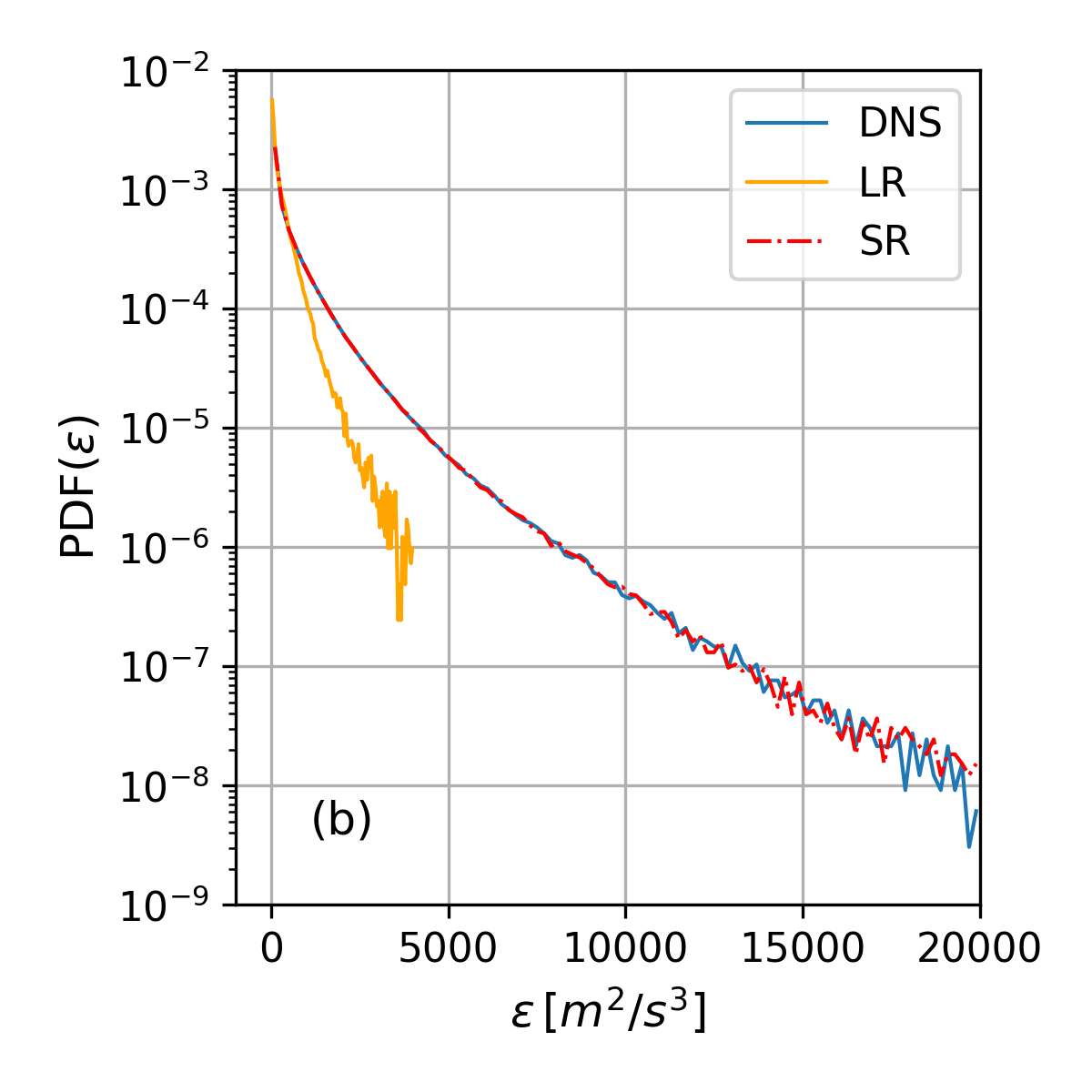}
	}

	\caption{PDF of subgrid  (\textbf{a}) and total (\textbf{b}) dissipation rates for 100 randomly selected 2-D slices from the entire test datasets.  }
	\label{fig:01004-00H_PFT-Prt_ArchTest_condSRGANModel/inference_testOnfx_1/pdf_eps_k_sgs}
\end{figure}

In Fig.\,\ref{fig:01004-00H_PFT-Prt_ArchTest_condSRGANModel/inference_testOnfx_1/pdf_eps_k_sgs}, the PDF of subgrid, and total  dissipation rates are depicted. 
The PDF plots are evaluated using 100 randomly selected 2-D slices  from the entire test dataset that include 16,000 2-D $128 \times 128$ samples from all particle-free and particle-laden cases (decaying- and forced-turbulence). 
The predictions of the Smagorinsky model for the subgrid dissipation rate is shown as well. It is computed as  
\begin{equation}\label{eq:subgrid_diss_smag}
	{\varepsilon}_{sgs,SMAG} = 2 \left(C_s \Delta_s \right)^2 |\overline{{S}}|^3,
\end{equation}
where $C_s (=0.17)$ is the Smagorinsky constant, $\Delta_s=4\Delta_{DNS}$ is  the effective filter width, and $|\overline{{S}}|= 2\sqrt{\overline{{S}}_{ij}\,\overline{{S}}_{ij}}$ is the modulus of the filtered strain rate. It can be seen in  Fig.\,\ref{fig:01004-00H_PFT-Prt_ArchTest_condSRGANModel/inference_testOnfx/test4_4PDF_eps_sgs__x_gt_with_particle128_720000} that backscatter (negative subgrid dissipation)~\cite{PiomelliU1991_PFA} exists in the DNS which is well reproduced in the SR fields, however, the Smagorinsky model is by definition positive and cannot predict the backscatter.   Furthermore, the predicted positive subgrid dissipation rate from the SR fields match the DNS values better than the Smagorinsky model. 

In Fig.\,\ref{fig:01004-00H_PFT-Prt_ArchTest_condSRGANModel/inference_testOnfx/test4_4PDF_eps__x_gt_with_particle128_720000}, the PDF of the total dissipation rate is shown. In the DNS, it is evaluated as  
$	{\varepsilon} = 2\nu {{S}}_{ij}{{S}}_{ij}$, 
where $\nu$ is the viscosity. The predictions by the SR fields are evaluated by using $\hat{u_i}$ in the ${{S}}_{ij}$ evaluation, while the LR predictions are calculated by directly using filtered velocities in the ${{S}}_{ij}$ evaluation, which represents a ``no-model'' approach.  As can be seen, the no-model approach under-predicts the high dissipation rates, i.e.  the rare events, and over predicts the low dissipation rates, while the SR  matches very well the PDF of the total dissipation rate from the DNS values.

\begin{figure}[!htb] 
	\centering
	
	\subfigure[]
	{
		\label{fig:01004-00H_PFT-Prt_ArchTest_condSRGANModel/inference_testOnfx_masked_prt_ema/vel_vort_4x_test_4x_gt_128_720000_img13200_full_256}
		\includegraphics[width=\textwidth,trim=0cm 0.0cm 0.0cm 0cm, clip]{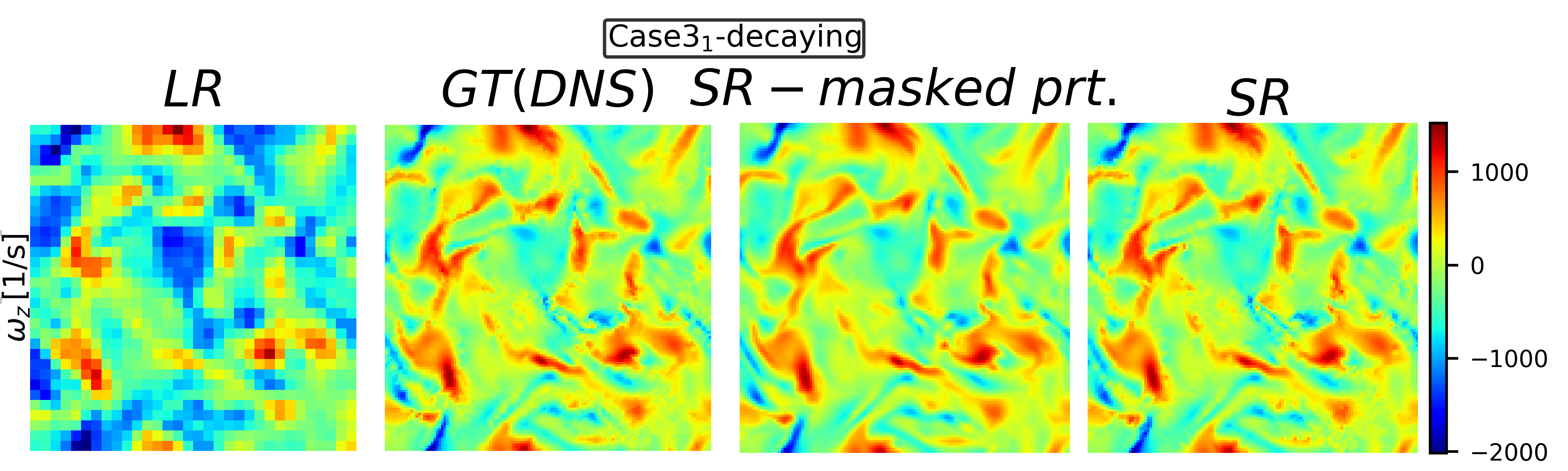}	 
	}

	\subfigure[]
	{
		\label{fig:01004-00H_PFT-Prt_ArchTest_condSRGANModel/inference_testOnfx_masked_prt_ema/E_1D_4x_test_4x_gt_128_720000_img13200}
		\includegraphics[width=.42\textwidth,trim=0.5cm 0.5cm 0.3cm 0.3cm, clip]{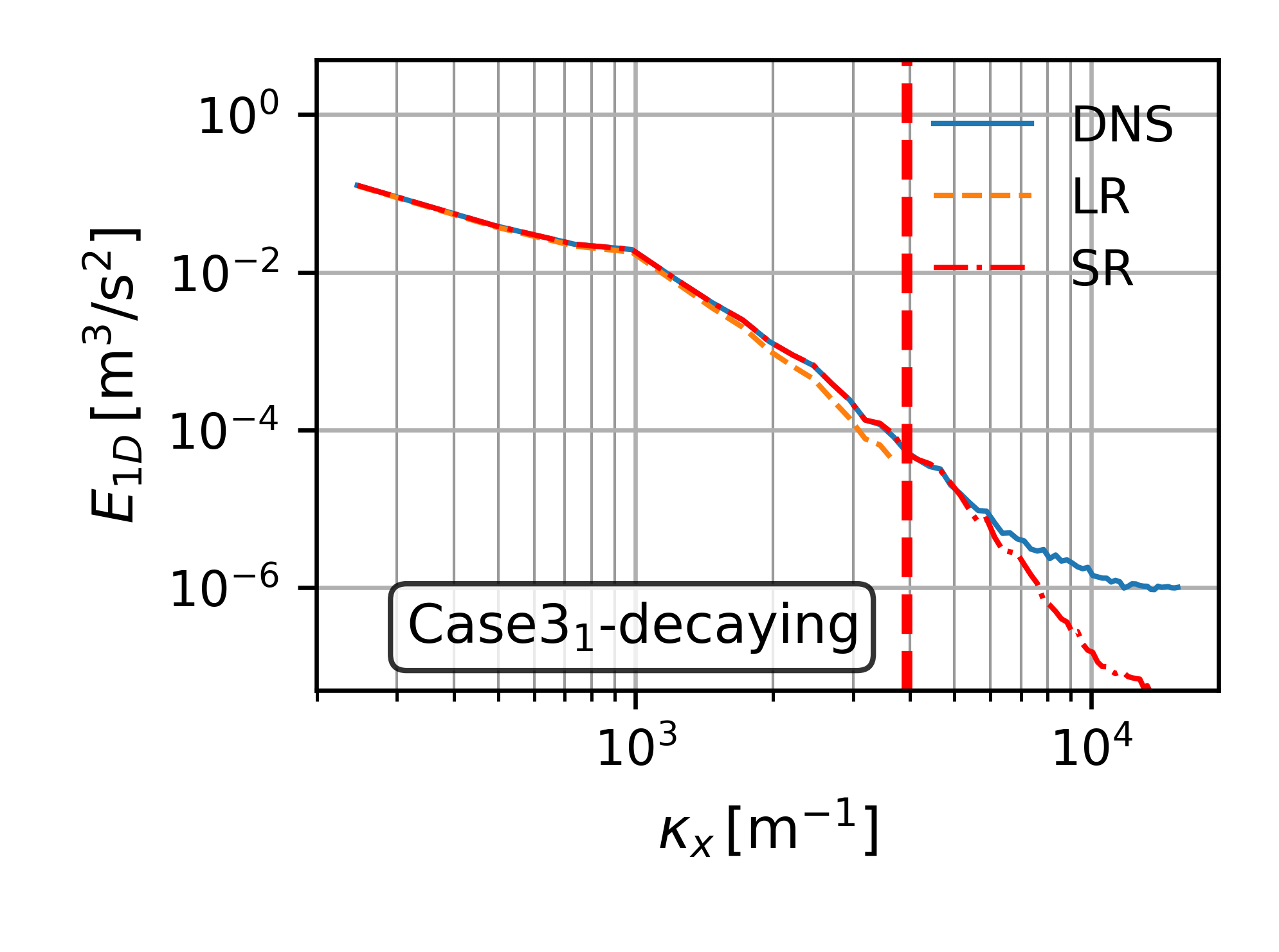}
	}
	\subfigure[]
	{
		\label{fig:01004-00H_PFT-Prt_ArchTest_condSRGANModel/inference_testOnfx_ema/E_1D_4x_test_4x_gt_128_720000_img13200}
		\includegraphics[width=.42\textwidth,trim=0.5cm 0.5cm 0.3cm 0.3cm, clip]{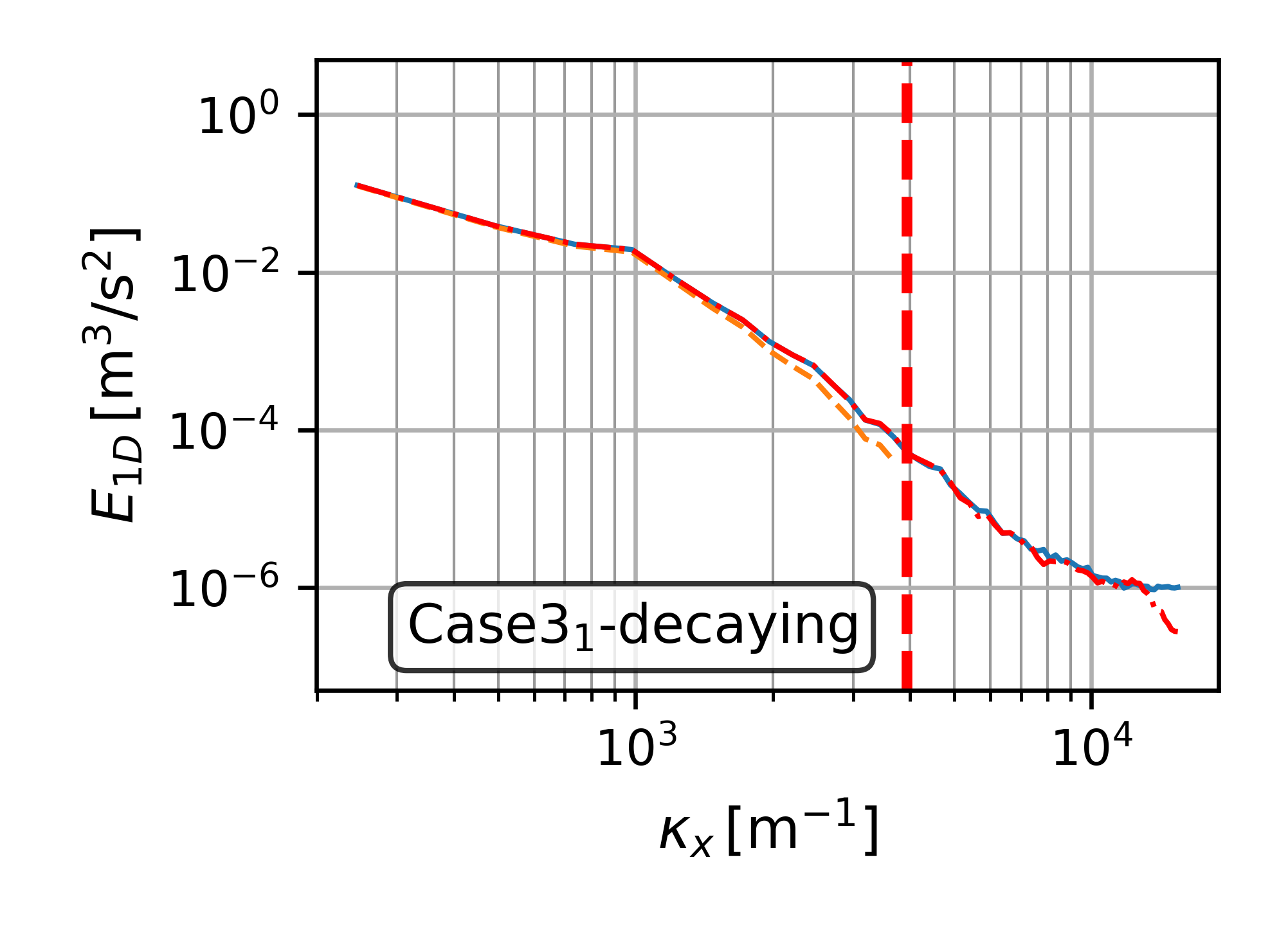}
	}

	\caption{Ablation study by masking input particle-related data: Contours of the  $z$-component of the vorticity vector (\textbf{a}) and 1-D kinetic  energy spectra when particle input data are masked (\textbf{b}) and not masked (\textbf{c}) during the inference. The slice is taken randomly from the test dataset for the  Case3$_1$ in decaying-turbulence mode.  }
	\label{fig:01004-00H_PFT-Prt_ArchTest_condSRGANModel/inference_testOnfx_masked_prt_ema}
\end{figure}

{We now assess  whether the generator network  exploits the input particle information. For this, an ablation study during the inference can be carried out by masking (zeroing) the particle data channel. In Fig.\,\ref{fig:01004-00H_PFT-Prt_ArchTest_condSRGANModel/inference_testOnfx_masked_prt_ema/vel_vort_4x_test_4x_gt_128_720000_img13200_full_256}, the contours of vorticity are shown. }
{The third panel ("SR-masked prt.") shows the output of the generator network when the particle-related input channel is masked  during inference. In this case, while the network reconstructs some large-scale features, much of the small-scale turbulence is lost, and the vorticity field appears overly smoothed compared to the DNS.  In contrast, the fourth panel ("SR") presents the super-resolved vorticity field generated when the full input including the particle data channel is provided to the network. Here, the small-scale structures are  better recovered, closely resembling the DNS ground truth and demonstrating the added value of the particle input channel. }
{This qualitative observation is supported in Figs.\,\ref{fig:01004-00H_PFT-Prt_ArchTest_condSRGANModel/inference_testOnfx_masked_prt_ema/E_1D_4x_test_4x_gt_128_720000_img13200} and  \ref{fig:01004-00H_PFT-Prt_ArchTest_condSRGANModel/inference_testOnfx_ema/E_1D_4x_test_4x_gt_128_720000_img13200}, which show the one-dimensional kinetic  energy spectra for the masked  and unmasked  cases, respectively. In both plots, the energy spectrum of the DNS (ground truth), LR input, and SR output are shown. When the particle channel is masked (Fig.\,\ref{fig:01004-00H_PFT-Prt_ArchTest_condSRGANModel/inference_testOnfx_masked_prt_ema/E_1D_4x_test_4x_gt_128_720000_img13200}), the SR spectrum decays rapidly at high wave-numbers, indicating a lack of small-scale energy. When the particle channel is included (Fig.\,\ref{fig:01004-00H_PFT-Prt_ArchTest_condSRGANModel/inference_testOnfx_ema/E_1D_4x_test_4x_gt_128_720000_img13200}), the SR spectrum follows the DNS much more closely across a broader range of wave-numbers, particularly beyond the LR cut-off. This confirms  that the network effectively utilizes the particle data to reconstruct subgrid-scale content and shows that the modulated subgrid-scale structures by the particles,  require particle information for a proper reconstruction. 
}

{To quantify the effect of particle data,  the normalized root mean square error (NRMSE)~\cite{ChengR2025_PF,ChungW2023_CONF} of unresolved ($NRMSE_{sp}(E_{1D,sgs})$) and resolved ($NRMSE_{sp}(E_{1D,res})$) parts of the kinetic energy spectrum, the stress tensor ($NRMSE(\mathcal{T}_{ij})$) and the velocity vector ($NRMSE(\bm{U})$) with respect to the DNS values  are reported in Tab.\,\ref{tab:mask_comparison}.  The statistics have been evaluated using the entire test dataset with 16,000 samples. } 
\begin{table}[h!] \footnotesize
	\centering
	\caption{Comparison of error metrics when including vs. masking particle information in the input of the generator during inference. The errors are calculated over the whole test dataset including 16,000 sampled 2-D planes.}
	{
		\begin{tabular}{llccc}
			
			\hline
			& $NRMSE_{sp}(E_{1D,sgs})$     &$NRMSE_{sp}(E_{1D,res})$ &$NRMSE(\mathcal{T}_{ij})$ & $NRMSE(\bm{U})$      \\
			\hline
			Including particle information  &  $ 4.6\cdot 10^{-2}$ &  $4.6\cdot 10^{-4}$   & $1.1\cdot 10^{-1}$   &  $4.4\cdot 10^{-2}$  \\
			Masking particle information  &  $ 6.8\cdot 10^{-2}$ &  $4.7\cdot 10^{-4}$   & $1.3\cdot 10^{-1}$   &  $4.8\cdot 10^{-2}$           \\
		\end{tabular}
	}
	
	\label{tab:mask_comparison}
\end{table}
{Note that the unresolved ($E_{1D,sgs}$) and resolved ($E_{1D,res}$) parts of the spectra are separated by the cut-off wave-number ($\kappa_{c}=\frac{\pi}{\Delta\times\Delta_{DNS}}$)~\cite{ChengR2025_PF}. 
When particle information is included, all error metrics are systematically lower compared to the masked case. For example, $NRMSE_{sp}(E_{1D,sgs})$ drops almost $30\%$, and $NRMSE(\mathcal{T}_{ij})$ improves from $1.3\cdot 10^{-1}$ to $1.1\cdot 10^{-1}$. These reductions demonstrate that providing particle data as a conditioning channel enables the generator network to more accurately reconstruct both subgrid and under-resolved energy scales, as well as the stress tensor and velocity fields. 
This quantitative improvement aligns with the qualitative and spectral results shown   (Fig.\,\ref{fig:01004-00H_PFT-Prt_ArchTest_condSRGANModel/inference_testOnfx_masked_prt_ema}),  confirming that the network is able to utilize particle information to achieve superior performance, especially for reconstructing small-scale features that are otherwise lost when the conditioning is masked.} 

Finally, in Tab.\,\ref{tab:error_u_at_prt}, the error in the prediction of $u_x$ at the particle positions is shown. The prediction of  velocity at the particle positions, i.e.  the velocity of the cells where particles are located is an important factor since it determines the computed drag force in the Euler-Lagrange simulations and thus it is directly related to the accuracy of the modeling of particle-turbulence interactions. In Tab.\,\ref{tab:error_u_at_prt}, mean and standard deviation of the mean square error between SR and DNS, and also between LR and DNS are reported. The error in the LR data  represents the no-model approach, i.e.  using the filtered velocity in the computation of the drag force. Moreover, the metric is evaluated on the whole test dataset, as well as a subset of the test dataset related to Case1 which has the lowest Stokes number based on Kolmogorov scales. As can be seen in both tests the error of the no-model approach   is approximately 10 times the ones of the SR. This shows that the SR model can effectively decrease the error of neglecting the subgrid dispersion.

\begin{table}[h!] \footnotesize
	\centering
	\caption{The mean and standard deviation of the error in the prediction of $u_x$ at the particle position for the whole test dataset including 16,000 sampled 2-D planes and a subset of the test dataset related to Case1 including 800 sampled 2-D planes.}
	{
		\begin{tabular}{llc|cc}
		 
			\hline
			&\multicolumn{2}{c|}{Entire test dataset} &\multicolumn{2}{c}{Case1 test dataset}    \\  
			\hline                     
			& SR     &LR& SR     &LR \\
			\hline
			Mean error  & 0.0039& 0.0445   & 0.0099& 0.0928             \\
			Std. error  & 0.0039& 0.0360  & 0.0011   & 0.0083            \\
		\end{tabular}
	}
	
	\label{tab:error_u_at_prt}
\end{table}

\subsubsection{Tests on the unseen physics}\label{sec:prt-data-unseen}
In the previous sections the test dataset included the samples which were not seen during the training, however, the nominal or target Stokes numbers were similar for both the training and test datasets. In this section the aim is to perform out of distribution tests. This is carried out by using a test dataset which is made based on the DNS cases in Tab.\,\ref{tab::OF128setup-particle_test} with different Stokes numbers than the training data. 
\begin{figure}[!htb]
	\centering
	
	\subfigure[$\mathrm{St}_{\eta} = 3\,(Case4)$]
	{
		\label{fig:01004-00H_PFT-Prt_ArchTest_condSRGANModel/inference_interpf41_st3/E_1D_4x_test_4x_gt_128_720000_img400}
		\includegraphics[width=.42\textwidth,trim=0.3cm 0.3cm 0.3cm 0.3cm, clip]{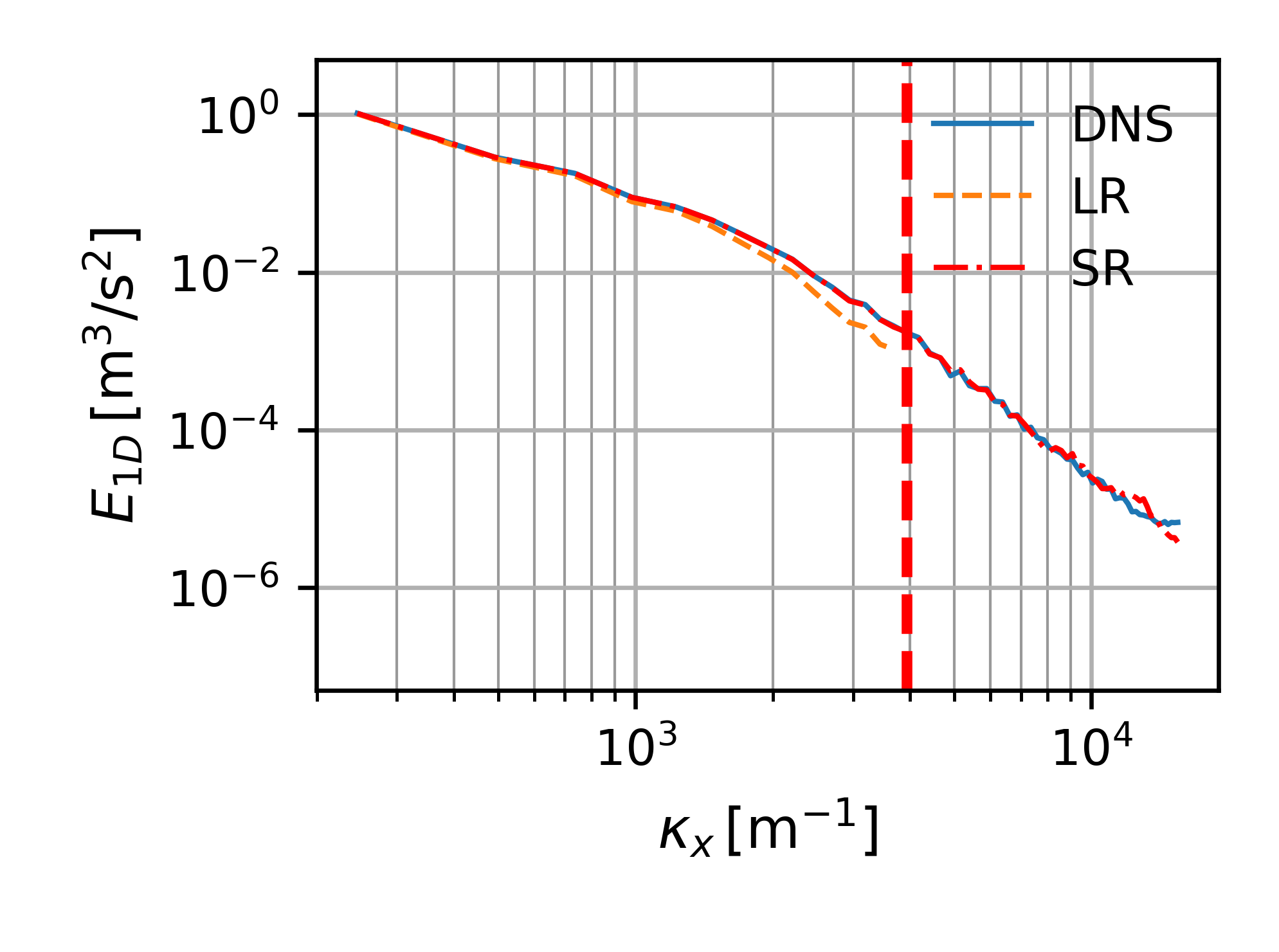}
	}
	\subfigure[$\mathrm{St}_{\eta} = 10\,(Case5)$]
	{
		\label{fig:01004-00H_PFT-Prt_ArchTest_condSRGANModel/inference_exterpf51_st10/E_1D_4x_test_4x_gt_128_720000_img0}
		\includegraphics[width=.42\textwidth,trim=0.3cm 0.3cm 0.3cm 0.3cm, clip]{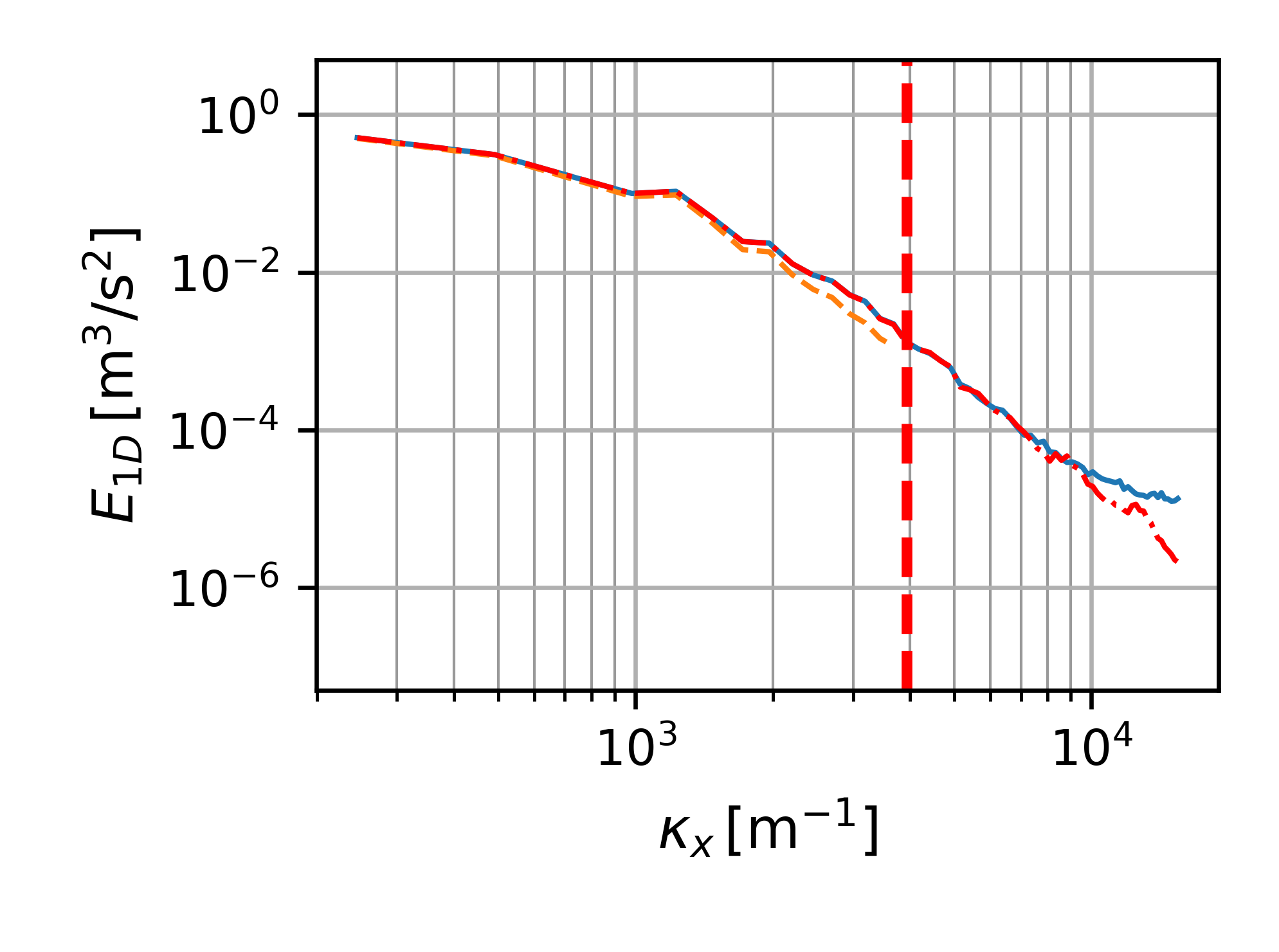}
	}
	
	\caption{One-dimensional kinetic energy spectrum: The model performance tested on cases with different Stokes numbers which were not included in the training. }
	\label{fig:01004-00H_PFT-Prt_ArchTest_condSRGANModel/inference_testOn41_51}
\end{figure}
In Figs.\,\ref{fig:01004-00H_PFT-Prt_ArchTest_condSRGANModel/inference_interpf41_st3/E_1D_4x_test_4x_gt_128_720000_img400}, and \ref{fig:01004-00H_PFT-Prt_ArchTest_condSRGANModel/inference_exterpf51_st10/E_1D_4x_test_4x_gt_128_720000_img0} the kinetic energy spectra for a randomly sampled plane from Case4 and Case5 with Stokes numbers 3 and 10, respectively, are shown. The model performs well for both interpolation and extrapolation in a wide range of wave-numbers.  Deviations can be seen at high wave-numbers for the extrapolation test (Fig.\,\ref{fig:01004-00H_PFT-Prt_ArchTest_condSRGANModel/inference_exterpf51_st10/E_1D_4x_test_4x_gt_128_720000_img0}) containing energies below 0.0001~$\mathrm{m^3/s^2}$.  This can be explained with the wavelet transformation.

\begin{figure}[!htb]
	\centering
	
	\subfigure[$St_{\eta} = 3\,(Case4)$]
	{
		\label{fig:01004-00H_PFT-Prt_ArchTest_condSRGANModel/inference_interpf41_st3/wavelet_2_test_4x_gt_128_720000_img400}
		\includegraphics[width=\textwidth,trim=0.3cm 0.3cm 0.3cm 0.3cm, clip]{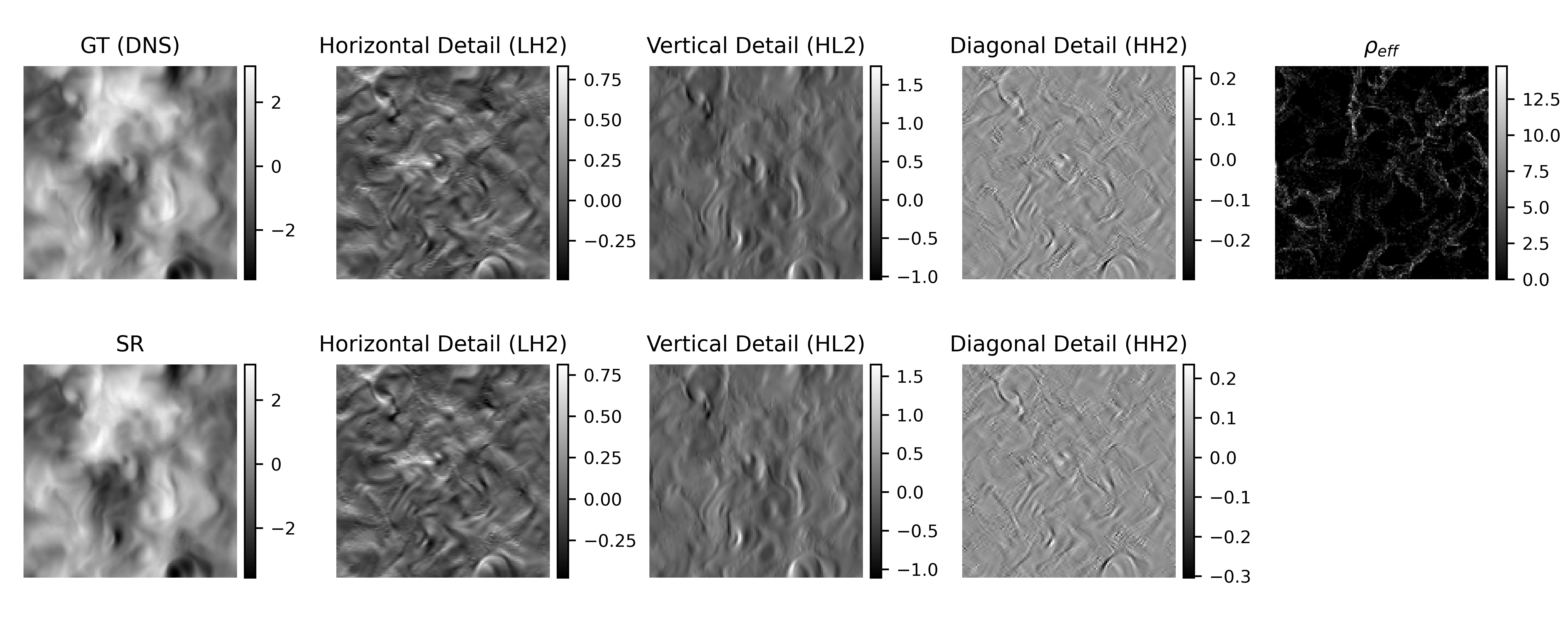}
	}
	\subfigure[$St_{\eta} = 10\,(Case5)$]
	{
		\label{fig:01004-00H_PFT-Prt_ArchTest_condSRGANModel/inference_exterpf51_st10/wavelet_2_test_4x_gt_128_720000_img0}
		\includegraphics[width=\textwidth,trim=0.3cm 0.3cm 0.3cm 0.3cm, clip]{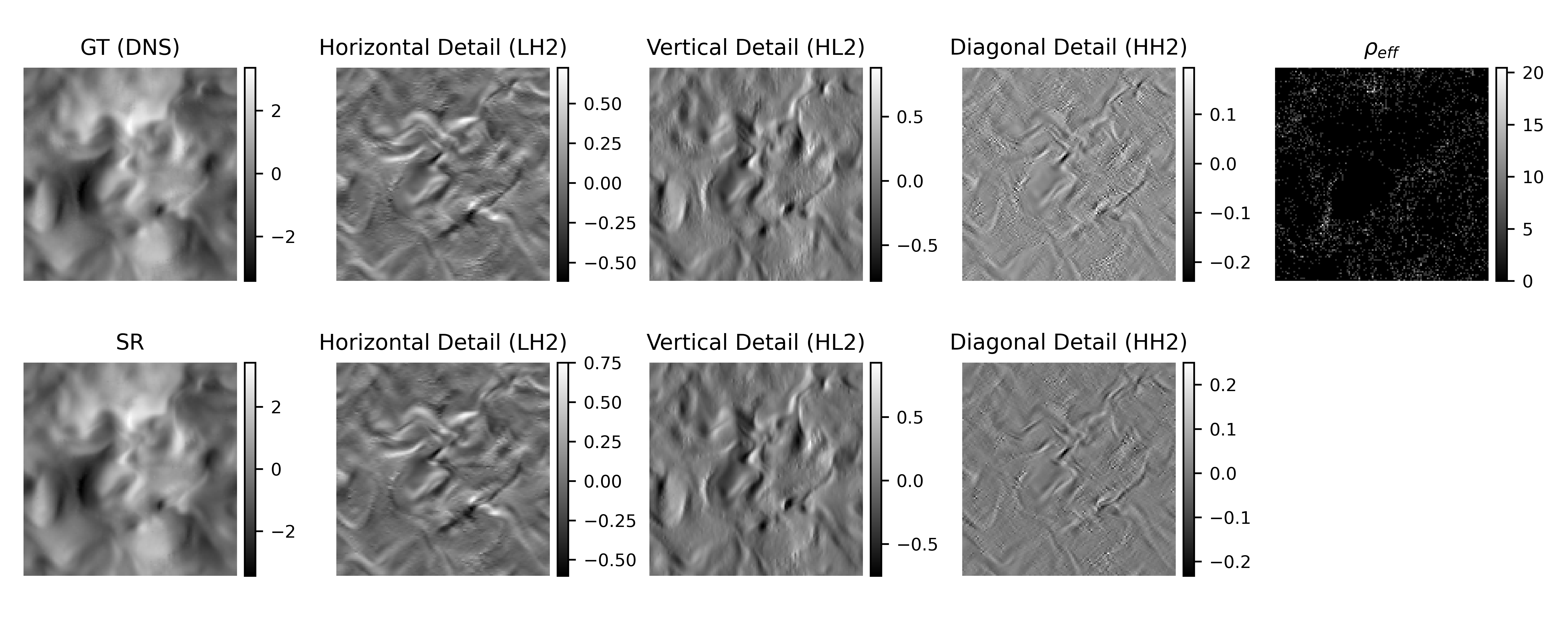}
	}
	
	\caption{Wavelet analysis on cases with different Stokes numbers not included in the training set. The first row of each figure displays the ground-truth DNS values while the second row shows the corresponding super-resolved fields. The first column shows the first component of velocity field, i.e. $u$ in $\mathrm{m/s}$, columns 2-4 show how the Haar wavelet detail coefficients at level two of the decomposition in $\mathrm{m/s}$, and the last column shows the dimensionless effective particle mass density from DNS. }
	\label{fig:01004-00H_PFT-Prt_ArchTest_condSRGANModel/inference_interpf41_st3/wavelet_2}
\end{figure}

In Fig.\,\ref{fig:01004-00H_PFT-Prt_ArchTest_condSRGANModel/inference_interpf41_st3/wavelet_2},  the coefficients of the wavelet transform of $u_x$ from the sampled planes analyzed in  Fig.\,\ref{fig:01004-00H_PFT-Prt_ArchTest_condSRGANModel/inference_testOn41_51} are shown. The particle information at the HR level from GT is also shown to highlight the local manipulation of small-scale structures of the gas velocity field.  
It is seen that in the interpolation test, Case4 with $\mathrm{St}_{\eta}=3$ shown in Fig.\,\ref{fig:01004-00H_PFT-Prt_ArchTest_condSRGANModel/inference_interpf41_st3/wavelet_2_test_4x_gt_128_720000_img400}, the location and morphology of  the extracted horizontal, vertical, and diagonal details at the second level of the wavelet transform of $u_x$ from SR agree  qualitatively well with the DNS counterparts. Furthermore, the magnitudes of the coefficients show a good quantitative match which confirms the finding in Fig.\,\ref{fig:01004-00H_PFT-Prt_ArchTest_condSRGANModel/inference_interpf41_st3/E_1D_4x_test_4x_gt_128_720000_img400} regarding the match of energy distribution  at different scales. In Fig.\,\ref{fig:01004-00H_PFT-Prt_ArchTest_condSRGANModel/inference_exterpf51_st10/wavelet_2_test_4x_gt_128_720000_img0}, however,  the magnitude of the diagonal coefficients (HH$_2$) shows discrepancies with respect to the DNS (peak values of around 0.2 as compared to 0.1 in the DNS) which may explain the deviations observed at high wave-numbers in Fig.\,\ref{fig:01004-00H_PFT-Prt_ArchTest_condSRGANModel/inference_exterpf51_st10/E_1D_4x_test_4x_gt_128_720000_img0}.


\section{Conclusions}

In this study, a deep learning-based super-resolution (SR) framework tailored for two-way coupled particle-laden turbulent flows has been introduced and validated. Employing conditional generative adversarial networks (cGANs), this framework effectively reconstructs high-resolution turbulent velocity fields from low-resolution (LR) data by explicitly conditioning on effective particle mass density and subgrid kinetic energy. Furthermore, the conditional discriminator network, conditioned on high-frequency structures extracted by stationary wavelet transformation and LR data forces the generator to achieve high fidelity in high frequencies while preserving the consistency with the LR input.  

The performance of the proposed SR framework was assessed using high-fidelity DNS datasets covering a broad range of particle Stokes numbers ($\mathrm{St}_{\eta} = 0.6, 1, 6$), mass loadings (ML$=0.49, 0.75$), and flow regimes, including  forced and decaying isotropic turbulence. 

Results from the validation analyses demonstrated the model's capability to accurately reconstruct energy spectra across a range of scales. 
Additionally, PDFs of turbulent quantities, such as the velocity components and vorticity, were closely reproduced, further validating the physical consistency of the reconstructed fields. 
The scatter plot of subgrid dissipation rate also illustrated the model's superiority in the prediction of subgrid dissipation rate and specifically the backscattering. 
{The effect of the particle data channel at low resolution, used as input to the generator, was investigated through an ablation study in which the particle data channel was masked. The results showed that the network utilizes the particle data to enhance subgrid structures that are modulated by the particles.  }
Wavelet-based  decomposition analyses showed that the proposed model is capable of resolving localized turbulent features and accurately predicting particle-induced modifications to turbulence structures. 
Particularly, good accuracies were obtained in test cases representing previously unseen physics, including cases with particle Stokes numbers ($\mathrm{St}_{\eta}$) of 3 and 10, demonstrating the generalization capability of the proposed SR framework.

\section*{Acknowledgments}
The authors acknowledge the financial support by the German Research Foundation (DFG) project  No. 513858356. A. Kronenburg and R. Chen  acknowledge the financial support from the China Scholarship Council (CSC) (Grant No. 202206020071). 
O. T. Stein  acknowledges the financial support by the Helmholtz Association of German Research Centers (HGF), within the research field \textit{Energy}, program \textit{Materials and Technologies for the Energy Transition} (MTET), topic \textit{Resource and Energy Efficiency}. The training was performed on the HoreKa supercomputer funded by the Ministry of Science, Research and the Arts Baden-W\"urttemberg and by the Federal Ministry of Education and Research. The HPC resources are provided by HLRS Stuttgart. 


\section*{Data Availability Statement}
The code and training/testing datasets are available upon request. 



\bibliography{ML-DATABASE.bib} 

\begin{thebibliography}{47}%
\makeatletter
\providecommand \@ifxundefined [1]{%
 \@ifx{#1\undefined}
}%
\providecommand \@ifnum [1]{%
 \ifnum #1\expandafter \@firstoftwo
 \else \expandafter \@secondoftwo
 \fi
}%
\providecommand \@ifx [1]{%
 \ifx #1\expandafter \@firstoftwo
 \else \expandafter \@secondoftwo
 \fi
}%
\providecommand \natexlab [1]{#1}%
\providecommand \enquote  [1]{``#1''}%
\providecommand \bibnamefont  [1]{#1}%
\providecommand \bibfnamefont [1]{#1}%
\providecommand \citenamefont [1]{#1}%
\providecommand \href@noop [0]{\@secondoftwo}%
\providecommand \href [0]{\begingroup \@sanitize@url \@href}%
\providecommand \@href[1]{\@@startlink{#1}\@@href}%
\providecommand \@@href[1]{\endgroup#1\@@endlink}%
\providecommand \@sanitize@url [0]{\catcode `\\12\catcode `\$12\catcode
  `\&12\catcode `\#12\catcode `\^12\catcode `\_12\catcode `\%12\relax}%
\providecommand \@@startlink[1]{}%
\providecommand \@@endlink[0]{}%
\providecommand \url  [0]{\begingroup\@sanitize@url \@url }%
\providecommand \@url [1]{\endgroup\@href {#1}{\urlprefix }}%
\providecommand \urlprefix  [0]{URL }%
\providecommand \Eprint [0]{\href }%
\providecommand \doibase [0]{http://dx.doi.org/}%
\providecommand \selectlanguage [0]{\@gobble}%
\providecommand \bibinfo  [0]{\@secondoftwo}%
\providecommand \bibfield  [0]{\@secondoftwo}%
\providecommand \translation [1]{[#1]}%
\providecommand \BibitemOpen [0]{}%
\providecommand \bibitemStop [0]{}%
\providecommand \bibitemNoStop [0]{.\EOS\space}%
\providecommand \EOS [0]{\spacefactor3000\relax}%
\providecommand \BibitemShut  [1]{\csname bibitem#1\endcsname}%
\let\auto@bib@innerbib\@empty
\bibitem [{\citenamefont {Sagaut}(2006)}]{SagautP2006_BOOK}%
  \BibitemOpen
  \bibfield  {author} {\bibinfo {author} {\bibfnamefont {P.}~\bibnamefont
  {Sagaut}},\ }\href {\doibase 10.1007/b137536} {\emph {\bibinfo {title}
  {{Large Eddy Simulation for Incompressible Flows}}}},\ \bibinfo {edition}
  {3rd}\ ed.\ (\bibinfo  {publisher} {Springer Berlin, Heidelberg},\ \bibinfo
  {year} {2006})\BibitemShut {NoStop}%
\bibitem [{\citenamefont {Stolz}\ and\ \citenamefont
  {Adams}(1999)}]{StolzS1999_PF}%
  \BibitemOpen
  \bibfield  {author} {\bibinfo {author} {\bibfnamefont {S.}~\bibnamefont
  {Stolz}}\ and\ \bibinfo {author} {\bibfnamefont {N.~A.}\ \bibnamefont
  {Adams}},\ }\bibfield  {title} {\enquote {\bibinfo {title} {{An approximate
  deconvolution procedure for large-eddy simulation}},}\ }\href {\doibase
  10.1063/1.869867} {\bibfield  {journal} {\bibinfo  {journal} {Phys. Fluids}\
  }\textbf {\bibinfo {volume} {11}},\ \bibinfo {pages} {1699--1701} (\bibinfo
  {year} {1999})}\BibitemShut {NoStop}%
\bibitem [{\citenamefont {Shotorban}\ and\ \citenamefont
  {Mashayek}(2005)}]{ShotorbanB2005_PF}%
  \BibitemOpen
  \bibfield  {author} {\bibinfo {author} {\bibfnamefont {B.}~\bibnamefont
  {Shotorban}}\ and\ \bibinfo {author} {\bibfnamefont {F.}~\bibnamefont
  {Mashayek}},\ }\bibfield  {title} {\enquote {\bibinfo {title} {{Modeling
  subgrid-scale effects on particles by approximate deconvolution}},}\ }\href
  {\doibase 10.1063/1.2001691} {\bibfield  {journal} {\bibinfo  {journal}
  {Phys. Fluids}\ }\textbf {\bibinfo {volume} {17}},\ \bibinfo {pages} {1--4}
  (\bibinfo {year} {2005})}\BibitemShut {NoStop}%
\bibitem [{\citenamefont {Domingo}\ and\ \citenamefont
  {Vervisch}(2015)}]{DomingoP2015_PCI}%
  \BibitemOpen
  \bibfield  {author} {\bibinfo {author} {\bibfnamefont {P.}~\bibnamefont
  {Domingo}}\ and\ \bibinfo {author} {\bibfnamefont {L.}~\bibnamefont
  {Vervisch}},\ }\bibfield  {title} {\enquote {\bibinfo {title} {{Large Eddy
  Simulation of premixed turbulent combustion using approximate deconvolution
  and explicit flame filtering}},}\ }\href {\doibase
  10.1016/j.proci.2014.05.146} {\bibfield  {journal} {\bibinfo  {journal}
  {Proc. Combust. Inst.}\ }\textbf {\bibinfo {volume} {35}},\ \bibinfo {pages}
  {1349--1357} (\bibinfo {year} {2015})}\BibitemShut {NoStop}%
\bibitem [{\citenamefont {Domingo}\ and\ \citenamefont
  {Vervisch}(2017)}]{DomingoP2017_CF}%
  \BibitemOpen
  \bibfield  {author} {\bibinfo {author} {\bibfnamefont {P.}~\bibnamefont
  {Domingo}}\ and\ \bibinfo {author} {\bibfnamefont {L.}~\bibnamefont
  {Vervisch}},\ }\bibfield  {title} {\enquote {\bibinfo {title} {{DNS and
  approximate deconvolution as a tool to analyse one-dimensional filtered flame
  sub-grid scale modelling}},}\ }\href {\doibase
  10.1016/j.combustflame.2016.12.008} {\bibfield  {journal} {\bibinfo
  {journal} {Combust. Flame}\ }\textbf {\bibinfo {volume} {177}},\ \bibinfo
  {pages} {109--122} (\bibinfo {year} {2017})}\BibitemShut {NoStop}%
\bibitem [{\citenamefont {Bassenne}\ \emph {et~al.}(2019)\citenamefont
  {Bassenne}, \citenamefont {Esmaily}, \citenamefont {Livescu}, \citenamefont
  {Moin},\ and\ \citenamefont {Urzay}}]{BassenneM2019_IJMF}%
  \BibitemOpen
  \bibfield  {author} {\bibinfo {author} {\bibfnamefont {M.}~\bibnamefont
  {Bassenne}}, \bibinfo {author} {\bibfnamefont {M.}~\bibnamefont {Esmaily}},
  \bibinfo {author} {\bibfnamefont {D.}~\bibnamefont {Livescu}}, \bibinfo
  {author} {\bibfnamefont {P.}~\bibnamefont {Moin}}, \ and\ \bibinfo {author}
  {\bibfnamefont {J.}~\bibnamefont {Urzay}},\ }\bibfield  {title} {\enquote
  {\bibinfo {title} {{A dynamic spectrally enriched subgrid-scale model for
  preferential concentration in particle-laden turbulence}},}\ }\href {\doibase
  10.1016/j.ijmultiphaseflow.2019.04.025} {\bibfield  {journal} {\bibinfo
  {journal} {Int. J. Multiph. Flow}\ }\textbf {\bibinfo {volume} {116}},\
  \bibinfo {pages} {270--280} (\bibinfo {year} {2019})}\BibitemShut {NoStop}%
\bibitem [{\citenamefont {Oberle}, \citenamefont {Pruett},\ and\ \citenamefont
  {Jenny}(2020)}]{OberleD2020_PF}%
  \BibitemOpen
  \bibfield  {author} {\bibinfo {author} {\bibfnamefont {D.}~\bibnamefont
  {Oberle}}, \bibinfo {author} {\bibfnamefont {C.~D.}\ \bibnamefont {Pruett}},
  \ and\ \bibinfo {author} {\bibfnamefont {P.}~\bibnamefont {Jenny}},\
  }\bibfield  {title} {\enquote {\bibinfo {title} {{Temporal large-eddy
  simulation based on direct deconvolution}},}\ }\href
  {https://doi.org/10.1063/5.0006637} {\bibfield  {journal} {\bibinfo
  {journal} {Phys. Fluids}\ }\textbf {\bibinfo {volume} {32}},\ \bibinfo
  {pages} {065112} (\bibinfo {year} {2020})}\BibitemShut {NoStop}%
\bibitem [{\citenamefont {Wang}\ and\ \citenamefont
  {Ihme}(2017)}]{WangQ2017_CF}%
  \BibitemOpen
  \bibfield  {author} {\bibinfo {author} {\bibfnamefont {Q.}~\bibnamefont
  {Wang}}\ and\ \bibinfo {author} {\bibfnamefont {M.}~\bibnamefont {Ihme}},\
  }\bibfield  {title} {\enquote {\bibinfo {title} {{Regularized deconvolution
  method for turbulent combustion modeling}},}\ }\href {\doibase
  10.1016/j.combustflame.2016.09.023} {\bibfield  {journal} {\bibinfo
  {journal} {Combust. Flame}\ }\textbf {\bibinfo {volume} {176}},\ \bibinfo
  {pages} {125--142} (\bibinfo {year} {2017})}\BibitemShut {NoStop}%
\bibitem [{\citenamefont {Wang}\ and\ \citenamefont
  {Ihme}(2019)}]{WangQ2019_CFa}%
  \BibitemOpen
  \bibfield  {author} {\bibinfo {author} {\bibfnamefont {Q.}~\bibnamefont
  {Wang}}\ and\ \bibinfo {author} {\bibfnamefont {M.}~\bibnamefont {Ihme}},\
  }\bibfield  {title} {\enquote {\bibinfo {title} {{A regularized deconvolution
  method for turbulent closure modeling in implicitly filtered large-eddy
  simulation}},}\ }\href {\doibase 10.1016/j.combustflame.2019.03.009}
  {\bibfield  {journal} {\bibinfo  {journal} {Combust. Flame}\ }\textbf
  {\bibinfo {volume} {204}},\ \bibinfo {pages} {341--355} (\bibinfo {year}
  {2019})}\BibitemShut {NoStop}%
\bibitem [{\citenamefont {Wang}, \citenamefont {Zhao},\ and\ \citenamefont
  {Ihme}(2019)}]{WangQ2019_CF}%
  \BibitemOpen
  \bibfield  {author} {\bibinfo {author} {\bibfnamefont {Q.}~\bibnamefont
  {Wang}}, \bibinfo {author} {\bibfnamefont {X.}~\bibnamefont {Zhao}}, \ and\
  \bibinfo {author} {\bibfnamefont {M.}~\bibnamefont {Ihme}},\ }\bibfield
  {title} {\enquote {\bibinfo {title} {{A regularized deconvolution model for
  sub-grid dispersion in large eddy simulation of turbulent spray flames}},}\
  }\href {\doibase 10.1016/j.combustflame.2019.05.032} {\bibfield  {journal}
  {\bibinfo  {journal} {Combust. Flame}\ }\textbf {\bibinfo {volume} {207}},\
  \bibinfo {pages} {89--100} (\bibinfo {year} {2019})}\BibitemShut {NoStop}%
\bibitem [{\citenamefont {Bardina}, \citenamefont {Ferziger},\ and\
  \citenamefont {Reynolds}(1980)}]{BARDINAJ1980_CONF}%
  \BibitemOpen
  \bibfield  {author} {\bibinfo {author} {\bibfnamefont {J.}~\bibnamefont
  {Bardina}}, \bibinfo {author} {\bibfnamefont {J.}~\bibnamefont {Ferziger}}, \
  and\ \bibinfo {author} {\bibfnamefont {W.}~\bibnamefont {Reynolds}},\
  }\bibfield  {title} {\enquote {\bibinfo {title} {{Improved subgrid-scale
  models for large-eddy simulation}},}\ }in\ \href {\doibase
  10.2514/6.1980-1357} {\emph {\bibinfo {booktitle} {13th Fluid and Plasma
  Dynamics Conference}}}\ (\bibinfo {year} {1980})\BibitemShut {NoStop}%
\bibitem [{\citenamefont {DesJardin}\ and\ \citenamefont
  {Frankel}(1998)}]{DesJardinP1998_PF}%
  \BibitemOpen
  \bibfield  {author} {\bibinfo {author} {\bibfnamefont {P.~E.}\ \bibnamefont
  {DesJardin}}\ and\ \bibinfo {author} {\bibfnamefont {S.~H.}\ \bibnamefont
  {Frankel}},\ }\bibfield  {title} {\enquote {\bibinfo {title} {{Large eddy
  simulation of a nonpremixed reacting jet: Application and assessment of
  subgrid-scale combustion models}},}\ }\href {\doibase 10.1063/1.869749}
  {\bibfield  {journal} {\bibinfo  {journal} {Phys. Fluids}\ }\textbf {\bibinfo
  {volume} {10}},\ \bibinfo {pages} {2298} (\bibinfo {year}
  {1998})}\BibitemShut {NoStop}%
\bibitem [{\citenamefont {Shamooni}\ \emph
  {et~al.}(2020{\natexlab{a}})\citenamefont {Shamooni}, \citenamefont {Cuoci},
  \citenamefont {Faravelli},\ and\ \citenamefont
  {Sadiki}}]{ShamooniA2020_FTCa}%
  \BibitemOpen
  \bibfield  {author} {\bibinfo {author} {\bibfnamefont {A.}~\bibnamefont
  {Shamooni}}, \bibinfo {author} {\bibfnamefont {A.}~\bibnamefont {Cuoci}},
  \bibinfo {author} {\bibfnamefont {T.}~\bibnamefont {Faravelli}}, \ and\
  \bibinfo {author} {\bibfnamefont {A.}~\bibnamefont {Sadiki}},\ }\bibfield
  {title} {\enquote {\bibinfo {title} {{New Dynamic Scale Similarity Based
  Finite-Rate Combustion Models for LES and a priori DNS Assessment in
  Non-premixed Jet Flames with High Level of Local Extinction}},}\ }\href
  {\doibase 10.1007/s10494-019-00060-w} {\bibfield  {journal} {\bibinfo
  {journal} {Flow, Turbul. Combust.}\ }\textbf {\bibinfo {volume} {104}},\
  \bibinfo {pages} {233--260} (\bibinfo {year}
  {2020}{\natexlab{a}})}\BibitemShut {NoStop}%
\bibitem [{\citenamefont {Shamooni}\ \emph
  {et~al.}(2020{\natexlab{b}})\citenamefont {Shamooni}, \citenamefont {Cuoci},
  \citenamefont {Faravelli},\ and\ \citenamefont {Sadiki}}]{ShamooniA2020_FTC}%
  \BibitemOpen
  \bibfield  {author} {\bibinfo {author} {\bibfnamefont {A.}~\bibnamefont
  {Shamooni}}, \bibinfo {author} {\bibfnamefont {A.}~\bibnamefont {Cuoci}},
  \bibinfo {author} {\bibfnamefont {T.}~\bibnamefont {Faravelli}}, \ and\
  \bibinfo {author} {\bibfnamefont {A.}~\bibnamefont {Sadiki}},\ }\bibfield
  {title} {\enquote {\bibinfo {title} {{An a priori DNS analysis of scale
  similarity based combustion models for LES of non-premixed jet flames}},}\
  }\href {\doibase 10.1007/s10494-019-00099-9} {\bibfield  {journal} {\bibinfo
  {journal} {Flow, Turbul. Combust.}\ }\textbf {\bibinfo {volume} {104}},\
  \bibinfo {pages} {605--624} (\bibinfo {year}
  {2020}{\natexlab{b}})}\BibitemShut {NoStop}%
\bibitem [{\citenamefont {Ferrante}\ and\ \citenamefont
  {Elghobashi}(2003)}]{FerranteA2003_PF}%
  \BibitemOpen
  \bibfield  {author} {\bibinfo {author} {\bibfnamefont {A.}~\bibnamefont
  {Ferrante}}\ and\ \bibinfo {author} {\bibfnamefont {S.}~\bibnamefont
  {Elghobashi}},\ }\bibfield  {title} {\enquote {\bibinfo {title} {{On the
  physical mechanisms of two-way coupling in particle-laden isotropic
  turbulence}},}\ }\href {\doibase 10.1063/1.1532731} {\bibfield  {journal}
  {\bibinfo  {journal} {Phys. Fluids}\ }\textbf {\bibinfo {volume} {15}},\
  \bibinfo {pages} {315--329} (\bibinfo {year} {2003})}\BibitemShut {NoStop}%
\bibitem [{\citenamefont {Abdelsamie}\ and\ \citenamefont
  {Lee}(2012)}]{AbdelsamieA2012_PF}%
  \BibitemOpen
  \bibfield  {author} {\bibinfo {author} {\bibfnamefont {A.~H.}\ \bibnamefont
  {Abdelsamie}}\ and\ \bibinfo {author} {\bibfnamefont {C.}~\bibnamefont
  {Lee}},\ }\bibfield  {title} {\enquote {\bibinfo {title} {{Decaying versus
  stationary turbulence in particle-laden isotropic turbulence: Turbulence
  modulation mechanism}},}\ }\href {\doibase 10.1063/1.3678332} {\bibfield
  {journal} {\bibinfo  {journal} {Phys. Fluids}\ }\textbf {\bibinfo {volume}
  {24}},\ \bibinfo {pages} {015106} (\bibinfo {year} {2012})}\BibitemShut
  {NoStop}%
\bibitem [{\citenamefont {Elghobashi}(2019)}]{ElghobashiS2019_ARFM}%
  \BibitemOpen
  \bibfield  {author} {\bibinfo {author} {\bibfnamefont {S.}~\bibnamefont
  {Elghobashi}},\ }\bibfield  {title} {\enquote {\bibinfo {title} {{Direct
  Numerical Simulation of Turbulent Flows Laden with Droplets or Bubbles}},}\
  }\href {\doibase 10.1146/annurev-fluid-010518-040401} {\bibfield  {journal}
  {\bibinfo  {journal} {Annu. Rev. Fluid Mech.}\ }\textbf {\bibinfo {volume}
  {51}},\ \bibinfo {pages} {217--244} (\bibinfo {year} {2019})}\BibitemShut
  {NoStop}%
\bibitem [{\citenamefont {Brunton}, \citenamefont {Noack},\ and\ \citenamefont
  {Koumoutsakos}(2020)}]{BruntonS2020_ARFM}%
  \BibitemOpen
  \bibfield  {author} {\bibinfo {author} {\bibfnamefont {S.~L.}\ \bibnamefont
  {Brunton}}, \bibinfo {author} {\bibfnamefont {B.~R.}\ \bibnamefont {Noack}},
  \ and\ \bibinfo {author} {\bibfnamefont {P.}~\bibnamefont {Koumoutsakos}},\
  }\bibfield  {title} {\enquote {\bibinfo {title} {{Machine Learning for Fluid
  Mechanics}},}\ }\href {\doibase 10.1146/annurev-fluid-010719-060214}
  {\bibfield  {journal} {\bibinfo  {journal} {Annu. Rev. Fluid Mech.}\ }\textbf
  {\bibinfo {volume} {52}},\ \bibinfo {pages} {477--508} (\bibinfo {year}
  {2020})}\BibitemShut {NoStop}%
\bibitem [{\citenamefont {Fukami}, \citenamefont {Fukagata},\ and\
  \citenamefont {Taira}(2019)}]{FukamiK2019_JFM}%
  \BibitemOpen
  \bibfield  {author} {\bibinfo {author} {\bibfnamefont {K.}~\bibnamefont
  {Fukami}}, \bibinfo {author} {\bibfnamefont {K.}~\bibnamefont {Fukagata}}, \
  and\ \bibinfo {author} {\bibfnamefont {K.}~\bibnamefont {Taira}},\ }\bibfield
   {title} {\enquote {\bibinfo {title} {{Super-resolution reconstruction of
  turbulent flows with machine learning}},}\ }\href {\doibase
  10.1017/jfm.2019.238} {\bibfield  {journal} {\bibinfo  {journal} {J. Fluid
  Mech.}\ }\textbf {\bibinfo {volume} {870}},\ \bibinfo {pages} {106--120}
  (\bibinfo {year} {2019})}\BibitemShut {NoStop}%
\bibitem [{\citenamefont {Duraisamy}(2021)}]{DuraisamyK2021_PRF}%
  \BibitemOpen
  \bibfield  {author} {\bibinfo {author} {\bibfnamefont {K.}~\bibnamefont
  {Duraisamy}},\ }\bibfield  {title} {\enquote {\bibinfo {title} {{Perspectives
  on machine learning-augmented Reynolds-averaged and large eddy simulation
  models of turbulence}},}\ }\href {\doibase 10.1103/PhysRevFluids.6.050504}
  {\bibfield  {journal} {\bibinfo  {journal} {Phys. Rev. Fluids}\ }\textbf
  {\bibinfo {volume} {6}},\ \bibinfo {pages} {050504} (\bibinfo {year}
  {2021})}\BibitemShut {NoStop}%
\bibitem [{\citenamefont {Kim}\ \emph {et~al.}(2020)\citenamefont {Kim},
  \citenamefont {Kim}, \citenamefont {Won},\ and\ \citenamefont
  {Lee}}]{KimH2020_JFM}%
  \BibitemOpen
  \bibfield  {author} {\bibinfo {author} {\bibfnamefont {H.}~\bibnamefont
  {Kim}}, \bibinfo {author} {\bibfnamefont {J.}~\bibnamefont {Kim}}, \bibinfo
  {author} {\bibfnamefont {S.}~\bibnamefont {Won}}, \ and\ \bibinfo {author}
  {\bibfnamefont {C.}~\bibnamefont {Lee}},\ }\bibfield  {title} {\enquote
  {\bibinfo {title} {{Unsupervised deep learning for super-resolution
  reconstruction of turbulence}},}\ }\href {\doibase 10.1017/jfm.2020.1028}
  {\bibfield  {journal} {\bibinfo  {journal} {J. Fluid Mech.}\ }\textbf
  {\bibinfo {volume} {910}},\ \bibinfo {pages} {A29} (\bibinfo {year}
  {2020})}\BibitemShut {NoStop}%
\bibitem [{\citenamefont {Bode}\ \emph {et~al.}(2021)\citenamefont {Bode},
  \citenamefont {Gauding}, \citenamefont {Lian}, \citenamefont {Denker},
  \citenamefont {Davidovic}, \citenamefont {Kleinheinz}, \citenamefont
  {Jitsev},\ and\ \citenamefont {Pitsch}}]{BodeM2021_PCI}%
  \BibitemOpen
  \bibfield  {author} {\bibinfo {author} {\bibfnamefont {M.}~\bibnamefont
  {Bode}}, \bibinfo {author} {\bibfnamefont {M.}~\bibnamefont {Gauding}},
  \bibinfo {author} {\bibfnamefont {Z.}~\bibnamefont {Lian}}, \bibinfo {author}
  {\bibfnamefont {D.}~\bibnamefont {Denker}}, \bibinfo {author} {\bibfnamefont
  {M.}~\bibnamefont {Davidovic}}, \bibinfo {author} {\bibfnamefont
  {K.}~\bibnamefont {Kleinheinz}}, \bibinfo {author} {\bibfnamefont
  {J.}~\bibnamefont {Jitsev}}, \ and\ \bibinfo {author} {\bibfnamefont
  {H.}~\bibnamefont {Pitsch}},\ }\bibfield  {title} {\enquote {\bibinfo {title}
  {{Using physics-informed enhanced super-resolution generative adversarial
  networks for subfilter modeling in turbulent reactive flows}},}\ }\href
  {\doibase 10.1016/j.proci.2020.06.022} {\bibfield  {journal} {\bibinfo
  {journal} {Proc. Combust. Inst.}\ }\textbf {\bibinfo {volume} {38}},\
  \bibinfo {pages} {2617--2625} (\bibinfo {year} {2021})}\BibitemShut {NoStop}%
\bibitem [{\citenamefont {Bode}\ \emph {et~al.}(2023)\citenamefont {Bode},
  \citenamefont {Gauding}, \citenamefont {Goeb}, \citenamefont {Falkenstein},\
  and\ \citenamefont {Pitsch}}]{BodeM2023_PCI}%
  \BibitemOpen
  \bibfield  {author} {\bibinfo {author} {\bibfnamefont {M.}~\bibnamefont
  {Bode}}, \bibinfo {author} {\bibfnamefont {M.}~\bibnamefont {Gauding}},
  \bibinfo {author} {\bibfnamefont {D.}~\bibnamefont {Goeb}}, \bibinfo {author}
  {\bibfnamefont {T.}~\bibnamefont {Falkenstein}}, \ and\ \bibinfo {author}
  {\bibfnamefont {H.}~\bibnamefont {Pitsch}},\ }\bibfield  {title} {\enquote
  {\bibinfo {title} {Applying physics-informed enhanced super-resolution
  generative adversarial networks to turbulent premixed combustion and
  engine-like flame kernel direct numerical simulation data},}\ }\href
  {\doibase https://doi.org/10.1016/j.proci.2022.07.254} {\bibfield  {journal}
  {\bibinfo  {journal} {Proc. Combust. Inst.}\ }\textbf {\bibinfo {volume}
  {39}},\ \bibinfo {pages} {5289--5298} (\bibinfo {year} {2023})}\BibitemShut
  {NoStop}%
\bibitem [{\citenamefont {Fukami}, \citenamefont {Fukagata},\ and\
  \citenamefont {Taira}(2023)}]{FukamiK2023_TCFD}%
  \BibitemOpen
  \bibfield  {author} {\bibinfo {author} {\bibfnamefont {K.}~\bibnamefont
  {Fukami}}, \bibinfo {author} {\bibfnamefont {K.}~\bibnamefont {Fukagata}}, \
  and\ \bibinfo {author} {\bibfnamefont {K.}~\bibnamefont {Taira}},\ }\bibfield
   {title} {\enquote {\bibinfo {title} {{Super-resolution analysis via machine
  learning: a survey for fluid flows}},}\ }\href {\doibase
  10.1007/s00162-023-00663-0} {\bibfield  {journal} {\bibinfo  {journal}
  {Theor. Comput. Fluid Dyn.}\ }\textbf {\bibinfo {volume} {37}},\ \bibinfo
  {pages} {421--444} (\bibinfo {year} {2023})}\BibitemShut {NoStop}%
\bibitem [{\citenamefont {Tofighian}, \citenamefont {Denev},\ and\
  \citenamefont {Kornev}(2024)}]{TofighianH2024_PF}%
  \BibitemOpen
  \bibfield  {author} {\bibinfo {author} {\bibfnamefont {H.}~\bibnamefont
  {Tofighian}}, \bibinfo {author} {\bibfnamefont {J.~A.}\ \bibnamefont
  {Denev}}, \ and\ \bibinfo {author} {\bibfnamefont {N.}~\bibnamefont
  {Kornev}},\ }\bibfield  {title} {\enquote {\bibinfo {title} {{A conditional
  deep learning model for super-resolution reconstruction of small-scale
  turbulent structures in particle-Laden flows}},}\ }\href {\doibase
  10.1063/5.0235192} {\bibfield  {journal} {\bibinfo  {journal} {Phys. Fluids}\
  }\textbf {\bibinfo {volume} {36}},\ \bibinfo {pages} {115173} (\bibinfo
  {year} {2024})}\BibitemShut {NoStop}%
\bibitem [{\citenamefont {Nista}\ \emph {et~al.}(2025)\citenamefont {Nista},
  \citenamefont {Schumann}, \citenamefont {Petkov}, \citenamefont {Pavlov},
  \citenamefont {Grenga}, \citenamefont {MacArt}, \citenamefont {Attili},
  \citenamefont {Markov},\ and\ \citenamefont {Pitsch}}]{NistaL2025_CF}%
  \BibitemOpen
  \bibfield  {author} {\bibinfo {author} {\bibfnamefont {L.}~\bibnamefont
  {Nista}}, \bibinfo {author} {\bibfnamefont {C.~D.}\ \bibnamefont {Schumann}},
  \bibinfo {author} {\bibfnamefont {P.}~\bibnamefont {Petkov}}, \bibinfo
  {author} {\bibfnamefont {V.}~\bibnamefont {Pavlov}}, \bibinfo {author}
  {\bibfnamefont {T.}~\bibnamefont {Grenga}}, \bibinfo {author} {\bibfnamefont
  {J.~F.}\ \bibnamefont {MacArt}}, \bibinfo {author} {\bibfnamefont
  {A.}~\bibnamefont {Attili}}, \bibinfo {author} {\bibfnamefont
  {S.}~\bibnamefont {Markov}}, \ and\ \bibinfo {author} {\bibfnamefont
  {H.}~\bibnamefont {Pitsch}},\ }\bibfield  {title} {\enquote {\bibinfo {title}
  {{Parallel implementation and performance of super-resolution generative
  adversarial network turbulence models for large-eddy simulation}},}\ }\href
  {\doibase 10.1016/j.compfluid.2024.106498} {\bibfield  {journal} {\bibinfo
  {journal} {Comput. Fluids}\ }\textbf {\bibinfo {volume} {288}},\ \bibinfo
  {pages} {106498} (\bibinfo {year} {2025})}\BibitemShut {NoStop}%
\bibitem [{\citenamefont {Cheng}\ \emph {et~al.}(2025)\citenamefont {Cheng},
  \citenamefont {Shamooni}, \citenamefont {Zirwes},\ and\ \citenamefont
  {Kronenburg}}]{ChengR2025_PF}%
  \BibitemOpen
  \bibfield  {author} {\bibinfo {author} {\bibfnamefont {R.}~\bibnamefont
  {Cheng}}, \bibinfo {author} {\bibfnamefont {A.}~\bibnamefont {Shamooni}},
  \bibinfo {author} {\bibfnamefont {T.}~\bibnamefont {Zirwes}}, \ and\ \bibinfo
  {author} {\bibfnamefont {A.}~\bibnamefont {Kronenburg}},\ }\bibfield  {title}
  {\enquote {\bibinfo {title} {{Improved super-resolution reconstruction of
  turbulent flows with spectral loss function}},}\ }\href {\doibase
  10.1063/5.0258090} {\bibfield  {journal} {\bibinfo  {journal} {Phys. Fluids}\
  }\textbf {\bibinfo {volume} {37}},\ \bibinfo {pages} {035208} (\bibinfo
  {year} {2025})}\BibitemShut {NoStop}%
\bibitem [{\citenamefont {Goodfellow}\ \emph {et~al.}(2020)\citenamefont
  {Goodfellow}, \citenamefont {Pouget-Abadie}, \citenamefont {Mirza},
  \citenamefont {Xu}, \citenamefont {Warde-Farley}, \citenamefont {Ozair},
  \citenamefont {Courville},\ and\ \citenamefont
  {Bengio}}]{GoodfellowI2020_CA}%
  \BibitemOpen
  \bibfield  {author} {\bibinfo {author} {\bibfnamefont {I.}~\bibnamefont
  {Goodfellow}}, \bibinfo {author} {\bibfnamefont {J.}~\bibnamefont
  {Pouget-Abadie}}, \bibinfo {author} {\bibfnamefont {M.}~\bibnamefont
  {Mirza}}, \bibinfo {author} {\bibfnamefont {B.}~\bibnamefont {Xu}}, \bibinfo
  {author} {\bibfnamefont {D.}~\bibnamefont {Warde-Farley}}, \bibinfo {author}
  {\bibfnamefont {S.}~\bibnamefont {Ozair}}, \bibinfo {author} {\bibfnamefont
  {A.}~\bibnamefont {Courville}}, \ and\ \bibinfo {author} {\bibfnamefont
  {Y.}~\bibnamefont {Bengio}},\ }\bibfield  {title} {\enquote {\bibinfo {title}
  {{Generative adversarial networks}},}\ }\href {\doibase 10.1145/3422622}
  {\bibfield  {journal} {\bibinfo  {journal} {Commun. ACM}\ }\textbf {\bibinfo
  {volume} {63}},\ \bibinfo {pages} {139--144} (\bibinfo {year}
  {2020})}\BibitemShut {NoStop}%
\bibitem [{\citenamefont {Mirza}\ and\ \citenamefont
  {Osindero}(2014)}]{MirzaM2014_arXiv}%
  \BibitemOpen
  \bibfield  {author} {\bibinfo {author} {\bibfnamefont {M.}~\bibnamefont
  {Mirza}}\ and\ \bibinfo {author} {\bibfnamefont {S.}~\bibnamefont
  {Osindero}},\ }\href {http://arxiv.org/abs/1411.1784} {\enquote {\bibinfo
  {title} {{Conditional Generative Adversarial Nets}},}\ } (\bibinfo {year}
  {2014})\BibitemShut {NoStop}%
\bibitem [{\citenamefont {Wang}\ \emph {et~al.}(2019)\citenamefont {Wang},
  \citenamefont {Yu}, \citenamefont {Wu}, \citenamefont {Gu}, \citenamefont
  {Liu}, \citenamefont {Dong}, \citenamefont {Qiao},\ and\ \citenamefont
  {Loy}}]{WangX2019_}%
  \BibitemOpen
  \bibfield  {author} {\bibinfo {author} {\bibfnamefont {X.}~\bibnamefont
  {Wang}}, \bibinfo {author} {\bibfnamefont {K.}~\bibnamefont {Yu}}, \bibinfo
  {author} {\bibfnamefont {S.}~\bibnamefont {Wu}}, \bibinfo {author}
  {\bibfnamefont {J.}~\bibnamefont {Gu}}, \bibinfo {author} {\bibfnamefont
  {Y.}~\bibnamefont {Liu}}, \bibinfo {author} {\bibfnamefont {C.}~\bibnamefont
  {Dong}}, \bibinfo {author} {\bibfnamefont {Y.}~\bibnamefont {Qiao}}, \ and\
  \bibinfo {author} {\bibfnamefont {C.~C.}\ \bibnamefont {Loy}},\ }\bibfield
  {title} {\enquote {\bibinfo {title} {{ESRGAN: Enhanced super-resolution
  generative adversarial networks}},}\ }in\ \href {\doibase
  10.1007/978-3-030-11021-5_5} {\emph {\bibinfo {booktitle} {Lect. Notes
  Comput. Sci. (including Subser. Lect. Notes Artif. Intell. Lect. Notes
  Bioinformatics)}}},\ Vol.\ \bibinfo {volume} {11133 LNCS}\ (\bibinfo {year}
  {2019})\ pp.\ \bibinfo {pages} {63--79}\BibitemShut {NoStop}%
\bibitem [{\citenamefont {Chung}\ \emph {et~al.}(2023)\citenamefont {Chung},
  \citenamefont {Akoush}, \citenamefont {Sharma}, \citenamefont {Tamkin},
  \citenamefont {Jung}, \citenamefont {Chen}, \citenamefont {Guo},
  \citenamefont {Brouzet}, \citenamefont {Talei}, \citenamefont {Savard},
  \citenamefont {Poludnenko},\ and\ \citenamefont {Ihme}}]{ChungW2023_CONF}%
  \BibitemOpen
  \bibfield  {author} {\bibinfo {author} {\bibfnamefont {W.~T.}\ \bibnamefont
  {Chung}}, \bibinfo {author} {\bibfnamefont {B.}~\bibnamefont {Akoush}},
  \bibinfo {author} {\bibfnamefont {P.}~\bibnamefont {Sharma}}, \bibinfo
  {author} {\bibfnamefont {A.}~\bibnamefont {Tamkin}}, \bibinfo {author}
  {\bibfnamefont {K.~S.}\ \bibnamefont {Jung}}, \bibinfo {author}
  {\bibfnamefont {J.~H.}\ \bibnamefont {Chen}}, \bibinfo {author}
  {\bibfnamefont {J.}~\bibnamefont {Guo}}, \bibinfo {author} {\bibfnamefont
  {D.}~\bibnamefont {Brouzet}}, \bibinfo {author} {\bibfnamefont
  {M.}~\bibnamefont {Talei}}, \bibinfo {author} {\bibfnamefont
  {B.}~\bibnamefont {Savard}}, \bibinfo {author} {\bibfnamefont {A.~Y.}\
  \bibnamefont {Poludnenko}}, \ and\ \bibinfo {author} {\bibfnamefont
  {M.}~\bibnamefont {Ihme}},\ }\bibfield  {title} {\enquote {\bibinfo {title}
  {{Turbulence in Focus: Benchmarking Scaling Behavior of 3D Volumetric
  Super-Resolution with BLASTNet 2.0 Data}},}\ }in\ \href {\doibase
  https://doi.org/10.48550/arXiv.2309.13457} {\emph {\bibinfo {booktitle} {37th
  Conf. Neural Inf. Process. Syst. (NeurIPS 2023)}}}\ (\bibinfo {year} {2023})\
  pp.\ \bibinfo {pages} {1--55}\BibitemShut {NoStop}%
\bibitem [{\citenamefont {Nista}\ \emph {et~al.}(2023)\citenamefont {Nista},
  \citenamefont {Schumann}, \citenamefont {Grenga}, \citenamefont {Attili},\
  and\ \citenamefont {Pitsch}}]{NistaL2023_PCI}%
  \BibitemOpen
  \bibfield  {author} {\bibinfo {author} {\bibfnamefont {L.}~\bibnamefont
  {Nista}}, \bibinfo {author} {\bibfnamefont {C.}~\bibnamefont {Schumann}},
  \bibinfo {author} {\bibfnamefont {T.}~\bibnamefont {Grenga}}, \bibinfo
  {author} {\bibfnamefont {A.}~\bibnamefont {Attili}}, \ and\ \bibinfo {author}
  {\bibfnamefont {H.}~\bibnamefont {Pitsch}},\ }\bibfield  {title} {\enquote
  {\bibinfo {title} {{Investigation of the generalization capability of a
  generative adversarial network for large eddy simulation of turbulent
  premixed reacting flows}},}\ }\href {\doibase 10.1016/j.proci.2022.07.244}
  {\bibfield  {journal} {\bibinfo  {journal} {Proc. Combust. Inst.}\ }\textbf
  {\bibinfo {volume} {39}},\ \bibinfo {pages} {5279--5288} (\bibinfo {year}
  {2023})}\BibitemShut {NoStop}%
\bibitem [{\citenamefont {Sundararajan}(2016)}]{SundararajanD2016_BOOK}%
  \BibitemOpen
  \bibfield  {author} {\bibinfo {author} {\bibfnamefont {D.}~\bibnamefont
  {Sundararajan}},\ }\href {\doibase https://doi.org/10.1002/9781119113119}
  {\emph {\bibinfo {title} {Discrete Wavelet Transform: A Signal Processing
  Approach}}},\ CourseSmart Series\ (\bibinfo  {publisher} {Wiley},\ \bibinfo
  {year} {2016})\BibitemShut {NoStop}%
\bibitem [{\citenamefont {Cotter}(2019)}]{CotterF2019_THESIS}%
  \BibitemOpen
  \bibfield  {author} {\bibinfo {author} {\bibfnamefont {F.}~\bibnamefont
  {Cotter}},\ }\emph {\bibinfo {title} {Uses of Complex Wavelets in Deep
  Convolutional Neural Networks}},\ \href {\doibase 10.17863/CAM.53748} {Ph.D.
  thesis},\ \bibinfo  {school} {Apollo - University of Cambridge Repository}
  (\bibinfo {year} {2019})\BibitemShut {NoStop}%
\bibitem [{\citenamefont {Wang}\ \emph {et~al.}(2023)\citenamefont {Wang},
  \citenamefont {Mahler}, \citenamefont {Steiglechner}, \citenamefont {Birk},
  \citenamefont {Scheffler},\ and\ \citenamefont {Lohmann}}]{WangQ2023_CONF}%
  \BibitemOpen
  \bibfield  {author} {\bibinfo {author} {\bibfnamefont {Q.}~\bibnamefont
  {Wang}}, \bibinfo {author} {\bibfnamefont {L.}~\bibnamefont {Mahler}},
  \bibinfo {author} {\bibfnamefont {J.}~\bibnamefont {Steiglechner}}, \bibinfo
  {author} {\bibfnamefont {F.}~\bibnamefont {Birk}}, \bibinfo {author}
  {\bibfnamefont {K.}~\bibnamefont {Scheffler}}, \ and\ \bibinfo {author}
  {\bibfnamefont {G.}~\bibnamefont {Lohmann}},\ }\bibfield  {title} {\enquote
  {\bibinfo {title} {{DISGAN: Wavelet-informed Discriminator Guides GAN to MRI
  Super-resolution with Noise Cleaning}},}\ }in\ \href {\doibase
  10.1109/ICCVW60793.2023.00259} {\emph {\bibinfo {booktitle} {2023 IEEE/CVF
  Int. Conf. Comput. Vis. Work.}}}\ (\bibinfo  {publisher} {IEEE},\ \bibinfo
  {year} {2023})\ pp.\ \bibinfo {pages} {2444--2453}\BibitemShut {NoStop}%
\bibitem [{\citenamefont {Xu}\ \emph {et~al.}(2025)\citenamefont {Xu},
  \citenamefont {Zhou}, \citenamefont {Ma}, \citenamefont {Yang}, \citenamefont
  {Wang}, \citenamefont {Zhang},\ and\ \citenamefont {Li}}]{XuY2025_KS}%
  \BibitemOpen
  \bibfield  {author} {\bibinfo {author} {\bibfnamefont {Y.}~\bibnamefont
  {Xu}}, \bibinfo {author} {\bibfnamefont {Y.}~\bibnamefont {Zhou}}, \bibinfo
  {author} {\bibfnamefont {H.}~\bibnamefont {Ma}}, \bibinfo {author}
  {\bibfnamefont {H.}~\bibnamefont {Yang}}, \bibinfo {author} {\bibfnamefont
  {H.}~\bibnamefont {Wang}}, \bibinfo {author} {\bibfnamefont {S.}~\bibnamefont
  {Zhang}}, \ and\ \bibinfo {author} {\bibfnamefont {X.}~\bibnamefont {Li}},\
  }\bibfield  {title} {\enquote {\bibinfo {title} {{Wavelet-based dual
  discriminator GAN for image super-resolution}},}\ }\href {\doibase
  10.1016/j.knosys.2025.113383} {\bibfield  {journal} {\bibinfo  {journal}
  {Knowledge-Based Syst.}\ }\textbf {\bibinfo {volume} {317}},\ \bibinfo
  {pages} {113383} (\bibinfo {year} {2025})}\BibitemShut {NoStop}%
\bibitem [{\citenamefont {Wang}\ \emph
  {et~al.}(2021{\natexlab{a}})\citenamefont {Wang}, \citenamefont {Xie},
  \citenamefont {Dong},\ and\ \citenamefont {Shan}}]{WangX2021_CONF}%
  \BibitemOpen
  \bibfield  {author} {\bibinfo {author} {\bibfnamefont {X.}~\bibnamefont
  {Wang}}, \bibinfo {author} {\bibfnamefont {L.}~\bibnamefont {Xie}}, \bibinfo
  {author} {\bibfnamefont {C.}~\bibnamefont {Dong}}, \ and\ \bibinfo {author}
  {\bibfnamefont {Y.}~\bibnamefont {Shan}},\ }\bibfield  {title} {\enquote
  {\bibinfo {title} {{Real-ESRGAN: Training Real-World Blind Super-Resolution
  with Pure Synthetic Data}},}\ }in\ \href {http://arxiv.org/abs/2107.10833}
  {\emph {\bibinfo {booktitle} {Proc. IEEE Int. Conf. Comput. Vis.}}},\ Vol.\
  \bibinfo {volume} {2021-Octob}\ (\bibinfo {year} {2021})\ pp.\ \bibinfo
  {pages} {1905--1914}\BibitemShut {NoStop}%
\bibitem [{\citenamefont {Miyato}\ \emph {et~al.}(2018)\citenamefont {Miyato},
  \citenamefont {Kataoka}, \citenamefont {Koyama},\ and\ \citenamefont
  {Yoshida}}]{MiyatoT2018_6ICLRI2-CTP}%
  \BibitemOpen
  \bibfield  {author} {\bibinfo {author} {\bibfnamefont {T.}~\bibnamefont
  {Miyato}}, \bibinfo {author} {\bibfnamefont {T.}~\bibnamefont {Kataoka}},
  \bibinfo {author} {\bibfnamefont {M.}~\bibnamefont {Koyama}}, \ and\ \bibinfo
  {author} {\bibfnamefont {Y.}~\bibnamefont {Yoshida}},\ }\bibfield  {title}
  {\enquote {\bibinfo {title} {{Spectral Normalization for Generative
  Adversarial Networks}},}\ }\href {http://arxiv.org/abs/1802.05957} {\bibfield
   {journal} {\bibinfo  {journal} {6th Int. Conf. Learn. Represent. ICLR 2018 -
  Conf. Track Proc.}\ } (\bibinfo {year} {2018})}\BibitemShut {NoStop}%
\bibitem [{\citenamefont {Nista}\ \emph {et~al.}(2022)\citenamefont {Nista},
  \citenamefont {Schumann}, \citenamefont {Scialabba}, \citenamefont {Grenga},
  \citenamefont {Pitsch},\ and\ \citenamefont {Attili}}]{NistaL2022_ASTFEASF2}%
  \BibitemOpen
  \bibfield  {author} {\bibinfo {author} {\bibfnamefont {L.}~\bibnamefont
  {Nista}}, \bibinfo {author} {\bibfnamefont {C.}~\bibnamefont {Schumann}},
  \bibinfo {author} {\bibfnamefont {G.}~\bibnamefont {Scialabba}}, \bibinfo
  {author} {\bibfnamefont {T.}~\bibnamefont {Grenga}}, \bibinfo {author}
  {\bibfnamefont {H.}~\bibnamefont {Pitsch}}, \ and\ \bibinfo {author}
  {\bibfnamefont {A.}~\bibnamefont {Attili}},\ }\bibfield  {title} {\enquote
  {\bibinfo {title} {{The influence of adversarial training on turbulence
  closure modeling}},}\ }\href {\doibase 10.2514/6.2022-0185} {\bibfield
  {journal} {\bibinfo  {journal} {AIAA Sci. Technol. Forum Expo. AIAA SciTech
  Forum 2022}\ ,\ \bibinfo {pages} {1--9}} (\bibinfo {year}
  {2022})}\BibitemShut {NoStop}%
\bibitem [{\citenamefont {Wang}\ \emph
  {et~al.}(2021{\natexlab{b}})\citenamefont {Wang}, \citenamefont {Xie},
  \citenamefont {Dong},\ and\ \citenamefont {Shan}}]{WangX2021_PIICCV}%
  \BibitemOpen
  \bibfield  {author} {\bibinfo {author} {\bibfnamefont {X.}~\bibnamefont
  {Wang}}, \bibinfo {author} {\bibfnamefont {L.}~\bibnamefont {Xie}}, \bibinfo
  {author} {\bibfnamefont {C.}~\bibnamefont {Dong}}, \ and\ \bibinfo {author}
  {\bibfnamefont {Y.}~\bibnamefont {Shan}},\ }\bibfield  {title} {\enquote
  {\bibinfo {title} {{Real-ESRGAN: Training Real-World Blind Super-Resolution
  with Pure Synthetic Data}},}\ }\href {\doibase 10.1109/ICCVW54120.2021.00217}
  {\bibfield  {journal} {\bibinfo  {journal} {Proc. IEEE Int. Conf. Comput.
  Vis.}\ }\textbf {\bibinfo {volume} {2021-Octob}},\ \bibinfo {pages}
  {1905--1914} (\bibinfo {year} {2021}{\natexlab{b}})}\BibitemShut {NoStop}%
\bibitem [{\citenamefont {Carroll}\ and\ \citenamefont
  {Blanquart}(2013)}]{CarrollP2013_PFa}%
  \BibitemOpen
  \bibfield  {author} {\bibinfo {author} {\bibfnamefont {P.~L.}\ \bibnamefont
  {Carroll}}\ and\ \bibinfo {author} {\bibfnamefont {G.}~\bibnamefont
  {Blanquart}},\ }\bibfield  {title} {\enquote {\bibinfo {title} {{A proposed
  modification to Lundgren's physical space velocity forcing method for
  isotropic turbulence}},}\ }\href {\doibase 10.1063/1.4826315} {\bibfield
  {journal} {\bibinfo  {journal} {Phys. Fluids}\ }\textbf {\bibinfo {volume}
  {25}},\ \bibinfo {pages} {105114} (\bibinfo {year} {2013})}\BibitemShut
  {NoStop}%
\bibitem [{\citenamefont {Maxey}\ and\ \citenamefont
  {Riley}(1983)}]{MaxeyM1983_PF}%
  \BibitemOpen
  \bibfield  {author} {\bibinfo {author} {\bibfnamefont {M.~R.}\ \bibnamefont
  {Maxey}}\ and\ \bibinfo {author} {\bibfnamefont {J.~J.}\ \bibnamefont
  {Riley}},\ }\bibfield  {title} {\enquote {\bibinfo {title} {{Equation of
  motion for a small rigid sphere in a nonuniform flow}},}\ }\href {\doibase
  10.1063/1.864230} {\bibfield  {journal} {\bibinfo  {journal} {Phys. Fluids}\
  }\textbf {\bibinfo {volume} {26}},\ \bibinfo {pages} {883--889} (\bibinfo
  {year} {1983})}\BibitemShut {NoStop}%
\bibitem [{\citenamefont {Bailly}\ and\ \citenamefont
  {Juve}(1999)}]{BaillyC1999_CONF}%
  \BibitemOpen
  \bibfield  {author} {\bibinfo {author} {\bibfnamefont {C.}~\bibnamefont
  {Bailly}}\ and\ \bibinfo {author} {\bibfnamefont {D.}~\bibnamefont {Juve}},\
  }\bibfield  {title} {\enquote {\bibinfo {title} {{A stochastic approach to
  compute subsonic noise using linearized Euler's equations}},}\ }in\ \href
  {\doibase 10.2514/6.1999-1872} {\emph {\bibinfo {booktitle} {5th AIAA/CEAS
  Aeroacoustics Conf. Exhib.}}},\ \bibinfo {series and number} {\bibinfo
  {number} {c}}\ (\bibinfo  {publisher} {American Institute of Aeronautics and
  Astronautics},\ \bibinfo {address} {Reston, Virigina},\ \bibinfo {year}
  {1999})\ pp.\ \bibinfo {pages} {496--506}\BibitemShut {NoStop}%
\bibitem [{\citenamefont {Saad}\ \emph {et~al.}(2017)\citenamefont {Saad},
  \citenamefont {Cline}, \citenamefont {Stoll},\ and\ \citenamefont
  {Sutherland}}]{SaadT2017_AJ}%
  \BibitemOpen
  \bibfield  {author} {\bibinfo {author} {\bibfnamefont {T.}~\bibnamefont
  {Saad}}, \bibinfo {author} {\bibfnamefont {D.}~\bibnamefont {Cline}},
  \bibinfo {author} {\bibfnamefont {R.}~\bibnamefont {Stoll}}, \ and\ \bibinfo
  {author} {\bibfnamefont {J.~C.}\ \bibnamefont {Sutherland}},\ }\bibfield
  {title} {\enquote {\bibinfo {title} {{Scalable Tools for Generating Synthetic
  Isotropic Turbulence with Arbitrary Spectra}},}\ }\href {\doibase
  10.2514/1.J055230} {\bibfield  {journal} {\bibinfo  {journal} {AIAA J.}\
  }\textbf {\bibinfo {volume} {55}},\ \bibinfo {pages} {327--331} (\bibinfo
  {year} {2017})}\BibitemShut {NoStop}%
\bibitem [{\citenamefont {Pope}(2000)}]{PopeS2000_BOOK}%
  \BibitemOpen
  \bibfield  {author} {\bibinfo {author} {\bibfnamefont {S.~B.}\ \bibnamefont
  {Pope}},\ }\href@noop {} {\emph {\bibinfo {title} {{Turbulent Flows}}}},\
  \bibinfo {edition} {1st}\ ed.\ (\bibinfo  {publisher} {Cambridge University
  Press},\ \bibinfo {year} {2000})\BibitemShut {NoStop}%
\bibitem [{\citenamefont {Ristorcelli}(2003)}]{RistorcelliJ2003_PF}%
  \BibitemOpen
  \bibfield  {author} {\bibinfo {author} {\bibfnamefont {J.~R.}\ \bibnamefont
  {Ristorcelli}},\ }\bibfield  {title} {\enquote {\bibinfo {title} {{The
  self-preserving decay of isotropic turbulence: Analytic solutions for energy
  and dissipation}},}\ }\href {\doibase 10.1063/1.1604780} {\bibfield
  {journal} {\bibinfo  {journal} {Phys. Fluids}\ }\textbf {\bibinfo {volume}
  {15}},\ \bibinfo {pages} {3248--3250} (\bibinfo {year} {2003})}\BibitemShut
  {NoStop}%
\bibitem [{\citenamefont {Piomelli}\ \emph {et~al.}(1991)\citenamefont
  {Piomelli}, \citenamefont {Cabot}, \citenamefont {Moin},\ and\ \citenamefont
  {Lee}}]{PiomelliU1991_PFA}%
  \BibitemOpen
  \bibfield  {author} {\bibinfo {author} {\bibfnamefont {U.}~\bibnamefont
  {Piomelli}}, \bibinfo {author} {\bibfnamefont {W.~H.}\ \bibnamefont {Cabot}},
  \bibinfo {author} {\bibfnamefont {P.}~\bibnamefont {Moin}}, \ and\ \bibinfo
  {author} {\bibfnamefont {S.}~\bibnamefont {Lee}},\ }\bibfield  {title}
  {\enquote {\bibinfo {title} {{Subgrid-scale backscatter in turbulent and
  transitional flows}},}\ }\href {\doibase 10.1063/1.857956} {\bibfield
  {journal} {\bibinfo  {journal} {Phys. Fluids A}\ }\textbf {\bibinfo {volume}
  {3}},\ \bibinfo {pages} {1766--1771} (\bibinfo {year} {1991})}\BibitemShut
  {NoStop}%
\end{thebibliography}%

\end{document}